\documentclass[12pt]{article}
\usepackage{geometry} % see geometry.pdf on how to lay out the page. There's lots.
\usepackage{fullpage}
\usepackage{amsmath}
\usepackage{amssymb}
\usepackage{graphicx}
\usepackage{color}
\usepackage{epstopdf}
\usepackage{mathrsfs}
\usepackage{braket}

\usepackage[comma,authoryear]{natbib}
\usepackage{endnotes}

\usepackage{setspace}
\begin{document}
\title{Kuhn Losses Regained: Van Vleck\\ from Spectra to Susceptibilities}

\author{Charles Midwinter and Michel Janssen}

\begin{singlespacing}
\maketitle
\end{singlespacing}

\begin{singlespacing}
\section{Van Vleck's Two Books and the Quantum Revolution}
\end{singlespacing}

\begin{singlespacing}
\subsection{Van Vleck's Trajectory from Spectra to Susceptibilities, 1926-1932}
\end{singlespacing}

\noindent
``The chemist is apt to conceive of the physicist as some one who is so entranced in spectral lines that he closes his eyes to other phenomena'' \citep[p.\ 493]{VanVleck_1928-the-new2}. This observation was made by the  American theoretical physicist John  H.\  Van Vleck (1899--1980) in an article on the new quantum mechanics in {\it Chemical Reviews}.  Only a few years earlier, Van Vleck himself would have fit this characterization of a physicist to a tee. Between 1923 and 1926, as a young assistant professor in Minneapolis, he spent much of his time writing a book-length {\it Bulletin} for the National Research Council (NRC) on the old quantum theory \citep{VanVleck_1926-quantum}. As its title, {\it Quantum Principles and Line Spectra}, suggests, this book deals almost exclusively with spectroscopy.  Only after a seemingly jarring change of focus in his research, a switch to the theory of electric and magnetic susceptibilities in gases, did he come to consider his previous focus myopic.  In 1927--28, now a full professor in Minnesota, he published a three-part paper on susceptibilities in {\it Physical Review} \citep{VanVleck_1927_on-dielectric1, VanVleck_1927_on-dielectric2, VanVleck_1928_on-dielectric}.  This became the basis for a second book, {\it The Theory of Electric and Magnetic Susceptibilities} \citep{VanVleck_1932-the-theory}, which he started to write shortly after he moved to Madison, Wisconsin, in the fall of 1928.

By the time he wrote his article in {\it Chemical Reviews}, Van Vleck had come to recognize that a strong argument against the old and in favor of the new quantum theory could be found in the theory of susceptibilities, a subject of marginal interest during the reign of the old quantum theory.  As he wrote in the first sentence of the preface of his 1932 book:

\begin{singlespacing}
\begin{quotation}
\noindent
The new quantum mechanics  is perhaps most noted for its triumphs in the field of spectroscopy, but its less heralded successes in the theory of electric and magnetic susceptibilities must be regarded as one of its great achievements \citep[p.\ vii]{VanVleck_1932-the-theory}.
\end{quotation}
\end{singlespacing}

\noindent
What especially struck Van Vleck was that, to a large extent, the new quantum mechanics made sense of susceptibilities not by offering new results, but by reinstating classical expressions that the old quantum theory had replaced with erroneous ones. Both in his articles of the late 1920s and in his 1932 book, Van Vleck put great emphasis on this point.%which was only obvious in hindsight.

His favorite example was the value of what he labeled $C$, a constant in the so-called Langevin-Debye formula used for both magnetic and electric susceptibilities \citep{Langevin_1905-magnetisme, Langevin_1905-sur-la-theorie,Debye_1912-einige}.
%that multiplies the term giving their temperature-dependence. 
Its classical value in the case of electric susceptibilities is $1/3$. This turns out to be a remarkably robust result in the classical theory, in the sense that it is largely independent of the model used for molecules with permanent electric dipoles. In the old quantum theory, the value of $C$ was much larger and, more disturbingly, as no experimental data were available to rule out values substantially different from the classical one, extremely sensitive to the choice of model and to the way quantum conditions were imposed. By contrast, the new quantum theory, like the classical theory, under very general conditions gave  $C = 1/3$. Van Vleck saw this regained robustness as an example of what he called ``spectroscopic stability" \citep[p.\ 740]{VanVleck_1927_on-dielectric1}. New experiments now also began to provide empirical evidence for this value and Van Vleck produced new and better proofs for the generality of the result, both in classical theory and in the new quantum mechanics. From this new vantage point, Van Vleck clearly recognized that the instability of the value for $C$ in the old quantum theory had been a largely unheeded indication of its shortcomings. 

The constant $C$ also comes into play if we want to determine the dipole moment $\mu$ of a polar molecule such as HCl.  Given a gas of these molecules, one can calculate $\mu$ using a measurement of the dielectric constant: the greater the value of $C$, the smaller the value of $\mu$. Because of the instability of the value of $C$, \citet{VanVleck_1928-the-new2} pointed out that, ``[t]he electrical moment of the HCl molecule \ldots has had quite a history" (p.\ 494).
\begin{figure}[h]
   \centering
   \includegraphics[width=5in]{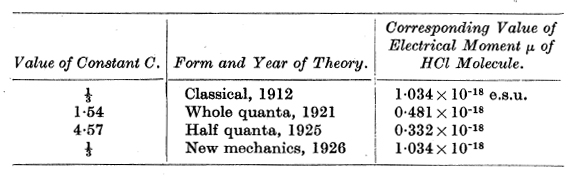} 
   \caption{The values of the constant $C$ in the Langevin-Debye formula and of the electric moment $\mu$ of HCl in classical theory, the old quantum theory, and quantum mechanics \citep[p.\ 494]{VanVleck_1928-the-new2}.}
   \label{fig1}
\end{figure}

Fig.\ \ref{fig1} shows the table with which Van Vleck illustrated this checkered history. The result for whole quanta was found by Wolfgang \citet{Pauli_1921-zur-theorie} while finishing his doctorate in Munich at age 21 \citep[p.\ 61]{Enz_2002-no-time}. Van Vleck, one year older than Pauli, read this paper as a graduate student at Harvard, but, indicative of the prevailing obsession with spectroscopy of the day, it did not make a big impression on him at that time \citep[p.\ 136]{Fellows_1985-j}. The  entry for half quanta is due to Linus \citet{Pauling_1926-the-quantum}, one year younger than Pauli. Although the paper was submitted  in February 1926, Pauling was still using the old quantum theory, which is probably why the year is given as 1925 in Van Vleck's table. As the table shows, $C$ increased by a factor of almost 14 between 1912 and 1926, reducing $\mu$ to a third of its classical value. ``Fortunately [in the new quantum mechanics] the electrical moment of the HCl molecule reverts to its classical 1912 value" \citep[p.\ 494]{VanVleck_1928-the-new2}.  

These observations, including the table, are reprised in his book on susceptibilities \citep[p.\ 107]{VanVleck_1932-the-theory}.  In fact, these fluctuations in the values of $C$ and $\mu$ so impressed Van Vleck that the first two columns of this table can still be found in his 1977 Nobel lecture \citep[p.\ 356]{VanVleck_1992-quantum}.

Van Vleck's 1932 book on susceptibilities was much more successful than his {\it Bulletin} on the old quantum theory, which was released just after the quantum revolution of 1925--26. The {\it Bulletin}, as its author liked to say with characteristic self-deprecation, ``{\it in a sense} was obsolete by the time it was off the press" \citep[p.\ 6, our emphasis]{VanVleck_1971-reminiscences}. The italicized qualification is important. In the late 1920s and early 1930s, physicists could profitably use the {\it Bulletin}  {\it despite} the quantum revolution.
The 1932 book, however, became a classic in the field it helped spawn. Interestingly, given that it grew out of work on susceptibilities in gases, that field is solid-state physics. In a biographical memoir about Van Vleck for the National Academy of Sciences (NAS), condensed-matter icon Philip W.\ Anderson, one of Van Vleck's students, wrote that the book ``set a standard and a style for American solid-state physics that greatly influenced its development during decades to come---for the better" \citep[p.\ 524]{Anderson_1987-john}.\footnote{See also, e.g., \citet[p.\ 1131]{Stevens_1995-magnetism}.} This book and the further research it stimulated would eventually earn Van Vleck  the informal title of ``father of modern magnetism" as well as part of the 1977 Nobel prize, which he shared with Anderson and  Sir Nevill Mott.

In this paper we follow Van Vleck's trajectory from his 1926 {\it Bulletin} on spectra to his 1932 book on susceptibilities.  Both books, as we will see, loosely qualify as textbooks.  As such, they provide valuable insights about the way pedagogical texts written in the midst (the 1926 {\it Bulletin}) or the aftermath (the 1932 book) of a scientific revolution reflect such dramatic upheavals.
%reflect a scientific revolution, while it is in progress (the 1926 {\it Bulletin}) or in its aftermath (the 1932 book).

\begin{singlespacing}
\subsection{Kuhn Losses, Textbooks, and Scientific Revolutions}
\end{singlespacing}

\noindent
The old quantum theory's trouble with susceptibilities, masked by its success with spectra, is a good example of what is known in the history and philosophy of science literature as a {\it Kuhn loss}. Roughly, a Kuhn loss is a success, empirical or theoretical, of a prior theory---or {\it paradigm} as Kuhn would have preferred---that does not carry over to the theory or paradigm that replaced it. As illustrated by the recovery in the new quantum theory of the robust value for the constant $C$ in the Langevin-Debye formula, a feature of the classical theory lost in the old quantum theory, Kuhn losses need not be permanent. As Kuhn himself recognized, they can be regained in subsequent theories or paradigms.

Incidentally, both Thomas S.\ Kuhn and Philip W.\ Anderson completed their Ph.D.'s at Harvard in 1949 with Van Vleck as their advisor.  In the memoir about Van Vleck mentioned above, \citet[p.\ 524]{Anderson_1987-john} wrote that ``[t]he decision to work with him was one of the wiser choices of my life." By contrast,   Kuhn, when asked  in an interview in 1995 why he had chosen to work with Van Vleck, answered: ``I was quite certain that I was not going to take a career in physics \ldots\ Otherwise I would have shot for a chance to work with Julian Schwinger" \citep[p.\ 274]{Baltas_2000_a-discussion}. This is particularly unkind when one recalls that in 1961, the year before the publication of {\it The Structure of Scientific Revolutions}, it was Van Vleck who suggested that his student-turned-historian-and-philosopher-of-science be appointed director of the project that led to the establishment of the Archive for the History of Quantum Physics (AHQP) \citep[p.\ viii; see also Baltas et al., 2000, pp.\ 302--303]{Kuhn_1967_sources}. 

In 1963, Kuhn interviewed his former teacher for the AHQP project. Van Vleck once again emphasized the importance of quantum mechanics having regained the Kuhn losses sustained by the old quantum theory in the area of susceptibilities, this time invoking no less an authority than Niels Bohr:

\begin{singlespacing}
\begin{quotation}
\noindent
I showed that the factor one-third [in the Langevin-Debye formula for susceptibilities] got restored in quantum mechanics, whereas in the old quantum theory, it had all kinds of horrible oscillations \ldots\  you got some wonderful nonsense, whereas it made sense with the new quantum mechanics. I think that was one of the strong arguments for quantum mechanics. One always thinks of its effect and successes in connection with spectroscopy, but I remember Niels Bohr saying that one of the great arguments for quantum mechanics was its success in these non-spectroscopic things such as magnetic and electric susceptibilities (AHQP interview, %October 1963, 
session 2, p.\ 5).\footnote{See also the opening sentence of the preface of Van Vleck's 1932 book quoted in sec.\ 1.1.}

\end{quotation}
\end{singlespacing}

To the best of our knowledge, Kuhn never used the ``wonderful nonsense" Van Vleck is referring to here as an example of a Kuhn loss. Still, one can ask whether the example bears out Kuhn's general claims about Kuhn losses. We will find that it does in some respects but not in others. For instance, contrary to claims by Kuhn in {\it Structure} about how scientific revolutions are papered over in subsequent textbooks, the prehistory of the theory of susceptibilities, including the Kuhn loss  the old quantum theory suffered in this area, is dealt with {\it at length} in Van Vleck's 1932 book.  However, we will also see that, in at least one important respect, Van Vleck's version of this prehistory is a little misleading and perhaps even a tad self-serving,  
%one might even say a tad Whiggish. 
which is just what Kuhn would have led us to expect. In general, there is much of value in Kuhn's account, which thus provides a good starting point for our analysis. Ultimately, our goal is not to argue for or against Kuhn but to use the fine structure of the quantum revolution to learn more about the structure of scientific revolutions in general.\footnote{We thus use Kuhn's work in the same spirit as Michael \citet[p.\ 62]{Ruse1989} in an essay on the plate-tectonics revolution in geology.}

\begin{singlespacing}
\subsubsection{Kuhn Losses}
\end{singlespacing}

\noindent
The concept of a Kuhn loss, though obviously not the term, is introduced in Ch.\ 9 of {\it Structure} \citep[pp.\ 103--110; page numbers refer to the 3rd edition]{Kuhn_1996_the-structure}.  To underscore that science does not develop cumulatively, Kuhn noted that in going from one paradigm to another there tend to be gains as well as losses.  ``[P]aradigm debates," he wrote, ``always involve the question: Which problems is it more significant to have solved?" (ibid., 110). 

In the transition from classical theory to the old quantum theory, gains in spectroscopy apparently outweighed losses in the theory of susceptibilities just as, at least until the early 1920s, they outweighed losses in dispersion theory. The former Kuhn loss was only regained in the new quantum theory,\footnote{As we will see in sec.\ 4, there were four papers published in 1926 all reporting the recovery of $C = 1/3$ in the new quantum theory. As Van Vleck wrote in the conclusion of the one submitted first but published last: ``This is a much more satisfactory result than in the older version of the quantum theory, in which both the calculations of Pauli [1921] with whole quanta \ldots\ and of Pauling [1926] with half quanta yielded results diverging from the classical Langevin theory even at high temperatures" \citep[p.\ 227]{VanVleck_1926_magnetic}.\label{van vleck 1926}} while the latter was recovered in the dispersion theory of Hendrik A.\ (Hans) \citet{Kramers_1924-the-law, Kramers_1924-the-quantum}.  Kramers' dispersion theory was formulated in the context of the old quantum theory of Bohr and Arnold Sommerfeld but quickly incorporated into the infamous BKS theory of Bohr, Kramers, and John C.\ Slater (1924), a short-lived quantum theory of radiation in which a number of fundamental tenets of the Bohr-Sommerfeld theory were abandoned \citep[secs.\ 3--4]{Duncan_2007-on-the-verge}.

Strictly speaking, of course, when we talk about Kuhn losses and their recovery, we should be talking about paradigms rather than theories. Kuhn exegesists, however, will forgive us, we hope, for proceeding on the assumption that a theory can be construed as a key component of a paradigm or a {\it disciplinary matrix}, the term Kuhn in his 1969 postscript to {\it Structure}  proposed to substitute for the term `paradigm' when used in the sense in which we need it here \citep[p.\ 182]{Kuhn_1996_the-structure}. Granted that assumption, we can continue to talk about Kuhn losses in transitions from one {\it theory} to another.

Although they are both Kuhn losses of the old quantum theory, the one in susceptibility theory is of a different kind than the one in dispersion theory. In the case of dispersion, there was clear experimental evidence all along for the key feature of the classical theory that was lost in the old quantum theory and recovered in the Kramers dispersion theory. In the case of susceptibility theory, as we mentioned above, experimental evidence for the key feature of the classical theory that was lost in the old quantum theory only became available {\it after} it was recovered in the new quantum theory. 

The key feature in the case of dispersion is that anomalous dispersion---the phenomenon that in certain frequency ranges the index of refraction gets smaller rather than larger with increasing frequency\footnote{For a brief discussion of this phenomenon and its discovery in the 19th century, see \citet[p.\ 233, note 1]{Buchwald_1985-from-maxwell}.}---occurs around the absorption frequencies of the dispersive medium. This is in accordance with the classical dispersion theories of Hermann von Helmholtz, Hendrik A.\ Lorentz, and Paul Drude \citep[pp.\ 575--576]{Duncan_2007-on-the-verge}. However, in the dispersion theories of  \citet{Sommerfeld_1915-die-allgemeine, Sommerfeld_1917-die-drudesche}, Peter \citet{Debye_1915-die-konstitution}, and Clinton J.\ \citet{Davisson_1916-the-dispersion}, based on the Bohr model of the atom, dispersion is anomalous around the orbital frequencies of the electrons, which differ sharply from the absorption frequencies of the atom except in the limit of high quantum numbers. As one would expect in the case of a Kuhn loss, proponents of the Sommerfeld-Debye-Davisson theory had a tendency to close their eyes to this problem.  Others, however, including Bohr himself, raised it as serious objection early on. A few years before  \citet{Kramers_1924-the-law, Kramers_1924-the-quantum}, building on work by Rudolf Ladenburg and Fritz Reiche \citep{Ladenburg_1921-untersuchungen, Ladenburg_1923-absorption}, eventually solved the problem, Paul S.\ Epstein sharply criticized the  Sommerfeld-Debye-Davisson theory on this score in a paper with the subtitle ``Critical comments on dispersion:"

\begin{singlespacing}
\begin{quotation}
\noindent
[T]he positions of maximal dispersion and absorption do not lie at the position of the emission lines of hydrogen but at the position of the mechanical frequencies of the model \ldots\ {\it the conclusion seems  unavoidable to us  that the foundations of the Debye-Davysson} [sic] {\it theory are incorrect} \citep[pp.\ 107--108; emphasis in the original; quoted and discussed by Duncan and Janssen, 2007, pp.\ 580--581]{Epstein_1922-die-stoerungsrechnung}.
\end{quotation}
\end{singlespacing}

By contrast, it was only {\it after} the new quantum theory had restored the classical value $C=1/3$ in the Langevin-Debye formula for electric susceptibilities that the ``horrible oscillations" in the old quantum theory came to be seen as the ``wonderful nonsense" Van Vleck made them out to be. When \citet{Pauli_1921-zur-theorie}, for instance, first found a deviation from  $C=1/3$, he did not blink an eye. He just stated matter-of-factly that ``the numerical factor in the final formula for the polarization depends on the specific model \ldots\ while in the classical theory the Maxwell distribution and with it the numerical factor $1/3$ hold generally" \citep[p.\ 325]{Pauli_1921-zur-theorie}. In the conclusion of his paper, Pauli exhorted experimentalists to measure the temperature-dependence of the dielectric constant of hydrogen halides such as HCl, adding that this ``should not pose any particular difficulties" (ibid., p.\ 327). Noting that his quantum theory predicted a much smaller value for the electric dipole moment $\mu$ of HCl than the classical theory ($\mu_{\rm classical} = 2.1471 \, \mu_{\rm quantum}$; cf.\ the table in Fig.\ \ref{fig1}), he suggested that this might provide a way to decide between the two theories. The distance between the two nuclei in, say, a HCl molecule could accurately be determined on the basis of 
%infrared absorption 
spectroscopic data. This distance, Pauli argued, gives an upper bound on the dipole length $d = \mu/e$ between the charges $+e$ and $-e$ forming the dipole in this case. Hence, he concluded, ``{\it if the classical formula for the dielectric constant gives a dipole length that is greater than the nuclear separation extracted from infrared spectra, the formula must be rejected}\," \citep[p.\ 327, emphasis in the original]{Pauli_1921-zur-theorie}. Three years later, the experimentalist C.\ T.\ \citet{Zahn_1924-the-electric} took up Pauli's challenge, but came to the disappointing conclusion that ``[t]he upper limit for the moment given by the infrared absorption data for HCl \ldots\ is 6 times the classical value and 13 times the quantum value and hence does not decide between the two theories" (p.\ 400). Van Vleck's own citations to the experimental literature in his 1932 book strongly suggest that it was only in the period following the quantum revolution of 1925--26 that reliable data in favor of the value $C=1/3$ became available \citep[p.\ 61]{VanVleck_1932-the-theory}. The Kuhn loss in the theory of susceptibilities emphasized by Van Vleck is thus the loss of a theoretical feature that in hindsight proved to be empirically correct, not, as in the case of the Kuhn loss in dispersion theory, a loss of empirical adequacy in some area. Van Vleck's most persuasive argument against the results of Pauli and Pauling was  that they deviated from the classical result even at high temperatures.
%where the old quantum theory was expected to agree with the classical theory. 
As he put it in his 1932 book: ``the correspondence principle led us to expect usually an asymptotic connection of the classical and quantum results at high temperatures" \citep[p.\ 107, see also the quotation in note \ref{van vleck 1926}]{VanVleck_1932-the-theory}.\footnote{Incidentally, Zahn, who concluded in 1924 that experiment could not decide between the classical formula for the temperature-dependence of electric susceptibilities and Pauli's new quantum formula, is one of the two physicists who showed over a decade later that experiments on the velocity-dependence of the electron mass in the early years of the century could not decide between the theoretical predictions of Albert Einstein's special theory of relativity and Lorentz's ether theory, on the one hand, and Max Abraham's so-called electromagnetic view of nature, on the other \citep{Zahn_1938-a-critical}. As one of us has argued, the proponents of these competing theories, though paying lip service to the experimental results, especially when they favored their own theories, put much more stock in theoretical arguments \citep[pp.\ 105--108]{Janssen_2007-from-classical}. When, for instance, Alfred H.\ Bucherer presented new data favoring Lorentz and Einstein at the same annual meeting of German Physical Scientists and Physicians  in Cologne in 1908 where Hermann Minkowski  gave his now famous talk on the geometrical underpinnings of special relativity, Minkowski, while welcoming Bucherer's new data, dismissed Abraham's theory on purely theoretical grounds. He called Abraham's model of a rigid electron, not subject to length contraction, a ``monster" and ``no working hypothesis, but a working hindrance," and described Abraham's insertion of this model into classical electrodynamics as going to a concert wearing ear plugs (ibid., p.\ 88)! This is reminiscent of how Van Vleck dismissed results derived by the likes of  Pauli and Pauling in the old quantum theory  as ``wonderful nonsense." As we will see in sec.\ 5.2, Van Vleck heaped more scorn on the treatment of susceptibilities in the old quantum theory in his 1932 book \citep[Ch.\ V]{VanVleck_1932-the-theory}.\label{minkowski}}
%Abusive language heaped on old quantum theory treatment of susceptinilities by Van Vleck "the bugbear" (used earlier, I noticed, in VV 1927b, p. 37) of weak vs. strong spatial quantization in sec. 9 (with the coup de grace in the last paragraph on p. 111: mystifying, hazy, alarmed). Same story in the next section (sec. 31): "ludicrously large", negative "an absurdity"

Kuhn losses come in a variety of forms. In most of Kuhn's own examples,  what is lost (and sometimes regained) in successive paradigm shifts are certain types of accounts of phenomena deemed acceptable in a paradigm. In the one example to which he devotes more than a paragraph, \citet[pp.\ 104--106]{Kuhn_1996_the-structure} argues, for instance, that the Newtonian notion of gravity as an innate attraction between particles can be seen as a ``reversion (which is not the same as a retrogression)" to the kind of scholastic essences that proponents of the mechanical tradition earlier in the 17th century thought they had banished from science for good. Although our examples involve different components of the disciplinary matrix (empirical adequacy, features attractive on theoretical grounds), quantum mechanics can likewise be said to have brought about a reversion but not a retrogression to classical theory in the cases of dispersion and susceptibilities.

\begin{singlespacing}
\subsubsection{Textbooks and Kuhn Losses}
\end{singlespacing}

\noindent
\citet[Ch.\ 11]{Kuhn_1996_the-structure} famously identified textbooks as the main culprit in rendering the disruption of normal science by scientific revolutions invisible. Textbooks, he argued, by their very nature must present science as a cumulative enterprise. This means that Kuhn losses must be swept under the rug. Textbooks, he wrote,

\begin{singlespacing}
\begin{quotation}
\noindent
address themselves to an already articulated body of problems, data, and theory, most often to  the particular set of paradigms$^{\rm [ }$\footnote{As we will see below, `paradigm' is used here in the sense for which \citet[187]{Kuhn_1996_the-structure} later introduced the term `exemplar'.}$^{\rm ]}$   to which the scientific community is committed at the time they are written \ldots\  [B]eing pedagogic vehicles for the perpetuation of normal science \ldots\  [they] have to be rewritten in the aftermath of each scientific revolution, and, once rewritten, they inevitably disguise not only the role but the very existence of the revolutions that produced them \ldots\  [thereby] truncating the scientist's sense of his discipline's history \citep[pp.\ 136--137]{Kuhn_1996_the-structure}.
\end{quotation}
\end{singlespacing}

\noindent
When he wrote this passage, Kuhn was probably thinking first and foremost of modern science textbooks at both the undergraduate and the graduate level. Given the scope of the general claims in {\it Structure}, however, his claims about textbooks had better hold up for books used as such in the period and the field we are considering. 

The two monographs by Van Vleck examined in this paper would seem to qualify as (graduate) textbooks even though under a strict and narrow definition of the genre they might not. Most of their actual readers may have been research scientists but they were written with the needs of students in mind and both books saw classroom use, albeit limited. Student notes for a two-semester course on quantum mechanics that Van Vleck offered in Wisconsin in 1930--31 show that, despite the quantum revolution that had supposedly made it obsolete four years earlier, Van Vleck was still using his NRC {\it Bulletin} as the main reference for almost two-thirds of the first semester.\footnote{These notes, taken by Ralph P. Winch, have been deposited at the Niels Bohr Library \& Archives of the American Institute of Physics in College Park, Maryland. Notes for a course in 1927--28 in Minnesota, taken by Robert B.\ Whitney and not nearly as meticulous as Winch's, also contain numerous references to Van Vleck's {\it Bulletin}. A full photocopy of these notes was obtained by Roger Stuewer, who kindly made them available to us (accompanying this photocopy is a letter from Barbara Buck to Roger Stuewer, December 9, 1977, detailing its provenance). \label{student notes}} It is unclear whether Van Vleck himself ever used his 1932 book on susceptibilities in his classes. However, one of his colleagues at Wisconsin, Ragnar Rollefson, told Van Vleck's biographer Fred \citet[p.\ 264]{Fellows_1985-j} that he had occasionally used the lengthy Ch.\ VI, ``Quantum-mechanical foundations," which includes a thorough discussion of quantum perturbation theory, in his courses on quantum mechanics.\footnote{In his  review of the 1932 book in {\it Die Naturwissenschaften}, \citet{Pauli_1933-review} wrote: ``One can say that it has the character in part of a handbook and in part of a textbook. The former aspect is expressed in the exhaustive discussion of all questions of detail, the latter in that the foundations of the theory are also presented."  In summary, he wrote: ``Both for learning the theory of the field covered and for an authoritative introduction to the details the book can be most warmly recommended" (ibid.). Similiarly, \citet[p.\ 4121]{Pauling_1932-review} wrote in his review: ``The book is characterized by clear exposition and interesting style, which combined with the sound and reliable treatment, should make it a valuable text for an advanced course, as well as the authoritative reference book in the field." \label{pauli pauling review}} 

So one can reasonably ask how well Van Vleck's two books fit with Kuhn's seductive picture of how the regrouping of a scientific community in response to a scientific revolution is reflected in the textbooks it produces.
%, or, in the case of the NRC {\it Bulletin}, produced while the revolution was still in progress. 
It will be helpful to separate two aspects of this picture: how textbooks delineate and orient further work in their (sub-)disciplines, and how, in doing so, they inevitably distort the prehistory of these (sub-)disciplines and paper over Kuhn losses. 

%Before we address this question, it will be helpful to separate two elements in Kuhn's picture. The first is about how textbooks delineate and orient further work in their (sub-)disciplines. The second is about how, in doing so, they  inevitably distort the prehistory of these (sub-)disciplines and paper over Kuhn losses. On the former count, Van Vleck's books by and large bear out Kuhn's expectations; on the latter, they do not, at least not in the short run. Interestingly, the 1932 book would have in the long run, if Van Vleck had followed through on his plan late in life to publish a revised edition. This last wrinkle notwithstanding, the final two clauses of the passage from {\it Structure} quoted above (``inevitably disguise \ldots" and ``truncating \ldots") are clearly too strong.

Van Vleck's NRC {\it Bulletin} confirms several of his former student's generalizations about textbooks. The {\it Bulletin} is organized around the correspondence principle as a strategy for tackling problems mostly in atomic spectroscopy. Van Vleck thus took the approach he, Kramers, Max Born and others at the research frontier of the old quantum theory had adopted around 1924 and fixated that approach in a book meant to initiate others in the field. Putting these correspondence-principle techniques and the problems amenable to them at the center of his presentation and relegating work along different lines or in other areas to the periphery, Van Vleck clearly identified and promoted what he thought was and should be the core pursuit of the old quantum theory. 

Those engaged in work that was marginalized in this way predictably took exception. In a review of the {\it Bulletin}, one such colleague, Adolf Smekal, complained about Van Vleck's organization of the material. Smekal recognized that some organizing principle was needed given the sheer quantity of material to be covered but he did not care for the choices Van Vleck had made:

\begin{singlespacing}
\begin{quotation}
\noindent
Selection of, arrangement of, and space devoted to the offerings is heavily influenced by subjective viewpoints and cannot win every reader's approval everywhere. Instead of the presumably available option of letting all fundamental connections emerge systematically, the author has preferred to put up front what is felt to be the internally most unified part of the quantum theory as it has developed so far, followed by more or less isolated applications to specific problems \citep[p.\ 63]{Smekal_1927-review}.
\end{quotation}
\end{singlespacing}

The way in which  correspondence-principle techniques take center stage in Van Vleck's book provides a nice example of how textbooks transmit what Kuhn in the postscript to {\it Structure} called {\it exemplars}, the ``entirely appropriate [meaning] both philologically and autobiographically" of the term  `paradigm' \citep[pp.\ 186--187]{Kuhn_1996_the-structure}. By an exemplar, Kuhn wrote,

\begin{singlespacing}
\begin{quotation}
\noindent
 I mean, initially, the concrete problem solutions that students encounter from the start of their scientific education, whether in laboratories, on examinations  or at the ends of chapters in science texts. To these shared examples should, however, be added at least some of the technical problem-solutions found in the periodical literature that scientists encounter during their post-educational research careers and that also show them by example how their job is to be done \citep[p.\ 187]{Kuhn_1996_the-structure}.
\end{quotation}
\end{singlespacing}

\noindent
Van Vleck's {\it Bulletin} presented such ``technical problem-solutions found in the periodical literature" in a more didactic text that should help its readers become active contributors to this literature themselves. 

Confirming another article of Kuhnian doctrine, the problem with susceptibilities, a Kuhn loss in the transition from classical theory to the old quantum theory, is not mentioned anywhere in the {\it Bulletin}. Van Vleck may have forgotten about the problem but there is clear evidence that he had been aware of it earlier. In a term paper of 1921,  entitled ``Theories of magnetism," for a course he took with Percy W.\ Bridgman as a graduate student at Harvard, Van Vleck touched on the paper in which \citet{Pauli_1921-zur-theorie} derived the entry $C=1.54$ for whole quanta in the table in Fig.\  \ref{fig1}  \citep[p.\ 136]{Fellows_1985-j}. 

Whereas the {\it Bulletin} passes over the Kuhn loss in the theory of susceptibilities in silence, the Kuhn loss in dispersion theory in that same transition is flagged prominently. It is easy to understand why. By the time Van Vleck wrote his {\it Bulletin}, \citet{Kramers_1924-the-law, Kramers_1924-the-quantum} had already recovered that Kuhn loss with his new dispersion formula. Moreover, as we will see in sec.\ 3.2, this formula was one of the striking successes of the correspondence-principle approach central to the book. 
%\citep[cf.\ note \ref{duncan janssen dispersion} below]{Duncan_2007-on-the-verge}. 
Van Vleck thus could and did use the recovered Kuhn loss in dispersion theory to promote this approach. 

In his 1932 book, as we will see in sec.\ 5.2, Van Vleck made even more elaborate use of the recovered Kuhn loss in susceptibility theory to promote his new quantum-mechanical treatment of susceptibilities.
%than he had done with the recovered Kuhn loss in dispersion theory in the {\it Bulletin}. 
He devoted a whole chapter of the book to the problems of the old quantum theory in this area. Of course, the Kuhn loss in susceptibility theory was regained only after a major theory change. The difference between the two cases, however, is smaller than one might initially think. The BKS theory \citep{Bohr_1924-the-quantum} into which Kramers' dispersion formula was quickly integrated constituted such a radical departure from Bohr's original theory that it might well have been remembered as a completely new theory had it not been so short-lived \citep[pp.\ 597--613]{Duncan_2007-on-the-verge}.

Like the {\it Bulletin}, the 1932 book provided its readers with all the tools they needed to become researchers in the field it so masterfully mapped out for them. Had the correspondence-principle approach to atomic physics been moribund by the time the {\it Bulletin} saw print, the approach to electric and magnetic susceptibilities championed in the 1932 book would prove to be remarkably fruitful.

\begin{singlespacing}
\subsubsection{Continuity and Discontinuity in Scientific Revolutions}
\end{singlespacing}

\noindent
A couple of Kuhn losses proudly displayed rather than swept under the rug in a pair of books that only broadly qualify as textbooks may not seem like much of a threat to Kuhn's general account of how textbooks make scientific revolutions invisible. But they do point, we believe, to a more serious underlying issue. Van Vleck managed to write two books that  equipped their readers with the tools they needed to start doing the kind of research their author envisioned themselves {\it without} the kind of wholesale distortion and suppression of the prehistory of their subject matter that Kuhn claimed are unavoidable. That is not to say that such distortion and suppression were or could have been completely avoided. 

The 1932 book provides the clearest example of this. As mentioned above, Van Vleck devoted an entire chapter to the old quantum theory, putting the problems it ran into with susceptibilities on full display.
%\citep[Ch.\ V]{VanVleck_1932-the-theory}. 
Yet he conveniently neglected to mention that there had been no clear empirical evidence exposing these problems.

Smekal's review of the NRC {\it Bulletin} suggests that in 1926 Van Vleck did not completely steer clear of distorting the history of his subject either. Smekal had been championing an alternative dispersion theory, which he complained was ``completely misunderstood and distorted" \citep[p.\ 63]{Smekal_1927-review} in the one paragraph  \citet[p.\ 159]{VanVleck_1926-quantum} devoted to it. Whether or not this complaint was well-founded, it would have been counterproductive in terms of Van Vleck's pedagogical objectives to cover Smekal's and other competing theories of dispersion to their proponents' satisfaction. 

That said, there were many elements in older theories that helped rather than hindered Van Vleck in achieving these objectives. As a result, much of the continuity that can be discerned in the discussions of classical theory and quantum theory in the NRC {\it Bulletin} is  {\it not}, as Kuhn would have it, an artifact of how history is inevitably rewritten in textbooks, but actually matches the historical record tolerably well. Despite its misleading treatment of the experimental state of affairs in the early 1920s, the same can be said about the 1932 book. The final two clauses of the passage from {\it Structure} quoted above (``inevitably disguise \ldots" and ``truncating \ldots") are clearly too strong.

On the Kuhnian picture of scientific revolutions as paradigm shifts akin to Gestalt switches, it is hard to understand how a post-revolutionary textbook could make the prehistory of its subject matter look  more or less continuous and thereby perfectly suitable to its pedagogical objectives {\it without} seriously  disguising, distorting, and truncating that prehistory. An important part of the explanation, at least in the case of these two books by Van Vleck, is the continuity of mathematical techniques through the conceptual upheavals that mark the transition from classical theory to the old quantum theory, and finally to modern quantum mechanics. 

In his recent book, {\it Crafting the Quantum}, on the Sommerfeld school in theoretical physics, Suman \citet{Seth_2010-crafting} makes a similar point. He reconciles the continuous and the discontinuous aspects of the development of quantum theory in the 1920s by emphasizing, as we do, the continuity of mathematical techniques. Scientific revolutions, he writes, ``are revolutions of conceptual foundations, not of puzzle-solving techniques. Most simply: Science sees revolutions of principles, not of problems" \citep[p.\ 268]{Seth_2010-crafting}. To illustrate his point, Seth quotes Arnold Sommerfeld, who wrote 
%in the first edition of his supplement on wave mechanics ({\it Wellenmechanischer Erg\"anzungsband}) to {\it Atombau und Spekrallinien} 
in 1929: ``The new development does not signify a revolution, but a joyful advancement of what was already in existence, with many fundamental clarifications and sharpenings" (ibid., p.\ 266).
%\label{seth 1}} 

Given the radical conceptual changes involved in the transition from classical physics to quantum physics, it is important to keep in mind that there was at the same time great continuity of mathematical structure in this transition. Both the old quantum theory and matrix mechanics, for instance, retain, in a sense, the laws of classical physics. The old quantum theory just put some additional constraints on the motions allowed by Newtonian mechanics. The basic idea of matrix mechanics, as reflected in the term {\it Umdeutung} (reinterpretation) in the title of the paper with which Werner \citet{Heisenberg_1925-ueber-die-quantentheoretische} laid the basis for the new theory, was not to {\it repeal} the laws of mechanics but to {\it reinterpret} them. Heisenberg took the quantities related by these laws to be arrays of numbers, soon to be recognized as matrices \citep{Duncan_2007-on-the-verge, Duncan_2008-Pascual}. It is this continuity of mathematical structure that undergirds the continued effectiveness of the mathematical tools wielded in the context of these different theories. 

In the old quantum theory, techniques from perturbation theory in celestial mechanics were used to analyze electron orbits in atoms classically as a prelude to the translation of the results into quantum formulas under the guidance of the correspondence principle  \citep[pp.\ 592--593, pp.\ 627--637]{Duncan_2007-on-the-verge}. This is the procedure that led Kramers to his dispersion formula. It is also the procedure that \citet{VanVleck_1924-the-absorption1,VanVleck_1924-the-absorption2} followed in his early research and made central to his exposition of the old quantum theory in the 1926 {\it Bulletin}. It inspired the closely related perturbation techniques in matrix mechanics developed in the famous {\it Dreim\"annerarbeit} of Born, Heisenberg, and Pascual Jordan (1926). In his papers of the late 1920s and in his 1932 book, Van Vleck adapted these perturbation techniques to the treatment of susceptibilities. A reader comparing Van Vleck's books of 1926 and 1932 is probably struck first by the shift from spectra to susceptibilities. Underlying that discontinuity, however, is the continuity in these perturbation techniques, made possible by the survival of much of the structure of classical mechanics in both the old and the new quantum theory. These techniques actually fit Kuhn's definition of an exemplar very nicely, even though they cut across what by Kuhn's reckoning are two major paradigm shifts.

One way to highlight the continuity of Van Vleck's trajectory from spectra to susceptibilities is to note that the derivation of the Kramers dispersion formula, a prime example of Van Vleck's approach in his NRC {\it Bulletin} on the old quantum theory, and the derivation of the Langevin-Debye formula for electric susceptibilities, central to his classic of early solid-state physics, both involve applications of canonical perturbation theory in action-angle variables to calculate the electric moment of a multiply-periodic system in an external electric field. The main difference is that in the case of dispersion we are interested in the instantaneous value of the electric moment of individual multiply-periodic systems in response to the periodically changing electric field of an incoming electromagnetic wave, whereas in the case of susceptibilities we are interested in thermal ensemble averages of the electric moments of many such systems averaged over the periods of their motion in response to a constant external field (cf.\ secs.\ 3.2 and 5.2 and note \ref{1927 Kramers dispersion formula}).

The remarkable continuity of mathematical structures and techniques in the transitions from classical theory to the old quantum theory to modern quantum mechanics makes it perfectly understandable that Van Vleck could still use his 1926 {\it Bulletin} in his courses on quantum mechanics in the early 1930s. It also explains how Van Vleck could make such rapid progress once he hit upon the problem of susceptibilities not long after he completed the {\it Bulletin} and mastered matrix mechanics.  

Kuhn had a tendency to see only discontinuity in paradigm shifts. This intense focus on discontinuity is what lies behind his fascination with Kuhn losses. It also made him overly suspicious of the seemingly continuous theoretical developments presented in science textbooks. The analysis of Van Vleck's 1926 and 1932 books and of his trajectory from one to the other provides an important corrective to the discontinuity bias in Kuhn's stimulating and valuable observations about Kuhn losses and textbooks and will thus, we hope, contribute to a more nuanced understanding of the role of the textbooks in shaping and sustaining (sub-)disciplines in science.

%One last observation. 
Whether one sees continuity or discontinuity in the transition from classical physics to quantum physics depends, to a large extent, on one's perspective. The historian trying to follow the events as they unfolded on the ground, will probably mainly see continuities. The historian who takes a bird's eye view and compares the landscapes before and after the transition will most likely be struck first and foremost by discontinuities.
%\footnote{As Suman \citet[p.\ 266]{Seth_2010-crafting} points out in the book mentioned in the preceding note, 
%\ref{seth 1}, 
%Kuhn, in the interviews he conducted for the AHQP, had a tendency to ask leading questions about the onset of a Kuhnian crisis in quantum theory in the early 1920s, thus obtaining an exaggerated picture of how widespread the feeling of crisis was among practitioners at the time. As an important corrective to this picture, Seth quotes Arnold Sommerfeld who wrote in the first edition of his supplement on wave mechanics ({\it Wellenmechanischer Erg\"anzungsband}) to {\it Atombau und Spekrallinien} in 1929: ``The new development does not signify a revolution, but a joyful advancement of what was already in existence, with many fundamental clarifications and sharpenings" (ibid.)\label{seth 2}} 
A final twist in our story about the recovered Kuhn loss in Van Vleck's 1932 book nicely 
%a twist  already alluded to above,
illustrates this difference in perspective.  

Van Vleck covered the troublesome recent history of its subject matter in Ch.\ V, ``Susceptibilities in the old quantum theory contrasted with the new." This chapter, as we will show in more detail in sec.\ 5.2, allows us to see important elements of continuity in the transition from the old to the new quantum theory. Toward the end of his life, Van Vleck began revising his 1932 classic with the idea of publishing a new edition \citep[p.\ 258, pp.\ 262--263, p.\ 266]{Fellows_1985-j}.\footnote{We are grateful to David Huber and Chun Lin at the University of Wisconsin--Madison, two of Van Vleck's students,
%and co-authors (see, e.g., \citet{VanVleck_1977-absorption}), 
for providing us with copies of these revisions.\label{huber}} Wanting to add a chapter on modern developments without changing the total number of chapters, he intended to cut Ch.\ V on the grounds that  by then it only had historical value.\footnote{In the never completed manuscript of the revised edition, all of Ch.\ V was ``reduced to  a single section of four typewritten pages" \citep[p.\ 263]{Fellows_1985-j}.} Even in 1932 he began the chapter apologizing to his readers that ``it may seem like unburying the dead to devote a chapter to the old quantum theory" \citep[p.\ 105]{VanVleck_1932-the-theory}. Note also the one reservation \citet[p.\ 509]{Anderson_1987-john} expressed about the book in his NAS memoir: ``ÒIt is marked---perhaps even slightly marred, as a modern text for physicists poorly trained in classical mechanics---by careful discussion of the ways in which quantum mechanics, the old
quantum theory, and classical physics differ." As it happened, the new edition of the book never saw the light of day, but if it had, it would have been a confirming instance of an amended version of Kuhn's thesis, namely that, going through multiple editions, textbooks {\it eventually} suppress or at least sanitize the history of their subject matter and paper over Kuhn losses, especially those that turn out to have been only temporary.

\subsection{Van Vleck as Teacher}

%As we hope to have made clear in the preceding subsection, our main focus is not on Van Vleck's books as pedagogical tools, but on the specific ways he marshaled concepts and techniques in the service of new research programs.  Nevertheless, 
Although it will be clear from the preceding subsection that our main focus in this paper is not on Van Vleck's books as pedagogical tools,
it seems appropriate to devote a short subsection to Van Vleck as a teacher.

A good place to start is to compare testimony by Anderson and Kuhn, Van Vleck's unlikely pair of graduate students at Harvard in the late 1940s. In his NAS memoir about Van Vleck, Anderson offered the following somewhat back-handed compliment: 

\begin{singlespacing}
\begin{quotation}
\noindent
By the 1940s \dots\ his teaching style had become unique, and is remembered with fondness by everyone I spoke to. Most of the material was written in his inimitable scrawl on the board \ldots\  Especially in group theory [taught from \citet{Wigner_1931-Gruppentheorie} in the original German], his intuitive feeling for the subject often bewildered us as he 
scribbled \ldots\  in an offhand shorthand to demonstrate what we thought were exceedingly abstruse points  \citep[p.\ 524]{Anderson_1987-john}.
\end{quotation}
\end{singlespacing}

\begin{figure}[h]
   \centering
   \includegraphics[width=5.6in]{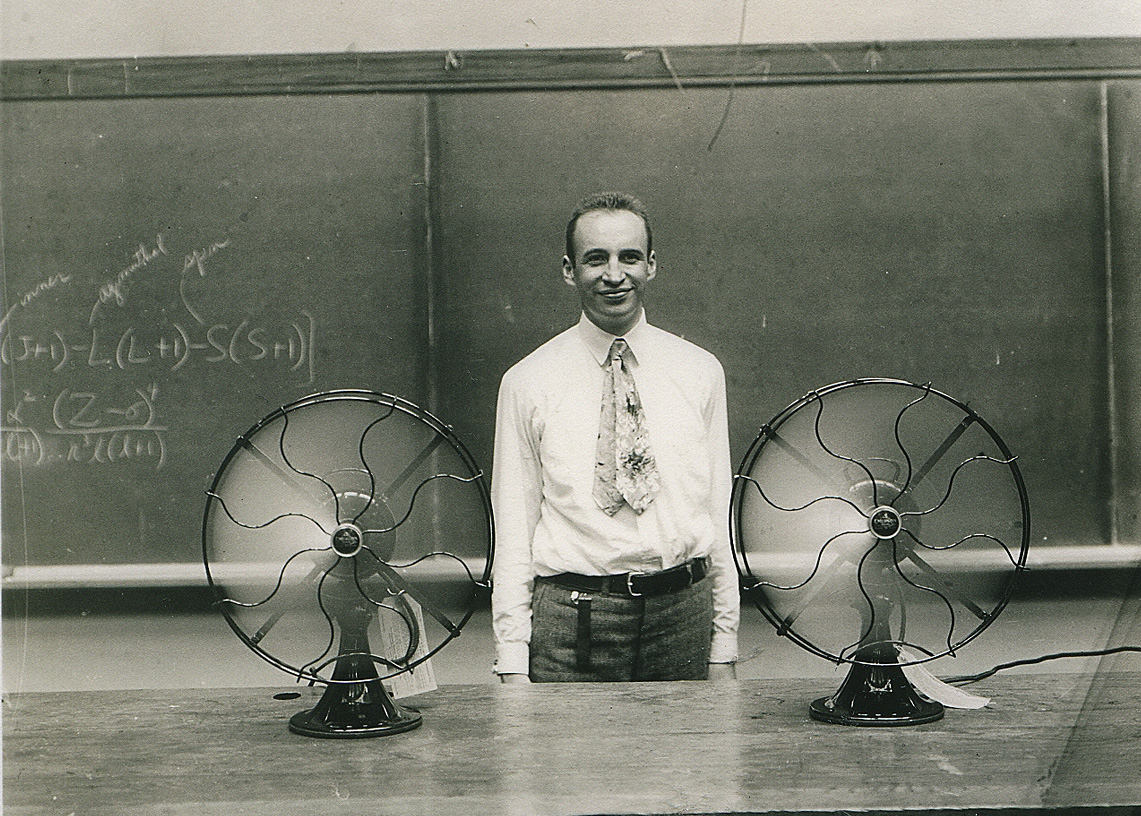} 
   \caption{Van between two fans at 1300 Sterling Hall, University of Wisconsin--Madison, ca.\ 1930 (picture courtesy of John Comstock).}
   \label{fig2}
\end{figure}

\noindent
Anderson's assessment is actually consistent with Kuhn's, even though the latter evidently did not share his fellow student's enthusiasm for the unique style of their advisor: ``One of the courses that I then took was group theory with Van Vleck. And I found that somewhat confusing \ldots\  Van Vleck was not a terribly good teacher" \citep[p.\ 272]{Baltas_2000_a-discussion}. 
%As Anderson surmised, 
Van Vleck's teaching style must have been less idiosyncratic in his earlier years. 
%This suspicion is confirmed by what 
As Robert Serber, who studied with Van Vleck in Madison in the early 1930s (cf.\ Fig.\ \ref{fig2}), wrote in the preface to his famous {\it Los Alamos Primer}:

%FF 294-295

\begin{singlespacing}
\begin{quotation}
\noindent
John Van Vleck was my professor at Wisconsin. The first year I was there he  gave a course in quantum mechanics.  No one wanted to take a degree that year. Everyone put it off because it was useless---there weren't any jobs. The next year Van had the same bunch of students, so he gave us advanced quantum mechanics, The year after that he gave us advanced quantum mechanics II. Van was extremely good, a good teacher and an outstanding physicist \citep[p.\ xxiv]{Serber_1992-the-los-alamos}.\footnote{Serber told Charles Weiner and Gloria B.\ Lubkin the same thing during an interview for the American Institute of Physics, February 10, 1967. As he put it in the interview, it was ``always the same gang hanging on"  \citep[p.\ 294]{Fellows_1985-j}. As \citet[p.\ 17]{VanVleck_1971-reminiscences} noted with obvious relish about Serber: ``One now identifies the present President of the American Physical Society with high energy physics, but before he fell under the influence of Oppenheimer at Berkeley, he worked on problems that today would be considered chemical physics."}
\end{quotation}
\end{singlespacing}

\noindent
Anderson offered the following explanation for Van Vleck's effectiveness as a teacher:

\begin{singlespacing}
\begin{quotation}
\noindent
In all of his classes  \ldots\  he used two basic techniques of the genuinely good teacher. First, he presented a set of carefully chosen problems  \ldots\  Second, he supplied a ``crib" for examination study, which we always thought was practically cheating, saying precisely what could be asked on the exam. It was only after the fact that you realized that it contained every significant idea of the course \citep[pp.\ 524--525]{Anderson_1987-john}.
\end{quotation}
\end{singlespacing}

\noindent
Even before the Great Depression,  students apparently took Van Vleck's quantum course more than once. Robert B.\ Whitney, whose notes for the 1927--28 edition of the course in Minnesota (see note \ref{student notes}) support the kinder of Anderson's two assessments  of Van Vleck's teaching quoted above, recalled that two advanced graduate students, Edward L.\ Hill and Vladimir Rojansky, attended the lectures the year he took the course, even though they both had to have taken it before \citep[pp.\ 175--176]{Fellows_1985-j}. Under Van Vleck's supervision, Hill and Rojansky wrote dissertations on topics in molecular and atomic spectroscopy, respectively, using the new quantum mechanics (ibid., p.\ 177, p.\ 181). Upon completion of his degree Hill went to Harvard as a postdoc to work with Van Vleck's Ph.D. advisor Edwin C. Kemble. Hill co-authored the second part of a review article on quantum mechanics with Kemble \citep{Kemble_1929-the-general, Kemble_1930-the-general}, which became the basis for the latter's quantum textbook \citep{Kemble_1937-the-fundamental}. In the preface, Kemble wrote that he was ``particularly indebted" to Van Vleck, by then his colleague at Harvard, ``for reading the entire manuscript and constant encouragement" (ibid.).

In his first year at Madison, 1928--29, Van Vleck immediately started supervising two postdocs, Kare Frederick Niessen and Shou Chin Wang, and two graduate students, probably J.\ V.\ Atanasoff and  Amelia Frank \citep[p.\ 230]{Fellows_1985-j}. He co-authored  papers with several of them, mostly related to his work on susceptibilities. Contributions by all four are acknowledged in his 1932 book. After its publication, Van Vleck continued to pursue research on susceptibilities, often in collaboration with students and postdocs  \citep[p.\ 13, p.\ 17]{VanVleck_1971-reminiscences}. In fact, in 1932, ten graduate students (among them Serber and Olaf Jordahl) and three postdocs (Fran\c coise Dony, William Penney, and Robert Schlapp) were working with Van Vleck \citep[pp.\ 294--295]{Fellows_1985-j}.

Physics 212, ``Quantum mechanics and atomic structure," was the only lecture course Van Vleck offered during his first few years in Wisconsin (ibid., p.\ 230). It was not until 1931--33, the period described by Serber, that Physics 232, ``Advanced Quantum Mechanics," and Physics 233, ``Continuation of Advanced Quantum Mechanics," were added (ibid., p.\ 294). Among the students taking the basic course in 1928--29 was John Bardeen (ibid., p.\ 230). Walter H.\ Brattain had taken the course in Minnesota the year before (ibid., p.\ 176). So two of the three men who won the 1956 Nobel Prize for the invention of the transistor, Bardeen and Brattain, as beginning graduate students took quantum mechanics  with Van Vleck. The Ph.D. advisor of the third, William B.\ Shockley, was Slater, Van Vleck's most important fellow graduate student at Harvard. This underscores the importance of the first generation of quantum physicists in the United States for the education of the next. 

\subsection{Structure of Our Paper}

The balance of this paper is organized as follows. In sec.\ 2, we sketch Van Vleck's early life against the backdrop of theoretical physics coming of age and maturing in the United States. Our main focus is on his years in Minneapolis leading up to the writing of his NRC {\it Bulletin} (1923--26).  Throughout the paper, but  especially in the more biographical secs.\ 2 and 4, we make heavy use of the superb dissertation on Van Vleck by Fred \citet{Fellows_1985-j}.  In sec.\ 3, we turn to the  {\it Bulletin} itself \citep{VanVleck_1926-quantum}. In sec.\ 3.1, we recount how what had originally been conceived as a review article of average length eventually ballooned into a 300-page book. In sec.\ 3.2 we give an almost entirely qualitative discussion of its contents, focusing on the derivation of Kramers' dispersion formula with the help of the correspondence-principle technique central to the book. For the details of this derivation we refer to \citet[cf.\ note \ref{duncan janssen dispersion} below]{Duncan_2007-on-the-verge}.  In sec.\ 4, we return to Van Vleck's biography.  We describe the years following the {\it Bulletin}'s publication, his move from Minneapolis to Madison, and the development of his expertise in the theory of susceptibilities.  In sec.\ 5, we discuss his book on susceptibilities \citep{VanVleck_1932-the-theory}.  The structure of sec.\ 5 mirrors that of sec.\ 3. In sec.\ 5.1, we recount how Van Vleck came to write his second book. In section 5.2, we discuss its content, not just qualitatively in this case but carefully going through various derivations. We focus on the vicissitudes of the Langevin-Debye formula in the transition from classical to quantum theory. In sec.\ 6, we briefly revisit the Kuhnian themes introduced above and summarize our findings.

\section{Van Vleck's Early life and Career}
%\section{Van Vleck as student and as teacher}

John Hasbrouck Van Vleck (1899--1980) was born in Middletown, Connecticut, to Edward Burr Van Vleck and Hester Laurence Van Vleck ({\it n\'ee} Raymond). In 1906, he moved to Madison, Wisconsin, where his father was appointed professor of mathematics.\footnote{Van Vleck Hall on the University of Wisconsin--Madison campus is named for E.\ B.\ Van Vleck.}  He had been named after his grandfather, John Monroe Van Vleck, but his mother, not fond of her father-in-law, called him Hasbrouck \citep[pp.\ 6--8]{Fellows_1985-j}.  To his colleagues, he would always be Van.  A nephew of Van's wife, Abigail June Pearson (1900--1989), recalls that a telegram from Japan congratulating Van Vleck on winning the Nobel prize was addressed to ``Professor Van" (John Comstock, private communication). 

In 1916, Van Vleck began his undergraduate studies at the University of Wisconsin, where he eventually majored in physics. In the fall of 1920, he enrolled at Harvard as a graduate student in physics.\footnote{In addition to three courses in physics, Van Vleck signed up for a course on railway operations in the Harvard Business School (AHQP interview, session 1, p.\ 3). As his wife Abigail recalled, Van Vleck abandoned the notion of pursuing a career in railroad management when the instructor asked him point blank whether he or anyone in his family actually owned a railroad \citep[p.\ 16]{Fellows_1985-j}. Van Vleck, however,  retained his fascination with railroads for the rest of his life. His knowledge of train schedules became legendary \citep[p.\ 503]{Anderson_1987-john}. Years later, now on the faculty at Harvard, he told a colleague, the renowned historian of science I.\ Bernard Cohen, which trains to take on an upcoming trip. Although the information Van Vleck supplied, apparently off the top of his head, turned out to be perfectly accurate, Cohen was puzzled when he reached his destination and was told by his host that he could have left an hour later, yet arrived an hour earlier, had he taken a different combination of trains. Upon his return to Cambridge, Cohen confronted Van Vleck with this intelligence. Van Vleck was undaunted. ``Of course," he is reported to have said, ``but wasn't that the best beef lunch you ever had?" (We are grateful to Roger Stuewer for telling us this story, which he heard from I.\ B.\ Cohen.)} He took Kemble's course on quantum theory and soon found himself working toward a doctorate under Kemble's supervision. In a biographical note accompanying the published version of his Nobel lecture, \citet[p.\ 351]{VanVleck_1992-john} noted that Kemble ``was the one person in America at that time qualified to direct purely theoretical research in quantum atomic physics." Indeed, it seems as though his course on quantum mechanics was the only one of its kind in America at the time. The course closely followed the ``Bible" of the old quantum theory, {\it Atombau und Spektrallinien} \citep{Sommerfeld_1919_atombau}.\footnote{Sommerfeld sent a copy of the English translation of the third edition of his book to the University of Minnesota. This copy is still in the university's library. He dedicated it to the graduate students of the University of Minnesota, which had been one of the earlier stops on his 1922--23  tour of American universities (see Michael Eckert's contribution to this volume). The dedication is signed Munich, October 16, 19[23] (the last two digits, unfortunately, have been cut off). By the time this copy of Sommerfeld's book arrived at the University of Minnesota, Van Vleck, as we will see, had joined its faculty.}  Kemble's 1917 dissertation had been the first predominantly theoretical dissertation in the United States.  Even Bohr and Sommerfeld had taken notice of Kemble's work by 1920.  When Van Vleck finished his doctorate just before the summer of 1922, he was solidly grounded in classical physics, especially in advanced techniques of celestial mechanics, but, more importantly, he had brought these skills to bear on quantum theory.  His dissertation, which was published in the {\it Philosophical Magazine} \citep{VanVleck_1922-the-normal}, was on a ``crossed-orbit'' model of the helium atom, and he had worked with Kemble to calculate the specific heat of hydrogen shortly afterward.  Neither of these calculations had agreed well with experiment, but at the time Van Vleck's results were among the best to be found.  It would take the advent of matrix mechanics in 1925 before the crossed orbit model was superseded, and before theoretical predictions for the specific heat of hydrogen could be brought into alignment with experiment \citep{Gearhart_2010-astonishing}.\footnote{As \citet{Gearhart_2010-astonishing} concludes, ``the story [of the specific heat of hydrogen] reminds us that the history of early quantum theory extends far beyond its better known applications in atomic physics" (p.\ 193). This underscores the remark by Van Vleck with which we opened our paper about physicists in the early 1920s focusing strongly on spectroscopy. Although, as Gearhart  shows, it drew much more attention in the old quantum theory than the problem of susceptibilities,  the problem of  specific heat is discussed only in passing by \citet[pp.\ 101--102]{VanVleck_1926-quantum} in his NRC {\it Bulletin}. There actually are some interesting connections between these two non-spectroscopic problems (see note \ref{clayton}).\label{gearhart}}

The following year, Van Vleck accepted a position as an instructor in Harvard's physics department.  This demanding job left him with little time for his own work.  Most of his time was spent preparing for lectures and lab sessions \citep[p.\ 49]{Fellows_1985-j}.    In the midst of this daily grind, the job offer that arrived from the University of Minnesota in early 1923 must have looked especially attractive. As \citet[p.\  351]{VanVleck_1992-john} would reflect later, it was an ``unusual move" for such an institution at that time---indicative, one may add, of the American physics community's growing recognition of the importance of quantum theory---to offer him an assistant professorship ``with purely graduate courses to teach."

At first, Van Vleck was hesitant to accept the position (AHQP interview, session 1, p.\ 14).  He and Slater had planned to tour Europe together on one of the fellowships then available to talented young American physicists.  In the end, however, and partly on the strength of his father's advice, he accepted the Minnesota offer.  After a summer in Europe with his parents (during which he managed to meet some of the most visible European theorists), he arrived in Minneapolis, ready for the fall semester in 1923.  His teaching load was indeed light. One might expect that he would thus have pursued his own research with a renewed focus. Initially, that is exactly what he did. 

In October 1924, after a preliminary report in the {\it Journal of the Optical Society of America} \citep{VanVleck_1924-a-correspondence}, a two-part paper appeared in {\it Physical Review} in which \citet{VanVleck_1924-the-absorption1, VanVleck_1924-the-absorption2} used correspondence-principle techniques to analyze the interaction between matter and radiation in the old quantum theory.  Its centerpiece was Van Vleck's own correspondence principle for absorption, but the paper also contains a detailed derivation of the Kramers dispersion formula.  Although Born had published a derivation of the formula that August, he and Van Vleck arrived at the result independently of one another  \citep[p.\ 590]{Duncan_2007-on-the-verge}.  
%Both submitted their papers in June.  
The quantum part of this paper by \citet{VanVleck_1924-the-absorption1} and the BKS paper \citep{Bohr_1924-the-quantum} are the only two papers with American authors that are included in a well-known anthology documenting the transition from the old quantum theory to matrix mechanics \citep{VanderWaerden_1968-sources}. The breakthrough \citet{Heisenberg_1925-ueber-die-quantentheoretische} achieved with his {\it Umdeutung} paper can be seen as a natural extension of the correspondence-principle techniques used by Kramers, Born, and Van Vleck (see sec.\ 3.2 below and Duncan and Janssen, 2007). 

After his 1924 paper, however, Van Vleck did not push this line of research any further. He had meanwhile been `invited' to produce the volume to which we now turn our attention.  Its completion would occupy nearly all of his available research time for the next two years. 

\section{The NRC {\it Bulletin}}
%\section{Van Vleck on the old quantum theory I: \\ the 1926 NRC {\it Bulletin}}

\subsection{Writing the {\it Bulletin}}

Later in life, when interviewed by Kuhn for the AHQP, Van Vleck recalled writing his NRC {\it Bulletin} over the course of about two years:

\begin{singlespacing}
\begin{quotation}
\noindent
I was already writing some chapters on that on rainy days in Switzerland in 1924. I would say I started writing that perhaps beginning in the spring of 1924, and finished it in late 1925. I worked on it very hard that summer \ldots\  I was sort of a ``rara avis'' at that time. I was a young theoretical physicist presumably with a little more energy than commitments than the older people interested in these subjects, so they asked me if I'd write this thing. I think it was by invitation rather than by my suggestion (AHQP interview, session 1, p.\ 21).
%Interview of John H. Van Vleck by Thomas S. Kuhn on 2 October, 1976 Niels Bohr Library and Archives, American Institute of Physics, College Park, MD USA}
\end{quotation}
\end{singlespacing}

\noindent
The invitation had come from Paul D. Foote of the U.S. Bureau of Standards, who was the chairman of the  NRC Committee on Ionization Potentials and Related Matters. Van Vleck served on this committee in the fall of 1922 \citep[p.\ 49]{Fellows_1985-j}. These NRC committees, Van Vleck recalled, had been created because ``there {\it was} a feeling among the more sophisticated of the American physicists that we were behind in knowing what was going on in theoretical physics in Europe'' (AHQP interview, session 1, p.\ 21, emphasis in the original). 

The committees organized the {\it Bulletins} of the NRC, which existed to present ``contributions from the National Research Council \ldots\  for which hitherto no appropriate agencies of publication [had] existed''   \citep[p. 173--174]{Swann_1922-electrodynamics}. This sounds rather vague and overly inclusive, and on reading the motley assortment of topics covered by the {\it Bulletins} through 1922, one finds that it {\it was} rather vague and overly inclusive.  The {\it Bulletins} served to disseminate whatever information the myriad committees deemed important.  A brief list of topics covered by these publications includes ``The national importance of scientific and industrial research,'' ``North American forest research,'' ``The quantum theory,'' ``Intellectual and educational status of the medical profession as represented in the United States Army,'' and ``The scale of the universe'' (ibid.).  The {\it Bulletins} tended to be short, averaging about 75 pages.  Several were even shorter, coming in under 50 pages.  The longest at the time Van Vleck was invited to write one on line spectra was a 172-page book, {\it Electrodynamics of Moving Media}  \citep{Swann_1922-electrodynamics}.  It had been written by four authors, including John T.\ (Jack) Tate, Van Vleck's senior colleague in Minnesota, and W.\ F.\ G.\ Swann, Van Vleck's predecessor in Minnesota 

Given the {\it Bulletin}'s publication history, Van Vleck was not making an unreasonable commitment when he accepted Foote's invitation.  Initially, his contribution was only to be a single paper in a larger volume on ``Ionization Potentials and Related Matters.'' (Foote to Van Vleck, March 22, 1924 [AHQP]).
%(49,6)} 
It is unclear exactly how the paper spiraled out of control and became the quagmire of a project that consumed over two years of his available research time, but an interesting story is suggested by his correspondence.

As we saw, Van Vleck later recalled having begun his {\it Bulletin} in the spring of 1924, but he must have started much earlier than that.  In March 1924, Foote returned a draft to Van Vleck along with extensive comments.  ``This has been read very carefully by Arthur E. Ruark,'' Foote wrote, ``who has prepared a long list of suggestions as enclosed.  These are merely suggestions for your consideration.  On some of them I do not agree with Ruark but many of his suggestions are of considerable interest'' (Foote to Van Vleck, March 22, 1924 [AHQP]). Foote was probably distancing himself from Ruark's remarks not only because of their severity, but also because of their sheer volume.  The ``suggestions'' amounted to 33 pages of typed criticism.  Van Vleck's handwritten reactions are recorded in the margins of Ruark's commentary (preserved in the AHQP).  Exclamation points and question marks abound, often side by side, punctuating Van Vleck's surprise and confusion.  Here and there, he makes an admission when a suggestion seems prudent.  For the most part, however, Ruark's suggestions are calls for additional details and clarification, more derivations, in short, a significant broadening of the ``article.''  As one reads on, Van Vleck's annotations become less and less frequent.  When they appear at all, they often amount to a single question mark.  One gets the impression of a young physicist brow-beaten into submission.  This is likely what precipitated the transformation of Van Vleck's {\it Principles} from review article to full-fledged book.
%textbook (in the loose sense of the term that we discussed in the introduction).

Perhaps Foote was still expecting a paper, but Van Vleck was producing something much more comprehensive.  By November, Foote was becoming impatient.  Van Vleck wrote to reassure him:

\begin{singlespacing}
\begin{quotation}
\noindent Like you I ``am wondering'' when my paper for the Research Council will ever be ready.  I am sorry to be progressing so slowly but I hope you realize that I am devoting to this report practically all of my time not occupied with teaching duties.  I still hope to have the manuscript ready by Christmas except for finishing touches (Van Vleck to Foote, November 21, 1924  [AHQP]).
%(49, 6)}
\end{quotation}
\end{singlespacing}

Van Vleck would blow the Christmas deadline as well.  It was not until August that he submitted a new draft:

\begin{singlespacing}
\begin{quotation}
\noindent I hope the bulletin will be satisfactory, as with the exception of one three-month period it has taken all my available research time for two years.

You wrote me that the bulletin should be ``fairly complete.''  My only fear is that it may be too much so.  I made sure to include references to practically all the important theoretical papers touching on the subjects covered in the various chapters.  Four new chapters have been included since an early draft of the manuscript was sent to you a year ago  \ldots\

You will note that I have used a new title ``Quantum Principles and Line-Spectra'' as this is much briefer and perhaps more a-propos than ``The Fundamental Concepts of the Quantum Theory of Line-Spectra" (Van Vleck to Foote, August 10, 1925 [AHQP]).
% (49, 6)}
\end{quotation}
\end{singlespacing}

\noindent
It is worth noting the change in title.  The old quantum theory was strongly focused 
%intently --sounds like something you'd only say about a person
on the phenomena of line spectra.  Van Vleck's new title conveys at once this focus even as he had significantly broadened the scope of his project.

Even when Foote sent him the galleys for inspection, Van Vleck could not resist making further additions to the {\it Bulletin}.  ``I have added 13 pages of manuscript  \ldots\ in which I have tried to summarize the work of Heisenberg, Pauli, and [Friedrich] Hund,''  Van Vleck wrote back.
% early February the following year.  
``I am sorry to make such an addition," he explained, ``but quantum theory progresses extremely rapidly, and I hope the new subject-matter will add materially to the value of the report'' (Van Vleck to Foote, February 2, 1926 [AHQP]).
% (49, 6)}

It is clear that however the project began, and whatever Van Vleck's initial expectations, in the end the {\it Bulletin} was intended by its author as a comprehensive and up-to-date review of  quantum theory.  This makes it useful not only as a review of the old quantum theory, but also as a window into Van Vleck's own perception and understanding of the field.
%evolving

Despite some critical notes,\footnote{See, e.g., the quotation from Smekal's (1927) review in sec.\ 1.2.2 above.} the {\it Bulletin} was ``on the whole, well-received" \citep[p.\ 88]{Fellows_1985-j}.  Van Vleck must have read Ruark's review of his {\it Bulletin} in the {\it  Journal of the Optical Society of America} with special interest, given Ruark's litany of complaints about an early draft of it. 
%Reading Ruark's review, one begins to suspect that these earlier reservations may have been due more to Ruark not being up on the latest developments in quantum theory than to any defects in Van Vleck's exposition of them. 
Ruark praised the final version as a thorough, clearly written, state-of the-art survey of a rapidly changing field:

\begin{singlespacing}
\begin{quotation}
\noindent 
This excellent bulletin will prove extremely useful to all who are interested in atomic physics \ldots\ [T]he fundamental theorems of Hamiltonian dynamics and perturbation methods of quantization are treated in a very readable fashion \ldots\ The chapter on the quantization of neutral helium is authoritative \ldots The author's treatment of the ``correspondence principle" is refreshingly clear \ldots\ The whole book is surprisingly up-to-date. Even the theory of spinning electrons and matrix dynamics are touched upon. It is to be hoped that this report will run through many revised editions as quantum theory progresses, for it fills a real need \citep{Ruark_1926-review}.
\end{quotation}
\end{singlespacing}

\noindent 
In fact, Ruark's main complaint was directed not at the author but at the publisher: ``Incidentally, many physicists would appreciate the opportunity of buying National Research Bulletins in a more durable binding" (ibid.). Yet, the review also hints at lingering disagreements between author and reviewer.  
%such as when Ruark wrote that Ch.\ VIII, ``The apparent non-conformity of quantum orbits to classical dynamics in atoms with more than one electron" \citep[pp.\ 105--117]{VanVleck_1926-quantum} ``is a valuable summary of a {\it very controversial} subject" \citep[our emphasis]{Ruark_1926-review}. 
%Most importantly, Ruark was clearly not as taken as Van Vleck was with the Kramers dispersion formula which was used in the {\it Bulletin} to showcase how fruitful the approach based on the correspondence principle was:
Most importantly, Ruark had his doubts about the Kramers dispersion theory which Van Vleck had used in the {\it Bulletin} to showcase the power of the correspondence principle:

\begin{singlespacing}
\begin{quotation}
\noindent 
Many readers will not agree with the author's conclusion that ``Kramers's dispersion theory \ldots furnishes by far the most satisfactory theory of dispersion" [Van Vleck, 1926b, pp.\ 156--157]  \ldots the reviewer believes that a final solution cannot be achieved until we have a much more thorough knowledge of the dispersion curves of monatomic gases and vapors  \citep{Ruark_1926-review}.
\end{quotation}
\end{singlespacing}

\noindent 
Subsequent developments would prove that Van Vleck's confidence in the Kramers dispersion formula was well-placed. It carried over completely intact to the new quantum mechanics \citep[p.\ 655]{Duncan_2007-on-the-verge}.

\subsection{The {\it Bulletin} and the Correspondence Principle}

The central element in Van Vleck's presentation of the old quantum theory in his NRC {\it Bulletin} is the correspondence principle. It forms the basis of 11 out of a total of 13 chapters.\footnote{The remaining chapters deal with ``Half quanta and the anomalous Zeeman effect" (Ch.\ XII) and ``Light-quants" [sic] (Ch. XIII).} As it says in the preface,

\begin{singlespacing}
\begin{quotation}
\noindent
Bohr's correspondence principle is used as a focal point for much of the discussion in Chapters I--X. In order to avoid introducing too much mathematical analysis into the discussion of the physical principles underlying the quantum theory, the proofs of certain theorems are deferred to Chapter XI, in which the dynamical technique useful in the quantum theory is summarized \citep[p.\ 3]{VanVleck_1926-quantum}.
\end{quotation}
\end{singlespacing}

\noindent
The correspondence principle first emerged in the paper in which \citet{Bohr_1913-on-the-constitution} introduced his model for the hydrogen atom \citep[p.\ 268, p.\ 274]{Heilbron_1969-the-genesis}. Perhaps the most radical departure from classical theory proposed in Bohr's paper was that the frequency of the radiation emitted or absorbed when an electron jumps from one orbit to another differs from the orbital frequency of the electron in both the initial and the final orbit \citep[pp.\ 571--572]{Duncan_2007-on-the-verge}. However, for high quantum numbers $N$, the orbital frequencies of the $N^{\rm th}$ and the  $(N-1)^{\rm th}$ orbit and the frequency of the radiation emitted or absorbed when an electron jumps from one to the other approach each other. This is the core of what later came to be called the correspondence principle. 

By the early 1920s, the correspondence principle had become a sophisticated scheme used by several researchers for connecting formulas in classical mechanics to formulas in the old quantum theory. The most important result of this approach was the Kramers dispersion formula, which \citet{Kramers_1924-the-law,Kramers_1924-the-quantum}  first introduced in two short notes in {\it Nature}. As we mentioned in sec.\  2, \citet{Born_1924-ueber-quantenmechanik} and \citet{VanVleck_1924-the-absorption1,VanVleck_1924-the-absorption2} independently of one another published detailed derivations of this result a few months later. Kramers himself would not publish the details of his dispersion theory until early 1925, in a paper co-authored with Heisenberg \citep{Kramers_1925-ueber-die-streuung}. This paper has widely been recognized as a decisive step toward Heisenberg's (1925a)  {\it Umdeutung} paper written in the summer of 1925 \citep[p.\ 554]{Duncan_2007-on-the-verge}. 

% with which \citet{Heisenberg_1925-ueber-die-quantentheoretische} laid the basis for matrix mechanics

As a concrete example of the use of the correspondence principle in the old quantum theory in the early 1920s, we sketch Van Vleck's derivation of the Kramers dispersion formula.\footnote{For a detailed reconstruction of this derivation, which follows Van Vleck's two-part paper of 1924 rather than his 1926 NRC {\it Bulletin}, see \citet{Duncan_2007-on-the-verge}: 
%The reconstruction takes up a number of subsections in this paper. 
in sec.\ 3.4 (pp.\ 591--593), an outline of the derivation is given;
%In secs.\ 5.1--5.2 (pp.\ 627--637), the result is derived for a simple harmonic oscillator. In sec.\ 6.2 (pp.\ 648--652), this derivation is generalized to an arbitrary multiply-periodic system. In sec.\ 7.1 (pp.\ 655--658), finally, it is shown that in modern quantum mechanics the Kramers dispersion formula holds for an even broader class of systems than in the old quantum theory.
in secs.\ 5.1--5.2 (pp.\ 627--637), the result is derived for a simple harmonic oscillator; in sec.\ 6.2 (pp.\ 648--652), this derivation is generalized to an arbitrary multiply-periodic system; finally, in sec.\ 7.1 (pp.\ 655--658),  it is shown that in modern quantum mechanics the Kramers dispersion formula holds for an even broader class of systems than in the old quantum theory.
\label{duncan janssen dispersion}} This formula and what \citet[p.\ 162]{VanVleck_1926-quantum} called the ``correspondence principle for dispersion" are presented in a section of only two and a half pages in Ch.\ X of the NRC {\it Bulletin} (ibid.,  sec.\ 51, pp.\ 162--164). The reason that Van Vleck could be so brief at this point is that the various ingredients needed for the derivation of the formula are all introduced elsewhere in the book, especially in Ch.\ XI on mathematical techniques. At 50 pages, this is by far the longest chapter of the {\it Bulletin}.

Consider some (multiply-)periodic system---ranging from a charged simple harmonic oscillator to an electron orbiting a nucleus---struck by an electromagnetic wave of a frequency $\nu$ not too close to that system's characteristic frequency $\nu_0$ or frequencies $\nu_k$. The Kramers dispersion formula is the quantum analogue of an expression in classical mechanics for the polarization of such a periodic system resulting from its interaction with the electric field of the incoming electromagnetic wave, multiplied by the number of such systems in the dispersive medium. This expression can easily be converted into an expression for the dependence of the index of refraction on the frequency of the refracted radiation. Optical dispersion, a phenomenon familiar from rainbows and prisms, is described by this frequency dependence of the refractive index \citep[sec.\ 3.1, pp.\ 573--578]{Duncan_2007-on-the-verge}. To obtain the Kramers dispersion formula in the old quantum theory, one has to derive an expression for the instantaneous dipole moment, induced by an external electromagnetic wave, of individual (multiply-)periodic systems in classical mechanics, multiply that expression by the number of such systems in the dispersive medium, and then translate the result into an expression in the old quantum theory under the guidance of the correspondence principle. 

As with all such derivations in the old quantum theory, the part involving classical mechanics called for advanced techniques borrowed from celestial mechanics. As we mentioned in sec.\ 2, Van Vleck had thoroughly mastered these techniques as a graduate student at Harvard. Decades later, when the Dutch Academy of Sciences  awarded him its prestigious Lorentz medal,  Van Vleck related an anecdote in his acceptance speech that demonstrates his early mastery of this material:

\begin{singlespacing}
\begin{quotation}
\noindent
In 1924 I was an assistant professor at the University of Minnesota. On an American trip, [Paul] Ehrenfest gave a lecture there  \ldots\ [He] said he would like to hear a colloquium by a member of the staff. I was selected to give a talk on my ``Correspondence Principle for Absorption" [Van Vleck, 1924a,b,c] \ldots\ I remember Ehrenfest being surprised at my being so young a man. The lengthy formulas for perturbed orbits in my publication on the three-body problem of the helium atom [Van Vleck, 1922] had given him the image of a venerable astronomer making calculations in celestial mechanics \citep[p.\ 9; quoted by Duncan and Janssen, 2007, p.\ 627]{VanVleck_1974}.
\end{quotation}
\end{singlespacing}

\noindent
Van Vleck put his expertise in classical mechanics to good use. Using  canonical perturbation theory in action-angle variables, he derived an expression in classical mechanics for the dipole moment of a charged multiply-periodic system hit by an electromagnetic wave of small amplitude that could then be translated into a quantum-theoretical expression. 

In general coordinates and their conjugate momenta $(q_k, p_k)$  (where $k = 1,  \ldots, f$, with $f$ the number of degrees of freedom), Hamilton's equations are:
\begin{equation}
\label{a}
    \dot{q}_k = \frac{\partial H}{\partial p_k},\;\;\;
    \dot{p}_k = -\frac{\partial H}{\partial q_k},
\end{equation}
where $H(q_k, p_k)$ is the Hamiltonian and dots indicate time derivatives. Given the Hamiltonian of some multiply-periodic system, one can often find special coordinates, $(w_k, J_k)$, called {\it action-angle variables}, such that the Hamiltonian in the new coordinates only depends on the new momenta, the action variables $J_k$, and not on the new coordinates, the angle variables $w_k$. In that case, Hamilton's equations take on the simple form:
\begin{equation}
    \dot{w}_k = \frac{\partial H}{\partial J_k} = \nu_k,\;\;\;
    \dot{J}_k = -\frac{\partial H}{\partial w_k} = 0.
\label{b}   
\end{equation}
The first of these equations shows what makes the use of action-angle variables so attractive in celestial mechanics. It makes it possible to extract the characteristic periods of the system from the Hamiltonian without having to know the details of the orbit. 

Action-angle variables played a central role in the old quantum theory. They are used to formulate the Sommerfeld-Wilson(-Ishiwara) quantum conditions  \citep[pp.\ 39--40]{VanVleck_1926-quantum}, which select the orbits allowed by the old quantum theory from all classically allowed ones. The relation between the new momenta $J_k$ and the original position and momentum variables $q_k$ and $p_k$ is: $J_k = \oint p_k dq_k$, where the integral is over one period of the motion. The Sommerfeld-Wilson quantum conditions restrict the classically allowed orbits to those satisfying 
\begin{equation}
J_k = \oint p_k dq_k = n_kh,
\label{Bohr-Sommerfeld}
\end{equation}
where $h$ is Planck's constant and the $n_k$'s are integers. 

For orbits with high values for all quantum numbers, there is only a small difference between the values of the Hamiltonian for $J_l = N_l  h$ and for $J_l = (N_l \pm 1) h$ (with the values of all $J_m$'s with $m \neq l$ fixed). The differential quotients in the first equation in Eq.\ \eqref{b} can then be approximated by difference quotients:
\begin{equation}
\nu_l = \frac{\partial H}{\partial J_l} \approx \frac{H(J_1, \ldots, J_l = (N_l + 1) h, \ldots, J_f) - H(J_1, \ldots, J_l = N_l h, \ldots, J_f)}{h}.
\label{corr princ for freq 2}
\end{equation}
The two values of the Hamiltonian in the numerator give the energies $E_{N_l}$ and $E_{N_l +1}$ of two orbits, close to each other, with high values for all quantum numbers (all, except for the $l^{\rm th}$ one, equal for the two orbits). Eq.\ \eqref{corr princ for freq 2} is thus of the form
\begin{equation}
h \nu_l = E_{N_l +1} - E_{N_l}.
\end{equation}
In the limit of high quantum numbers, this equation for the orbital frequency $\nu_l$ of the electron---and thereby, according to classical electrodynamics, the frequency of the radiation emitted because of the electron's acceleration in that orbit---coincides with Bohr's rule, 
\begin{equation}
h \nu_{i \rightarrow f} = E_{n_i} - E_{n_f}, 
\end{equation}
for the frequency $\nu_{i \rightarrow f}$ of the radiation emitted when an electron jumps from an initial orbit (quantum number $n_i$) to a final orbit (quantum number $n_f$). This asymptotic connection between this classical formula for the orbital frequencies $\nu_l$ and Bohr's quantum formula for the radiation frequencies $\nu_{i \rightarrow f}$ is what \citet[sec.\ 11, pp.\ 23--24]{VanVleck_1926-quantum} called ``the correspondence theorem for frequencies." 

Such asymptotic connections can be used in two ways, either to {\it check} that a given quantum formula reduces to its classical counterpart in the limit of high quantum numbers, or to make an educated {\it guess} on the basis of the classical formula assumed to be valid for high quantum numbers as to what its quantum-theoretical counterpart valid for all quantum numbers might be. While \citet{Born_1924-ueber-quantenmechanik,Born_1925-vorlesungen} emphasized the latter {\it constructive} use, \citet{VanVleck_1924-the-absorption1,VanVleck_1924-the-absorption2,VanVleck_1926-quantum} preferred the former {\it corroborative} use \citep[pp.\ 638--640]{Duncan_2007-on-the-verge}. The correspondence theorem for frequencies is a good example of the corroborative use of correspondence-principle arguments, the Kramers dispersion formula is the prime example of their constructive use.\footnote{\citet{Ruark_1926-review} picked up on this distinction in his review of the {\it Bulletin}. Elaborating on his praise for the ``refreshingly clear" treatment of the correspondence principle (see the quotation at the end of sec.\ 3.1), he explained that Van Vleck ``takes pains to point out that certain asymptotic connections between quantum theory and classical dynamics can be definitely proved, while other connections are only postulated. Thus he distinguishes carefully the correspondence {\it theorem} for frequencies, and the correspondence {\it postulates} for intensities and polarization."}

To derive a formula for the classical dipole moment from which its counterpart in the old quantum theory can be constructed (or against which it can be checked), one treats the electric field of the electromagnetic wave striking the periodic system under consideration as a small perturbation of the system in the absence of such disturbances. The full Hamiltonian $H$ is then written as the sum of an unperturbed part $H^0$ and a small perturbation $H^{\rm int} <\!\!< H^0$ (where `int' stands for `interaction').
%We will examine the procedure in more detail in sec.\ 5.2.1. 
Using action-angle variables in such perturbative calculations, one can derive the formula for the classical dipole moment without having to know anything about the dynamics of the unperturbed system other than that it is solvable in these variables.\footnote{The calculation of the effect of external fields on spectra, such as the Stark and Zeeman effects in atoms with one electron in external electric and magnetic fields, respectively, proceeds along similar lines \citep[Ch.\ V]{VanVleck_1926-quantum}.}

Once again, \citet{Born_1924-ueber-quantenmechanik,Born_1925-vorlesungen} and \citet{VanVleck_1924-the-absorption1,VanVleck_1924-the-absorption2,VanVleck_1926-quantum}  proceeded in slightly different ways. Born tried to find action-angle variables $(w,J)$ for the full Hamiltonian $H$,  Van Vleck continued to use action-angle variables $(w^0_k, J^0_k)$ for the unperturbed Hamiltonian $H^0$ even when dealing with the full Hamiltonian $H$. As \citet[p.\ 200]{VanVleck_1926-quantum} explicitly noted,  $H$ will in general depend on  both the $J^0_k$'s and the $w^0_k$'s, so $(w^0_k, J^0_k)$ are {\it not} action-angle variables for $H$, but one can still use them to describe the behavior of the full system with interaction.\footnote{This choice of variables is analogous to the one made in the Dirac interaction picture in time-dependent perturbation theory in modern quantum mechanics \citep[p.\ 655, note 204]{Duncan_2007-on-the-verge}.\label{dirac}} As we will see in sec.\ 5.2.1, \citet[p.\ 38]{VanVleck_1932-the-theory} likewise used action-angle variables for the unperturbed Hamiltonian in his later calculations of susceptibilities.\footnote{Those calculations, however, involve time-{\it independent} perturbation theory (cf.\ note \ref{dirac}).}

The classical formula Van Vleck eventually arrived at for the dipole moment of a multiply-periodic system has the form of a derivative with respect to the action variables $J^0_k$ of an expression involving squares of the amplitudes of the Fourier components and the characteristic frequencies $\nu^0_k = \dot{w}^0_k$ of the motion of the unperturbed system. The correspondence principle, as it was understood by Kramers, Born, Van Vleck and others in the early 1920s, amounted to the prescription to make three substitutions in this classical formula to turn it into a formula in the old quantum theory that is guaranteed to merge with the classical formula in the limit of high quantum numbers \citep[p.\ 635]{Duncan_2007-on-the-verge}:
\begin{enumerate}
\item Replace the characteristic frequencies $\nu_k^0$, the orbital frequencies of the motion in the unperturbed multiply-periodic systems under consideration, by the frequencies $\nu_{i \rightarrow f}$ of the radiation emitted in the transition from the $n_i^{\rm th}$ to the $n_f^{\rm th}$ orbit.
\item Replace squares of the amplitudes of the Fourier components of this motion by transition probabilities given by the $A$ coefficients for spontaneous emission in the quantum theory of radiation proposed by \citet{Einstein_1917-zur-quantentheorie}.
\item Replace the derivatives with respect to the action variables $J^0_k$ by difference quotients as in Eq.\ \eqref{corr princ for freq 2}. This last substitution is often attributed to Born but it was almost certainly discovered independently by Born, Kramers, and Van Vleck \citep[pp.\ 637--638, p.\ 668]{Duncan_2007-on-the-verge}.
\end{enumerate}
Although this construction guarantees that the quantum formula merges with the classical formula for high quantum numbers, it still took a leap of faith to assert that the quantum formula would continue to hold all the way down to small quantum numbers. In the case of the Kramers dispersion formula, however, there were other considerations, besides this correspondence-principle argument for it, that inspired confidence in the result. 

As mentioned in sec.\ 1.2, the Kramers dispersion formula amounted to the recovery of a Kuhn loss. Experiments clearly showed that the frequency ranges where dispersion becomes anomalous (i.e., where the index of refraction gets smaller rather than larger with increasing frequency) are around the absorption frequencies of the dispersive material. The classical dispersion theory of von Helmholtz, Lorentz, and Drude of the late-19th century was designed to capture this feature. The theory posited the existence of small charged harmonic oscillators inside matter with characteristic frequencies corresponding to the material's absorption frequencies. Instead of such harmonically bound electrons, the Bohr model of the atom had electrons orbit a nucleus as in a miniature solar system.  When \citet{Sommerfeld_1915-die-allgemeine, Sommerfeld_1917-die-drudesche}, \citet{Debye_1915-die-konstitution}, and \citet{Davisson_1916-the-dispersion} adapted the classical dispersion theory to this new model of matter, they were inexorably led to the conclusion that  dispersion should be anomalous in frequency ranges around the orbital frequencies $\nu^0_k$ of the Bohr atom, which, as noted above, differ sharply from the absorption and transition frequencies $\nu_{i \rightarrow f}$, at least for small quantum numbers. This is the Kuhn loss in dispersion theory mentioned in sec.\ 1.2. As long as the old quantum theory could boast of successes in spectroscopy, the problem with dispersion could be ignored. In the early 1920s, however, physicists started to take it more seriously (see, e.g., the comments by Epstein quoted in sec.\ 1.2.1). \citet{Ladenburg_1921-untersuchungen} and \citet{Ladenburg_1923-absorption} proposed a new quantum dispersion theory in which they simply assumed that dispersion would be anomalous in frequency intervals around the transition frequencies $\nu_{i \rightarrow f}$ rather than the orbital frequencies $\nu^0_k$. \citet{Kramers_1924-the-law, Kramers_1924-the-quantum}, in effect, generalized the formula that \citet{Ladenburg_1921-untersuchungen} had proposed and, in the process, provided it with the theoretical underpinnings it had been lacking before. And thus was the Kuhn loss in dispersion theory recovered.

The recovered Kuhn loss in dispersion theory is a good example of what \citet[p.\ 105]{Kuhn_1996_the-structure} described as a ``reversion (which is not the same as a retrogression)" to an older theory or paradigm (cf.\ sec.\ 1.2.1). Both in the classical theory and in the quantum theory, the dispersion formula contains a set of parameters, one for every absorption frequency, called the ``oscillator strengths." These parameters are adjusted to give the best fit with the experimental data. In the classical theory, the ``oscillator strength" for a given absorption frequency was interpreted as the number of harmonically-bound charges per atom with a resonance frequency equal to that absorption frequency. Unfortunately, this interpretation was strongly at odds with the experimental results. It was not uncommon to find values as low as 1 ``dispersion electron," as these charged oscillators were called, per 200 or even per 50,000 atoms! In quantum theory, as \citet{Ladenburg_1921-untersuchungen} first realized, the ``oscillator strength" for a given absorption frequency can be interpreted as the number of transitions with  transition frequencies $\nu_{i \rightarrow f}$ equal to that absorption frequency. The low values of these parameters then simply reflect that, for many frequencies $\nu_{i \rightarrow f}$, there will only be a small number of atoms in the initial excited state \citep[pp.\ 582--583]{Duncan_2007-on-the-verge}.

The correspondence-principle translation scheme outlined above was central to the research in the early 1920s of both \citet{VanVleck_1924-the-absorption1,VanVleck_1924-the-absorption2} and \citet{Born_1924-ueber-quantenmechanik}. In fact, their approaches were so similar 
%even though one was working in relative isolation in faraway Minneapolis while the other was working in G\"ottingen, a center for quantum theory in ascendance during this period, 
that the two men had a testy correspondence about the proper appropriation of credit for various results and insights \citep[pp.\ 569--571, pp.\ 638--639]{Duncan_2007-on-the-verge}. Moreover, both \citet{Born_1925-vorlesungen} and \citet{VanVleck_1926-quantum} wrote a book on the old quantum theory in which they organized the material covered around the correspondence principle as they had come to understand and use it in their research.\footnote{Born's book is analyzed in Domenico Giulini's contribution to this volume.} 

Both Born and Van Vleck missed the next step, which was to use the correspondence-principle translation scheme for the basic laws of classical mechanics rather than for individual formulas. That step would  be taken by \citet{Heisenberg_1925-ueber-die-quantentheoretische} in his {\it Um\-deu\-tung} paper \citep[sec.\ 3.5, pp.\ 593--596; sec.\ 8, p.\ 668]{Duncan_2007-on-the-verge}. In doing so, Heisenberg abandoned electron orbits altogether and formulated his theory entirely in terms of quantum transitions, accepting for the time being that there was nothing in the theory to represent the states between which such transitions were supposed to take place. \citet{Born_1925-Zur-Quantenmechanik}  first recognized that the two-index quantities thus introduced (referring to initial and final states of a transition) were nothing but matrices. 

Since the Sommerfeld-Wilson quantum condition \eqref{Bohr-Sommerfeld} refers to individual orbits, Heisenberg had to find a new quantum condition. Taking the difference in the values of $\oint p dq$ of two neighboring orbits and translating this using his {\it Umdeutung} scheme, he arrived at a corollary of the Kramers dispersion formula that had inspired his reinterpretation scheme. This corollary had been found independently by Werner \citet{Kuhn_1925-Ueber-die-Gesamtstaerke} and Willy \citet{Thomas_1925-Ueber-die-Zahl} and is known as the Thomas-Kuhn sum rule (which thus has nothing to do with Thomas S.\ Kuhn). \citet{Born_1925-Zur-Quantenmechanik} showed that the Thomas-Kuhn sum rule is equivalent to the diagonal elements of the basic commutation relations, $[p, q] = \hbar/i$, for position and momentum in matrix mechanics \citep[pp.\ 659--660]{Duncan_2007-on-the-verge}. The  Thomas-Kuhn sum rule, as \citet{VanVleck_1926-quantum} noted ruefully in his NRC {\it Bulletin}, ``appears to have first been incidentally suggested by the writer" (p.\ 152). It can be found in a footnote in the classical part of his two-part paper on his correspondence principle for absorption \citep[pp.\ 359--360; cf.\ Duncan and Janssen, 2007, pp.\ 595--596, p.\ 668]{VanVleck_1924-the-absorption2}. By 1924, Van Vleck thus had the two key physical ingredients of Heisenberg's {\it Umdeutung} paper, the Kramers dispersion formula and the Thomas-Kuhn sum rule. In a very real sense, he had been on the verge of {\it Umdeutung} \citep{Duncan_2007-on-the-verge}. 

Van Vleck apparently told his former student Kuhn in the early 1960s that, had he been ``a little more perceptive," he could have done what Heisenberg did. When Kuhn reminded him of that boast during the official interview for the AHQP in 1963, Van Vleck backed off and told his interviewer: ``Perhaps I should say {\it considerably} more perceptive" (AHQP interview, session 1, p.\ 24, quoted by Duncan and Janssen, 2007, pp.\ 555--556). 

Born was not that modest. In the preface to the 1927 English translation of his 1924 book, he claimed that ``discussions with my collaborators Heisenberg, Jordan, and Hund which attended the writing of this book have prepared the way for the critical step which we owe to Heisenberg" \citep[pp.\ xi--xii]{Born_1927-the-mechanics}. Even though it is not clear how much Heisenberg's {\it Umdeutung} paper owes to these discussions with Born, there is no doubt that Born   already recognized the limitations and the provisional character of the old quantum theory when he turned his lectures on `atomic mechanics'\footnote{The term `atomic mechanics' ({\it Atommechanik}) was chosen in analogy with the term `celestial mechanics' ({\it Himmelsmechanik})  \citep[preface]{Born_1925-vorlesungen} For the English translation, the title was rendered as {\it Mechanics of the Atom}, but in the text ``the clumsier expression atomic mechanics has often been employed" \citep[p.\ v, note]{Born_1927-the-mechanics}.\label{atomic mechanics}} of 1923/1924 into a book. In the preface, dated November 1924, he wrote:

\begin{singlespacing}
\begin{quotation}
\noindent
[T]he work is deliberately conceived as an attempt \ldots\ to ascertain the limit within which the present principles of atomic and quantum theory are valid and  \ldots\  to explore the ways by which we may hope to proceed \ldots\  [T]o make this program clear in the title, I have called the present book ``Vol. I;" the second volume is to contain a closer approximation to the ``final" atomic mechanics  \ldots\  The second volume may, in consequence, remain for many years unwritten. In the meantime let its virtual existence serve to make clear the aim and spirit of this book \citep[p.\ v]{Born_1925-vorlesungen}.
\end{quotation}
\end{singlespacing}

\noindent
By the time the English translation of Born's book was ready to be sent to press two years later, both matrix mechanics and wave mechanics had arrived on the scene. In the preface to the translation, dated January 1927, Born addressed the question whether, given these developments, ``the appearance of an English translation is justified" \citep[p.\ xi]{Born_1927-the-mechanics}. He believed it was, on three grounds:  

\begin{singlespacing}
\begin{quotation}
\noindent
[I]t seems to me that the time is [sic] not yet arrived when the new mechanics can be built up on its own foundations, without any connection with classical theory  \ldots\ Further, I can state with a certain satisfaction that there is practically nothing in the book which I wish to withdraw. The difficulties are always openly acknowledged \ldots
%and the applications of the theory to empirical details are so carefully formulated that no objection can be made from the point of view of the newest theory. 
Lastly, I believe that this book itself has contributed in some small measure to the promotion of the new theories, particularly those parts which have been worked out here in G\"ottingen\footnote{The next sentence is the one referring to Born's discussions with Heisenberg and others quoted above.} \citep[p.\ xi]{Born_1927-the-mechanics}.
\end{quotation}
\end{singlespacing}

\indent
%The sequel would be written much sooner than Born had anticipated. 
Quantum mechanics continued to develop rapidly in the late 1920s \citep{Duncan_2012-never}. Only three years after the English translation of his 1924 book, the sequel Born had promised in the preface to the original German edition appeared. The book, co-authored with his former student Jordan, who had meanwhile emerged as one of the leading young quantum theorists, is entitled {\it Elementary Quantum Mechanics}: {\it Lectures on Atomic Mechanics,} Vol.\ 2. In the preface, Born and Jordan explained that

\begin{singlespacing}
\begin{quotation}
\noindent
[t]his book is the continuation of the ``Lectures on atomic mechanics" published in 1925; it is the ``second volume" that was announced in the preface, of which ``the virtual existence should serve  to make clear the aim and spirit of this book." The hope that the veil that was still hanging over the real structure of the laws of the atom would soon be parted has been realized in a surprisingly fast and thorough fashion \citep[p. v]{Born_1930-elementare}.
\end{quotation}
\end{singlespacing}

\noindent
The authors then warned their readers that they had made a conscious effort to see how much could be done with ``elementary, i.e., predominantly algebraic means" (ibid., p.\ vi). In other words, elementary quantum mechanics, for Born and Jordan, was essentially matrix mechanics. They relegated wave-theoretical methods to a future book they promised to write ``as soon as time and energy permit" (ibid). 

In his review of {\it Elementary Quantum Mechanics} in {\it Die Naturwissenschaften}, Pauli took Born and Jordan to task for their decision to restrict themselves to matrix mechanics, adding pointedly that ``one cannot reproach the reviewer on the grounds that he finds the grapes sour because they are hanging too high for him"  \citep{Pauli_1930-review}. Pauli, after all, had solved the hydrogen atom in matrix mechanics before wave mechanics was available. The authors' promise of a future volume on wave mechanics (which never saw the light of day) provided Pauli with the perfect opening line for his review: ``The book is the second volume in a series in which goal and purpose of the $n^{\rm th}$ volume is always made clear through the virtual existence of the $(n+1)^{\rm th}$ volume" (ibid.). Pauli's review famously ends with the observation that ``the production of the book in terms of print and paper is excellent" (ibid.). Born was angry enough about this scathing review to complain about it to Pauli's teacher Sommerfeld  \citep[p.\ 641]{Duncan_2008-Pascual}. Pauli's negative verdict on Born and Jordan's 1930 effort stands in marked contrast to his high and unqualified praise for Van Vleck's 1932 book in the same journal a few years later.\footnote{See the quotations in note \ref{pauli pauling review} in sec.\ 1.2.2 and at the end of sec.\ 5.1.}

Contrary to Born, Van Vleck only seems to have realized how serious the problems facing the old quantum theory were after its demise. Talking to Kuhn in 1963, he claimed that, as early as 1924, he had a clear premonition that a drastic conceptual change was imminent: ``I certainly realized, and that must have been in 1924 or possibly 1925, certainly before the academic year 1925--26---more likely 1924--25---that there was something rotten in the state of Denmark as regards the old classical quantum theory" (AHQP interview, session 2, pp.\ 2--3). This is not the impression one gets if one looks at the text of the NRC {\it Bulletin}. It is true that Van Vleck was perfectly candid about the theory's failures and short-comings. He devoted an entire chapter (Ch.\ VIII) to the problems one ran into as soon as one considered atoms with more than one electron. Van Vleck, however, remained optimistic that these problems could be solved without abandoning the basic conceptual framework of the old quantum theory. 

In one of the sections of Ch.\ VIII,  sec.\ 35, entitled ``Standard Quantum Conditions and  Correspondence Theorem for Frequencies Remain Valid Even if Classical Mechanics Break [sic] Down," he wrote:

\begin{singlespacing}
\begin{quotation}
\noindent
[T]o escape from the difficulties thus encountered [in the preceding section] it appeared necessary to assume that the classical mechanics do not govern the motions of the electrons in the stationary states of atoms with more than one electron. It might seem that this {\it bold proposal} would invalidate the considerable degree of success already sometimes attained in complicated atoms \ldots\ Such successful applications, however, need not be forfeited if only we assume that the Bohr frequency condition and the standard quantum conditions retain their validity, even though the motions quantized by the latter are not in accord with ordinary dynamics in atoms with more than one electron \citep[p.\ 108, our emphasis]{VanVleck_1926-quantum}.
\end{quotation}
\end{singlespacing} 

As bold as Van Vleck may have thought his proposal was, by the time his {\it Bulletin} was in print, Heisenberg's {\it Umdeutung} paper had already made it clear that much more radical measures were called for, even though, as we formulated it in sec.\ 1.2.3 {\it Umdeutung} meant that the laws of mechanics were not repealed but reinterpreted. Working on his {\it Bulletin} in relative isolation in Minnesota, Van Vleck had not been privy to the skepticism with which electron orbits had increasingly been viewed by his European colleagues. Heisenberg and others were prepared to abandon orbits altogether. Van Vleck, by contrast, remained convinced that the old quantum theory was essentially right, and only in error concerning the specific details of the orbits. 

By the time he wrote the article about the new quantum theory in the {\it Chemical Reviews} from which we quoted at the beginning of this paper, Van Vleck had certainly understood that the transition from the old to the new quantum theory required much more radical steps than he had contemplated in his NRC {\it Bulletin}. As he explained to his colleagues in chemistry,

\begin{singlespacing}
\begin{quotation}
\noindent
one cannot use a meter stick to measure the diameter of an atom, or an alarm clock to record when an electron is at the perihelion of its orbit. Consequently we must not be surprised \ldots\  that models cannot be constructed with the same kind of mechanics as Henry Ford uses in designing an automobile \citep[p.\ 468, quoted and discussed by Duncan and Janssen, 2007, p.\ 666]{VanVleck_1928-the-new2}.
\end{quotation}
\end{singlespacing}

In the years following the {\it Bulletin}'s publication, Van Vleck's perceptions of the old quantum theory would change a great deal.  Specifically, he would come to see its shortcomings through the lens of his subsequent work on susceptibilities and his own accomplishments in this area as providing powerful arguments against the old and in favor of the new quantum theory.

\section{New Research and the Move to Wisconsin}

Only after the {\it Bulletin} was sent to press was Van Vleck able to confront matrix mechanics.  By late March of 1926, he had no doubt caught up with current developments, in part through his own reading and in part through direct contact with Born who lectured in Madison that month \citep[p.\ 102]{Fellows_1985-j}.  In January of 1926, Jack Tate, Van Vleck's senior colleague in Minnesota, had become the new editor-in-chief of the {\it Physical Review} \citep[p.\ 7]{VanVleck_1971-reminiscences}. Van Vleck joined the editorial board and assumed the responsibilities of associate editor  \citep[p.\ 105]{Fellows_1985-j}.\footnote{See also AHQP interview, session 2, p.\ 5.} He suddenly had access to the 
papers of his American colleagues on fellowships overseas before they were published.

In April 1926, Van Vleck read a paper submitted to the {\it Physical Review} by \citet{Pauling_1926-the-quantum} with a calculation of the electric susceptibility of HCl gas in the old quantum theory \citep[p.\ 106]{Fellows_1985-j}. New experimental evidence indicated that the rotation of polar molecules like HCl ought to be quantized with half quanta rather than, as \citet{Pauli_1921-zur-theorie} had done, with whole quanta. Pauling closely followed Pauli's calculation otherwise, using the old quantum theory to quantize the angular momentum of a rotating dumbbell or rigid rotator, the model used for the diatomic molecules under consideration.\footnote{We will discuss these calculations by \citet{Pauli_1921-zur-theorie} and \citet{Pauling_1926-the-quantum} in detail in sec.\ 5.2.2 (see also Fellows, 1985, pp.\ 141--142).} The results of both calculations can be found in the table in Fig.\  \ref{fig1}. They deviated sharply from the classical value of 1/3 for the constant $C$ in the Langevin-Debye formula. 

About a month after Van Vleck read Pauling's paper, a paper  by David M.\ \citet{Dennison_1926-the-rotation} came across his desk.  It would be the first involving matrix mechanics to be published in the {\it Physical Review}. As Van Vleck recalled decades later:

\begin{singlespacing}
\begin{quotation}
\noindent
I remember in particular [Tate] showing me an article by Dennison written in Copenhagen [while on an International Education Board (IEB) fellowship] which had the matrix elements for the symmetrical top. I realized this was just what was needed to compute the dielectric constant of a simple diatomic molecule. I requested Dennison's permission to use them in advance of their appearing in print, and remember his wiring me permission to do so. I found that they made the factor $C$ in the Debye formula  \ldots\ for the susceptibility reacquire the classical value $1/3$, replacing the nonsensical values yielded by the old quantum theory \citep[p.\ 8]{VanVleck_1971-reminiscences}.\footnote{Cf. the passage from the AHQP interview with Van Vleck quoted at the beginning of sec.\ 1.2.} 
\end{quotation}
\end{singlespacing}

\noindent
Van Vleck's calculation was analogous to Pauli's and Pauling's but, relying on Dennison's results, he now quantized the angular momentum of  diatomic molecules using matrix mechanics rather than the old quantum theory. He sent a quick note to {\it Nature} to secure priority, and in June set off for Europe, where his parents were vacationing.  The summer would bring disappointment, though. In July he received a letter from the editors of {\it Nature}, who were  ``rather wary of publications by comparatively unknown authors'' and requested a significant reduction in the length of his note {\citep[p.\ 109]{Fellows_1985-j}. Van Vleck complied but the delay cost him his priority in publishing the result. He still vividly remembered his disappointment in 1963:

\begin{singlespacing}
\begin{quotation}
\noindent
I must confess that that rather burned me up because I felt it was quite a significant achievement in quantum theory. When I mentioned it to Bohr he said ``you should have got me to endorse it, it would have gone through quicker." As it was, I think [Lucy] Mensing and Pauli beat me to it on being the first to publish that factor one-third. It was essentially a triple tie, though [Ralph de Laer] Kronig had it too, all three of us$^[$\footnote{Drawing the veil of charity over his subject's 1921 paper on the topic, Pauli's biographer Charles P.\  \citet{Enz_2002-no-time} concluded: ``Thus Mensing and Pauli's paper brought a long and confusing development to a close and helped establish faith in the new quantum theory" (p.\ 63). Enz does not mention Van Vleck or Kronig. We will discuss the paper by \citet{Mensing_1926-uber-die-dielektrizitatkonstante} in sec.\ 5.2.2.}$^]$ (AHQP interview, session 2, p.\ 5).
\end{quotation}
\end{singlespacing}

\noindent
\citet{VanVleck_1971-reminiscences} later called it a ``quadruple tie" (p.\ 7), adding a  paper by Charles \citet{Manneback_1926-die-elektrizitaetskonstante}. The latter, however, actually cited  \citet{Mensing_1926-uber-die-dielektrizitatkonstante} and claimed priority only for having derived the result in wave rather than matrix mechanics \citep[p.\ 564]{Manneback_1926-die-elektrizitaetskonstante}.\footnote{\citet[p.\ 567]{Manneback_1926-die-elektrizitaetskonstante} acknowledged Debye's interest and encouragement in this effort to recover the formula first published by \citet{Debye_1912-einige}.} Still, the papers by \citet{Mensing_1926-uber-die-dielektrizitatkonstante}, \citet{Kronig_1926-the-dielectric}, and \citet{Manneback_1926-die-elektrizitaetskonstante} all made it into print  before the note by \citet{VanVleck_1926_magnetic} finally appeared in the issue of {\it Nature} of August 14.\footnote{Referring to the 1926 note in his 1932 book, he described it as an ``abstract only" \citep[p.\ 147]{VanVleck_1932-the-theory}. See \citet[pp.\ 143--148]{Fellows_1985-j} for detailed discussion of Van Vleck's note and a reconstruction of some of the derivations he suppressed for brevity.\label{nature note}}  However, as \citet{Pauli_1933-review} would concede in his review of Van Vleck's 1932 book, it would fall to \citet{VanVleck_1927_on-dielectric1, VanVleck_1932-the-theory} to show in full generality that the new quantum mechanics restored the value $1/3$ for the factor $C$. These 1926 papers only dealt with  the special case in which the rigid rotator was used to model the gas molecules.\footnote{In sec.\ 5.2.3, we will give the flavor of Van Vleck's general derivation but we will only cover the details of his derivation for this special case.}

%\citet[p.\ 510, p.\ 512]{Mensing_1926-uber-die-dielektrizitatkonstante}, for instance, explicitly mentioned that they recovered this value only for the special case of the dumbbell model of diatomic molecules such as HCl. They nonetheless clearly recognized that the deviations from $C=1/3$ found in the old quantum theory were implausible in hindsight, noting in the concluding paragraph of their paper that the new result ``is completely opposite to the results that were obtained on the basis of the earlier quantum theory$^[$\footnote{Here the authors cite \citet{Pauli_1921-zur-theorie} and \citet{Pauling_1926-the-quantum}.}$^]$ \ldots\ according to which the coefficient $C$ \ldots\ should have a numerical value substantially different from $1/3$ even in the limiting case of high temperatures" \citep[p.\ 512]{Mensing_1926-uber-die-dielektrizitatkonstante}. In September 1926, Pauling too redeemed himself for the ``wonderful nonsense"  he had contributed earlier \citep{Pauling_1926-the-quantum}, submitting a paper in which he showed that the old quantum theory predicts that a magnetic field will strongly affect a substance's electric susceptibility, whereas the new quantum mechanics, in accordance with experiment, predicts no such effect \citep{Pauling_1927-the-influence}.

While crossing the Atlantic in June 1926, Van Vleck finished another calculation in quantum mechanics only to discover upon reaching Copenhagen that he had been scooped by Heisenberg. He thereupon extended this perturbative calculation to higher order but was scooped again, this time by Ivar Waller \citep[p.\ 108]{Fellows_1985-j}. For the remainder of the summer, Van Vleck worked primarily on calculating the specific heat of hydrogen, another ill-fated endeavor.  Dennison would solve that puzzle \citep[sec.\ 12, pp.\ 183--188]{Gearhart_2010-astonishing}.  On a train from Switzerland to Paris, Van Vleck happened to run into Pauling, whom he had not met in person before. Pauling told Van Vleck that he had become interested in calculating electric susceptibilities for molecules of new shapes, specifically symmetrical tops. They resolved to write a joint paper on the subject, but \citet{Kronig_1926-the-dielectric-constant-of-symmetrical} once again beat them to the punch  \citep[pp.\ 111--112, pp.\ 114--115, pp.\ 148--150]{Fellows_1985-j}. Despite this losing streak, Van Vleck's work during this period did sow the seeds of further research. His 1926 note briefly mentions an application of his approach to paramagnetism \citep[p.\ 227, discussed by Fellows, 1985, p.\ 152]{VanVleck_1926_magnetic}.  Ultimately, a general treatment of susceptibilities, both electric and magnetic, would cement his reputation as a theorist.

When Van Vleck returned to the United States, he found that quantum theorists were in high demand and that the publication of his NRC {\it Bulletin} had earned him a reputation as one of the few in the United States who had a grasp of the theory.  He had also found time that summer to write a short report on the new quantum mechanics for the Progress Committee of the Optical Society of America.  Leonard R.\ Ingersoll at the University of Wisconsin called it ``the only readable synopsis of the present situation in this difficult subject'' \citep[p.\ 162]{Fellows_1985-j}

%That fall, he was invited as a guest lecturer at the University of Iowa where he held forth both on the new quantum theory and his own theory of electric and magnetic susceptibilities.  Stewart, who had invited him, wrote that ``they [were] the most satisfying theoretical lectures'' he had seen (ibid., p. 163).

As Van Vleck's fame increased, he found himself wooed more and more doggedly by other universities.  From the fall of 1926 through the spring of 1928, he declined offers from the University of Chicago, Princeton, and the Mellon Institute.  Many of these he rejected out of a sense of loyalty to the University of Minnesota which had been so generous to him.  The department continued to recognize Van Vleck's value, following up with raises and promotions. In June 1926 he had become an associate professor and only a year later he became a full professor. By the summer of 1927, having married Abigail June Pearson, a native Minnesotan, he had established family ties to the state as well. It took an offer from his {\it alma mater} to win him over, and even then he vacillated for over a year before accepting a position at the University of Wisconsin \citep[pp.\ 169--175]{Fellows_1985-j}.  He arrived at Madison in time for the fall semester of 1928.

Over the same period, Van Vleck had been busy pursuing the line of inquiry that would secure him fame as an expert in magnetism.  He published a three-part paper that advanced a general theory of susceptibilities \citep{VanVleck_1927_on-dielectric1, VanVleck_1927_on-dielectric2, VanVleck_1928_on-dielectric}.  This trilogy would form the basis for 
%his masterful 
{\it The Theory of Electric and Magnetic Susceptibilities} \citep{VanVleck_1932-the-theory}.

Before turning to that volume in the next section, we wrap up this section with some brief comments about Van Vleck's career after he left the Midwest. In early 1934, 
%as his father had predicted he would once he had made a name for himself, 
Van Vleck was offered an associate professorship at Harvard to replace Slater, 
%his former fellow graduate student,  
who had left Harvard for MIT  \citep[p.\ 343]{Fellows_1985-j}. Harvard offered conditions Wisconsin could not match, not the least important of which was the renewed proximity to Kemble and Slater. Although it initially bothered him that he was not offered a full professorship right away, Van Vleck was satisfied by Harvard's assurances that he would quickly be promoted, so he and Abigail moved to Cambridge in the fall of 1934 (ibid., p.\ 350). Within a year he was made a full professor. 

\begin{figure}[h]
   \centering
   \includegraphics[width=5.5in]{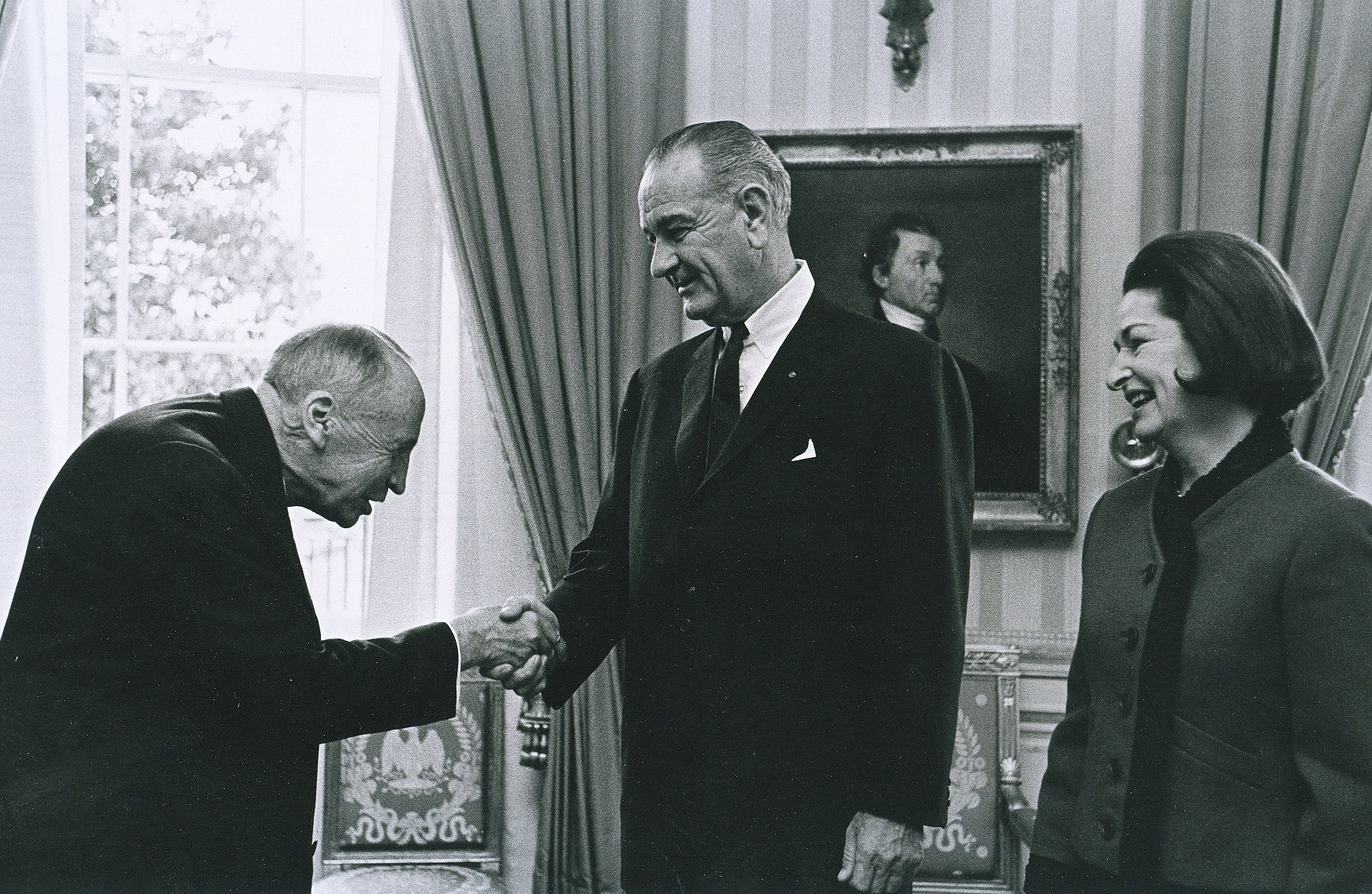} 
   \caption{Van Vleck receiving the National Medal of Science in 1966 from President Lyndon B.\ Johnson with Lady Bird Johnson looking on (picture courtesy of John Comstock).}
   \label{fig3}
\end{figure}

During World War II, Van Vleck was the head of the theory group at Harvard's Radio Research Laboratory, thinking about ways to jam enemy radar, and a consultant to MIT's much bigger Radiation Laboratory \citep[p.\ 514]{Anderson_1987-john}. 
%jamming enemy radar
From 1945 to 1949 he was chair of Harvard's physics department (ibid., p.\ 519). In 1951, he succeeded Bridgman
%, one of his teachers,
in the Hollis Chair of Mathematics and Natural Philosophy, a position he held until his retirement at the age of seventy in 1969. From 1951 to 1957, he served as Dean of Engineering and Applied Physics, and in 1952--53, he served a term as President of the American Physical Society (ibid.), something the more temperamental Slater never did, even though the two men were of comparable stature in the postwar American physics community. Of the many honors bestowed upon Van Vleck we already mentioned his share in the 1977 Nobel Prize and the Lorentz medal and will add only the National Medal of Honor, which he received out of the hands of President Lyndon B.\ Johnson in 1966 (see Fig.\ \ref{fig3}).

Even though Van Vleck spent the better part of his career at Harvard, he always retained a soft spot for Minnesota and Wisconsin. Together with Roger Stuewer (University of Minnesota) and Chun Lin (University of Wisconsin--Madison), he wrote an article on the origin of the popular fight songs ``On Wisconsin" and ``The Minnesota Rouser."
%for the football teams of Wisconsin and Minnesota. 
This article, which Van Vleck when talking to his co-authors would facetiously but affectionally call ``our {\it magnum opus}" (Roger Stuewer, private communication), appeared in slightly different versions in the alumni magazines of both universities \citep{Lin_1977-on-minnesota, Lin_1980-wisconsin}. As an undergraduate,  Van Vleck had been in the Wisconsin band, probably playing the flute \citep[p.\ 503]{Anderson_1987-john}. As a young boy, he had attended the game in Madison in November 1909 that saw the premiere of ``On Wisconsin." Unfortunately for young Van Vleck, the Badgers lost that game to the visiting Gophers \citep[p.\ 4]{Lin_1977-on-minnesota}. When many decades later he won the Nobel Prize, Stuewer sent him a one-word telegram: ``SKI-U-MAH." This is a Minnesota football cheer, which supposedly, as Stuewer had explained to Van Vleck earlier, is an old Native American war cry meaning ``victory." Van Vleck wrote back that of all the congratulatory messages he had received this one was ``the briefest and most to the point."\footnote{Stuewer to Van Vleck, October 11, 1977 (telegram); Van Vleck to  Stuewer, November 16, 1977. We are grateful to Roger Stuewer for providing us with copies of this correspondence.}

\begin{singlespacing}
\section{The Theory of Electric and Magnetic Susceptibilities}
\end{singlespacing}

\subsection{Writing the 1932 Book}

In 1928, Van Vleck had been thinking about writing his own book on quantum mechanics, but he became interested that fall when Ralph H.\ Fowler suggested that he write a book about susceptibilities for Oxford's International Series of Monographs on Physics instead. The idea of expanding his 1927--28 trilogy on susceptibilities  \citep{VanVleck_1927_on-dielectric1, VanVleck_1927_on-dielectric2, VanVleck_1928_on-dielectric} into a book appealed to him. As he wrote to Fowler: ``These papers would, in fact, be in a certain sense the backbone of what I would have to say." Fowler and the other editors of the international monographs series were eager to accept a volume on any theoretical subject Van Vleck might ``care to write about, and allow [them] to publish.'' Van Vleck liked the idea, but, the drawn-out process of writing the {\it Bulletin} still fresh in his mind,  warned Fowler of the ``adiabatic speed'' at which he wrote.\footnote{The quotations are from Fowler to Van Vleck, November 26, 1928, and Van Vleck to Fowler, November 28, 1928 \citep[pp.\ 233--234]{Fellows_1985-j}.} The caveat was well warranted.  It would take Van Vleck over three years to complete {\it The Theory of Electric and Magnetic Susceptibilities}.

The delays were of a different nature than the trials and tribulations that had prevented a slightly younger Van Vleck from publishing his completed ``article'' in the NRC {\it Bulletin}.  This time, he made his own original research a higher priority.  
%Before even beginning work on the new book, he finished a paper on molecular spectra that was eagerly awaited by Robert Mulliken (his new data were explained by Van Vleck's analysis).  
He also accepted several invitations to give talks in Iowa, Minneapolis, and New York.  This, and supervising the research of his graduate students and postdocs, took up most of his time during the 1928--29 school-year.  He did manage to squeeze in one chapter, however.  ``I have actually, mirab[i]le dictu, completed one chapter of my book,'' he wrote to Fowler in June, 1929.  ``At such a rate, you can calculate how long it will take me to write eleven more'' \citep[p.\ 238]{Fellows_1985-j}.  Clearly, Van Vleck was going miss his original Spring 1930 deadline.

%That summer, while teaching in Madison, he would make another original calculation concerning the likelihood of transitions in molecular spectra that involved changes in both the ``electronic'' and ``vibrational'' quantum numbers.  Some had thought that large changes in the vibrational quantum number might be possible in transitions of this kind (a change of 20 or more), but Van Vleck showed that while these transitions were not completely prohibited, they were extremely unlikely (ibid., p.\ 239).

After spending the summer on research, he devoted all of his free time in the fall to the book and completed another chapter.  The following spring, 1930, he negotiated a sabbatical leave in which he received half of his salary from Wisconsin, and made up the rest with a Guggenheim fellowship.  He and Abigail went to Europe, making stops in England, Holland, and Germany.  Finally, Van Vleck went to Switzerland while Abigail joined his parents for a tour of Italy.  Unfortunately, when Van Vleck arrived at the {\it Eidgen\"ossische Technische Hochschule} (ETH) in Zurich, he discovered that Pauli and other faculty were away on lengthy spring vacations (ibid, pp.\ 240--241). Van Vleck turned this to his advantage:

\begin{singlespacing}
\begin{quotation}
\noindent
The janitor at the ETH, fortunately, was very friendly and arranged for me to have the use of the library. I lived comfortably at the Hotel Waldhaus Dolder, and with a portable typewriter and no distractions by colloquia, social life or sight-seeing, I probably wrote more pages of my `Theory of Electric and Magnetic Susceptibilities' in my first month at Zurich than in any other comparable time interval \citep[p.\ 1236, quoted by Fellows, 1985, p.\ 242]{VanVleck_1968-my-swiss}.
\end{quotation}
\end{singlespacing}

%A kindly janitor gave him access to the library.  Van Vleck then holed up at the hotel {\it Waldhaus Dolder} with his portable typewriter and wrote uninterrupted for almost a month.  
\noindent
When Pauli returned from vacation and heard what Van Vleck had been up to, he was dismissive.  ``I don't republish my papers as a book,'' he said.  Partly in response to this criticism, Van Vleck resolved to include more original research \citep[p.\ 242]{Fellows_1985-j}.

In June 1930, Van Vleck received an invitation to the Sixth Solvay Congress, devoted to magnetism. In his contribution, \citet{VanVleck_1932-note} derived formulas for magnetic susceptibilities, using the same techniques  he had used in his 1927--28 trilogy and would use again in his book \citep{VanVleck_1927_on-dielectric1, VanVleck_1927_on-dielectric2, VanVleck_1928_on-dielectric, VanVleck_1932-the-theory}. He did not mention the failures of the old quantum theory with one word. It is possible that this was simply because he was talking about magnetic rather than electric susceptibilities, but it may have been, at least in part, because he did not want to incur the wrath of Pauli, whose 1921 paper, after all, was a prime example of the ``wonderful nonsense" the old quantum theory had produced on the subject.

After receiving permission from Wisconsin, he extended his trip into the fall, finally returning in October with the book almost complete. Serber had already begun as a graduate student. Van Vleck started him on a research problem immediately, not realizing he was only a first-year student.  The following spring, 1931, he enlisted the help of Serber and another graduate student, Amelia Frank \citep[p.\ viii]{VanVleck_1932-the-theory}, in proof-reading the galleys of the book.  True to form, Van Vleck continued to add material and make myriad corrections during these final phases.  Finally, in September, the publisher wrote to him, warning that he would be billed personally if he continued to ignore the usual limit of twenty corrections per proof-sheet.  He completed the corrections in December. The book was published in April 1932 \citep[ pp.\ 247--248]{Fellows_1985-j}.

Reviewers immediately recognized its importance.\footnote{For discussion of the book's reception, see \citet[pp.\ 282--284]{Fellows_1985-j}.}  
%J. E. Lennard-Jones predicted that it would become ``the standard book of reference on [the] subject'' (ibid., p.\ 283).
%\citep[p.\ 283]{Fellows_1985-j}.  
%Felix Bloch remarked on its exceptional clarity, and E. C. Stoner noted that the combination of wave mechanics and matrix methods wielded by Van Vleck ``proved a powerful weapon in dealing with the problems under consideration" (ibid.).  
%It was almost uniformly well-received.  
Even Pauli, whose caustic remarks about Born and Jordan's {\it Elementary Quantum Mechanics} we quoted in sec.\ 3.2, had nothing but praise for the volume that he had originally dismissed as a rehash of old papers.  This is all the more remarkable given that Van Vleck sharply criticized Pauli's (1921) own early contribution to the subject. \citet{Pauli_1933-review} called Van Vleck's book ``a careful and complete overview of the entire field \ldots\ of the dielectric constant and the magnetic susceptibility" (see also the quotations in note \ref{pauli pauling review}).
%in sec.\ 1.2.2). 
He recognized that many of the results reported in the book had first been found by Van Vleck himself, such as ``the general proof for the occurrence of the numerical factor 1/3 in the Langevin-Debye formula'' (ibid.). Pauling, who shared the responsibility for the ``wonderful nonsense" about susceptibilities in the old quantum theory with his near namesake, also wrote a glowing review, calling Van Vleck's book an ``excellent treatise \ldots\ written by the world's leading authority in the field" \citep[cf.\  note \ref{pauli pauling review}]{Pauling_1932-review}. Unlike the NRC {\it Bulletin} (recall Ruark's complaint quoted at the end of sec.\ 3.1), the 1932 book came in a durable binding, as Pauling noted in the last line of his review: ``The volume is handsomely printed, with pleasing typography and binding" (ibid.). It is tempting to read this as a tongue-in-cheek reference to the last line of Pauli's review of Born and Jordan's book (quoted in sec.\ 3.2), though Pauling's variation on this theme has none of the venom of Pauli's original.

\subsection{The 1932 Book and Spectroscopic Stability}
%the factor $1/3$ in the Langevin-Debye formula for susceptibilities lost and regained}

Van Vleck's \emph{The Theory of Electric and Magnetic Susceptibilities} is remarkable both for the wide range of concepts its covers and techniques it assembles, and for the amount of discussion devoted to the historical development of the theories under consideration. Even though the main focus of the book is on gases, it ended up, as we mentioned in the introduction, setting ``a standard and a style for American solid-state physics'' \citep[p.\ 524]{Anderson_1987-john}. As Van Vleck explained in the preface:

\begin{singlespacing}
\begin{quotation}
\noindent
At the outset I intended to include only gaseous media, but the number of paramagnetic gases is so very limited that any treatment of magnetism not applicable to solids would be rather unfruitful \citep[p.\ vii]{VanVleck_1932-the-theory}.
\end{quotation}
\end{singlespacing}

\noindent
In the book, Van Vleck clearly demonstrated how his general Langevin-Debye formula for susceptibilities in gases can be adapted to the study of magnetism in crystalline solids, sketching out the research program that would occupy him and his students for years to come.

The book can roughly be divided into two parts, separated by an interstitial aside concerning the defects and demise of the old quantum theory.  
%The first four chapters 
Chs.\ I--IV constitute the first part.  Here Van Vleck surveyed the classical theories of electric and magnetic susceptibilities.  In addition to marshaling resources that will be drawn from in later chapters, Van Vleck carefully examined the failings of the classical theories, motivating the quantum-mechanical approach that is developed in the book's second half.  
%The fifth chapter 
Ch.\ V is the interstitial aside, which we will discuss in more detail in section 5.2.2.  
%The sixth chapter 
Ch.\ VI begins the book's second half, which develops a quantum-theoretical approach to electric and magnetic susceptibilities. Like Ch.\ XI of the NRC {\it Bulletin} on mathematical techniques, this chapter on ``Quantum-Mechanical Foundations,'' is by far the longest of the book. It takes up 59 pages (Ch. XI of the {\it Bulletin} ran to 50 pages). It is so complete that, as we mentioned in sec.\ 1.2.2, it was sometimes used by itself as an introductory text in courses on the new theory.  Although Van Vleck's work had largely been in the tradition of matrix mechanics, his general exposition of quantum mechanics, in his book as well as in his lectures (as evidenced by the lecture notes mentioned in note \ref{student notes}), has none of the ``G\"ottingen parochialism" \citep[p.\ 641]{Duncan_2008-Pascual} of Born and Jordan's (1930)  {\it Elementary Quantum Mechanics}. As Van Vleck wrote about Chapter VI in the preface:

\begin{singlespacing}
\begin{quotation}
\noindent
I have tried to correlate and intermingle the use of wave functions and of matrices, rather than relying exclusively on the one or the other, as is too often done. It is hoped that this chapter may be helpful as a presentation of the perturbation machinery of quantum mechanics, quite irrespective of the magnetic applications \citep[p.\ viii]{VanVleck_1932-the-theory}.\footnote{As one of the reviewers of the book noted, ``particular attention [is] being paid to the relation between the wave and matrix methods, a combination of which, in Van Vleck's hands, has proved a powerful weapon in dealing with the problems under consideration" \citep[p.\ 490]{Stoner_1932-magnetism}.}
%quoted by Fellows, 1985, p.\ 283
\end{quotation}
\end{singlespacing}

Chs. VII--XII interrogate and extend Van Vleck's general Langevin-Debye formula, sometimes with impressive numerical accuracy, as in the case of para\-magnetism, where Van Vleck had made one of his most famous contributions to the field by 1932, and sometimes qualitatively with suggestions for future lines of research, as in the case of fields within crystals and ferromagnetism. Ch.\ XIII, finally, is devoted to some related optical phenomena. The first section of this chapter (sec.\ 82, pp.\ 361--365) is devoted to the Kramers dispersion formula.

The book does exactly what a good textbook ought to do according to \citet[see, e.g., the passages on pp.\ 136--137 and p.\ 187 quoted in sec.\ 1.2.2]{Kuhn_1996_the-structure}.\footnote{Cf.\ the characterizations of Van Vleck's book in the reviews by Pauli and Pauling quoted in note \ref{pauli pauling review}.} It not only set much of the agenda for the research program envisioned by its author, it did so in the form of a pedagogically carefully constructed text in which all the relevant theoretical and experimental literature is reviewed and all the required mathematical techniques are introduced, along with their canonical applications, all with the aim, ultimately, of preparing its readers to become active contributors to this research program themselves.

The book also reflects Van Vleck's own trajectory from his early work in the old quantum theory to the line of work in the new quantum theory that won him his reputation as one of the pioneering theorists of solid-state physics in the United States (cf.\ the remarks by Anderson quoted above). Although he changed fields in the process, Van Vleck's journey from spectra to susceptibilities shows a remarkable continuity. To highlight this continuity, we already drew attention (see sec.\ 1.2.3) to the connection between the derivation of the Kramers dispersion formula in his early work \citep{VanVleck_1924-the-absorption1, VanVleck_1924-the-absorption2, VanVleck_1926-quantum} and the derivation of the Langevin-Debye formula for electric susceptibilities, which played a central role in Van Vleck's work in quantum mechanics that began in 1926 and reached a milestone with his second book \citep{VanVleck_1926_magnetic, VanVleck_1927_on-dielectric1, VanVleck_1927_on-dielectric2, VanVleck_1928_on-dielectric, VanVleck_1932-the-theory}.  As we emphasized in sec.\ 1.2.3, it is the perturbation theory used in both derivations that provides the continuity in the transition from classical theory to the old quantum theory to modern quantum mechanics and in Van Vleck's career move from spectra to susceptibilities. As we will see in this section, it is the old quantum theory's problems with the quantization of specific periodic systems that is responsible for the  discontinuity and the Kuhn loss in the area of susceptibilities, and it is the new quantum theory's systematic solution to the problem of how to quantize such systems that is behind the recovery of that Kuhn loss.

The Langevin-Debye formula for the electric susceptibility $\chi$ of some gas is
\begin{equation}
\chi = N \left(\alpha + \frac{\mu^2}{3kT} \right),
\label{langevin debye formula}
\end{equation}
where $N$ is the number of molecules, $\alpha$ is a constant, $\mu$ is the permanent electric moment of the molecule under consideration, $k$ is Boltzmann's constant, and $T$ is the temperature
%\citep[p.\ 28]{VanVleck_1932-the-theory}.
(Van Vleck, 1927a, p.\ 727, 1932b, p.\ 28).\footnote{The electric susceptibility $\chi$ is related to the dielectric constant $\varepsilon$ of the gas via:
$\chi = (3/4 \pi)\, (\varepsilon - 1)/(\varepsilon + 2)$
%\chi = \frac{3}{4 \pi} \frac{\varepsilon - 1}{\varepsilon + 2}
(see. e.g., Pauli, 1921, p.\ 319; Pauling, 1926, p. 568, Van Vleck, 1927b, p.\ 32, 1932b, p.\ 28).} The first term comes from the induced moment of the molecule, resulting from the deformation of the molecule by the external electric field. The second term comes from the alignment of the permanent moment of the molecule with the field. Thermal motion will frustrate this alignment, which is expressed in the inverse proportionality to the temperature $T$. As Van Vleck noted when he introduced the formula in his book:

\begin{singlespacing}
\begin{quotation}
\noindent
The idea of induced polarization is an old one \ldots\ The suggestion that part of the electric susceptibility might be due to alinement [sic] of permanent moments, resisted by temperature agitation, does not appear to have been made until 1912 by Debye [1912].
%$^[$\footnote{At this point a  footnote is inserted with a reference to a paper by \citet{Debye_1912-einige}.}$^]$  
A magnetic susceptibility due entirely to the orientation of permanent moments was suggested some time previously, in 1905, by [Paul] Langevin [1905a,b],
%$^[$\footnote{At this point a  footnote is inserted with references to two papers by Paul \citet{Langevin_1905-magnetisme, Langevin_1905-sur-la-theorie}.}$^]$ 
and the second term of [Eq.\ \eqref{langevin debye formula}] is thus an adaptation to the electric case of Langevin's magnetic formula. (In the electric case, a formula such as [\eqref{langevin debye formula}] is commonly called just the Debye formula, but we use the compound title Langevin-Debye in order to emphasize that the mathematical methods which we use to derive the second term of [Eq.\ \eqref{langevin debye formula}] apply equally well to magnetic or electric dipoles.)
 \citep[p.\ 30]{VanVleck_1932-the-theory}.
 %\footnote{Van Vleck gave citations to papers of both Paul \citet{Langevin_1905-magnetisme, Langevin_1905-sur-la-theorie} and \citet{Debye_1912-einige}.}
\end{quotation}
\end{singlespacing}

%of this second term. 
%temperature-dependent term comes from the permanent moment of the molecule, as this moment's tendency to align itself perfectly with the external field competes with the molecule's thermal motion. 
\noindent
It is this  temperature-dependent second term that Van Vleck was most interested in. We can write this term as
\begin{equation}
\frac{NC \mu^2}{kT}.
\label{C}
\end{equation}
Both classical theory and quantum mechanics correctly predict that, under very general conditions, $C = 1/3$. The two theories agree except at very low temperatures, where the classical theory breaks down and where quantum mechanics gives deviations from $1/3$ \citep[p.\ 185, p.\ 197]{VanVleck_1932-the-theory}. Other than that, the factor $1/3$ is a remarkably robust prediction of both theories. It is true for a wide range of models (e.g., dumbbell, symmetrical top) and it is independent of the choice of a $z$-axis for the quantization of the $z$-component of the angular momentum in these models. The latter feature is an example of what Van Vleck called ``spectroscopic stability." As he put it in Part I of the trilogy that provided the backbone for his 1932 book:

\begin{singlespacing}
\begin{quotation}
\noindent
[T]he high spectroscopic stability characteristic of the new quantum mechanics is the cardinal principle underlying the continued validity of the Langevin-Debye formula. We shall not attempt a precise definition of the term ``spectroscopic stability."$^{[}$\footnote{Yet another illustration of the continuity of Van Vleck's research across the quantum revolution of 1925--26 is that the footnote inserted at this point refers to the subsection, ``The Hypothesis of Spectroscopic Stability," of sec.\ 54, ``The Polarization of Resonance Radiation," of his NRC {\it Bulletin}  \citep[pp.\ 171--173]{VanVleck_1926-quantum}.\label{spec stab 0}}$^{]}$  It means roughly that the effect of orientation or of degeneracy in general is no greater than in the classical theory, and this usually implies that summing over a discrete succession of quantum-allowed orientations gives the same result as a classical average over a continuous distribution \citep[p.\ 740]{VanVleck_1927_on-dielectric1}.\footnote{On the next page, before giving his general proof of spectroscopic stability, Van Vleck noted that a similar result for a special case had already been established in the {\it Dreim\"annerarbeit} \citep[p.\ 590]{Born_1926-zur-quantenmechanik} and that  he was ``informed that the more general result has also been obtained independently by Born (unpublished)" \citep[p.\ 741]{VanVleck_1927_on-dielectric1}. So, as in the old quantum theory (see sec.\ 3.2), Born and Van Vleck were pursuing similar lines of research in matrix mechanics. In the discussion of electric susceptibilities in their book, \citet[sec.\ 42, pp.\ 212--225]{Born_1930-elementare} followed Van Vleck, citing (ibid., p.\ 219) his note in {\it Nature} and the trilogy in {\it Physical Review} \citep{VanVleck_1926_magnetic, VanVleck_1927_on-dielectric1, VanVleck_1927_on-dielectric2, VanVleck_1928_on-dielectric}. Born and Jordan did not use the term `spectroscopic stability' in this context (see note \ref{spec stab 2}).\label{spec stab 1}}
\end{quotation}
\end{singlespacing}

The old quantum theory gave values for $C$ much greater than $1/3$, as \citet{Pauli_1921-zur-theorie} and \citet{Pauling_1926-the-quantum} discovered using the rigid rotator as their model for the gas molecules  
%Pauli and Pauling found $C = 1.54$ and $C=4.57$, respectively 
(see the table in Fig.\ \ref{fig1}). Redoing the calculation in matrix mechanics, \citet{Mensing_1926-uber-die-dielektrizitatkonstante} recovered the value $C=1/3$ for this special case, as did  \citet{Kronig_1926-the-dielectric}, \citet{Manneback_1926-die-elektrizitaetskonstante}, and  \citet{VanVleck_1926_magnetic} (see sec.\ 4). Van Vleck, however, was the only one who stated explicitly that this result is independent of the choice of the axis of quantization of the rigid rotator's angular momentum: ``in the matrix theory the susceptibility is the same with spacial$^[$\footnote{This is how Van Vleck consistently spelled `spatial'.}$^]$ quantization relative to the applied field as with random orientations" (ibid., p.\ 227; see Fellows, 1985, p.\ 144). Van Vleck managed to salvage a plausibility argument for this claim when he had to shorten his note for {\it Nature} (see sec.\ 5.2.2). In subsequent publications, he  gave the full proof, not just for the rigid rotator but for a broad class of models \citep{ VanVleck_1927_on-dielectric1, VanVleck_1932-the-theory}. 

That the susceptibility of a gas of rigid rotators does not depend on the axis of quantization is an example of spectroscopic stability. In his book, Van Vleck devoted considerable space to the ``principle" or the ``theorem" of spectroscopic stability \citep[p.\ 111, p.\ 139]{VanVleck_1932-the-theory}.  Before giving a mathematical proof (ibid., sec.\ 35, pp.\ 137--143), he explained the situation qualitatively in the chapter on the old quantum theory (ibid., sec.\ 30, 111--113). After conceding that the term, which he took from \citet[p.\ 85]{Bohr_1918-on-the-quantum}, ``is not a particularly happy one" (ibid., p.\ 111),\footnote{Bohr introduced the term in the context of the Zeeman effect: ``from a consideration of the necessary ``stability" of spectral phenomena, it follows that the total radiation of the components, in which a spectral line, which originally is unpolarized, is split up in the presence of a small external field, cannot show characteristic polarisation with respect to any direction" \citep[p.\ 85]{Bohr_1918-on-the-quantum}. \citet[p.\ 13, p.\ 106, p.\ 161]{Born_1930-elementare} attributed the term `spectroscopic stability' to Heisenberg, citing a paper submitted in November 1924 on the polarization of fluorescent light \citep{Heisenberg_1925-ueber-eine-anwendung}. Van Vleck also emphasized the connection between spectroscopic stability and the polarization of resonance radiation (Van Vleck, 1926b, p.171 [see note \ref{spec stab 0}]; 1927a, p.\ 730; 1932b, p.\ 111). The first example \citet[pp.\ 12--13]{Born_1930-elementare} gave of spectroscopic stability is the Thomas-Kuhn sum rule (see sec.\ 3.2).\label{spec stab 2}} he wrote:

\begin{singlespacing}
\begin{quotation}
\noindent
[I]t can for our purposes be considered identical with the idea that the susceptibility is invariant 
%[read: independent] NO: it's invariance of the quantization under rotation of the axes
of the type of quantization, or in the special case of spacial quantization, that summing over the various quantized orientations is equivalent, as far as results are concerned, to a classical integration over a random orientation of orbit. It is indeed remarkable that a discrete quantum summation gives exactly the same answers as a continuous integration. This was not at all true in the old quantum theory \citep[p.\ 111]{VanVleck_1932-the-theory}.
\end{quotation}
\end{singlespacing}

%``Spectroscopic Stability in the New Quantum Mechanics," he explained the situation qualitatively 

%\citet[pp.\ 111--113; sec.\ 35, pp.\ 137--143]{VanVleck_1932-the-theory} 
%This theorem guarantees that an observable's matrix elements will have the same squared average over the degenerate states in the unperturbed system, as they will over those same states once the degeneracy is lifted, for instance, as in the case of susceptibilities, by the application of an external field. Ultimately, it is this feature that restores the factor $1/3$ in the temperature-dependent term in the formula for electric susceptibilities.
%This theorem when applied to susceptibilities assures us that the results are independent of the choice of an ``axis of quantization''

In the three subsections that follow, we present derivations of the formula for the electric susceptibility in gases in classical theory (sec.\ 5.2.1), the old quantum theory (sec.\ 5.2.2), and quantum mechanics (sec.\ 5.2.3). In the quantum theory, old and new, we focus on the special case in which the gas molecules are modeled as rigid rotators. We will  see how the robustness of the value $C= 1/3$ was established, lost, and regained. In secs.\ 5.2.1 and 5.2.3, we follow \citet{VanVleck_1932-the-theory}. In sec.\ 5.2.2, we follow  \citet{Pauli_1921-zur-theorie}, \citet{Pauling_1926-the-quantum, Pauling_1927-the-influence}, and \citet{Mensing_1926-uber-die-dielektrizitatkonstante}, though we will also quote liberally from Ch.\ V of Van Vleck's 1932 book. In this chapter, ``Susceptibilities in the old quantum theory contrasted with the new," the author used some uncharacteristically strong language to describe the shortcomings of the old quantum theory in this area.

\subsubsection{Susceptibilities in Classical Theory}

The susceptibility of a gas, $\chi$, is a measure of how the gas responds to external fields.  We will consider the electric susceptibility in particular.  The field, $\bold{E}$, and polarization, $\bold{P}$, are assumed to be parallel, and the medium is assumed to be both isotropic and homogenous.  Predictions of $\chi$ require one to deal with the motions of the systems used as models for the gas molecules and their constituent atoms: the specific behavior of these systems in response to the external field will determine their electric moments, and in turn, the polarization of the medium.

Consider a small volume of a gas of molecules with permanent dipole moments, such as HCl.  When an electric field is applied, say in the $z$-direction of the coordinate system we are using, the molecules experience a torque that tends to align them with the field.  In addition, the charges in each molecule will rearrange themselves in response.  If the field is too weak to cause ionization, the charges will settle into equilibrium with the field, creating a temporary induced electric moment.  Both of these effects contribute to a molecule's electric moment $\bold{p}$.  Following Van Vleck, we largely focus on the first of these effects, which, as mentioned above, is responsible for the temperature-dependent term in the Langevin-Debye formula (see Eqs.\ \eqref{langevin debye formula}--\eqref{C}).

To find the polarization, $\bold{P}$, we need to take two averages over the component of these electric moments in the direction of the field $\bold{E}$, in this case the $p_z$ component. First, we need to average $p_z$ over the period(s) of the motion of the molecule (or in the case of quantum theory, over the stationary state). This is indicated by a single overbar: $\overline{p_z}$. Second, we need to average this time-average $\overline{p_z}$ over a thermal ensemble of a large number $N$ of such molecules. This is indicated by a double overbar: $\overline{\overline{p_z}}$. All derivations of expressions for the susceptibility call for this two-step averaging procedure.\footnote{\citet{Pauling_1926-the-quantum} gives a particularly clear statement of this procedure: ``[double bar] is the average value \ldots\ for all molecules in the gas, and [single bar] is the time average \ldots\ for one molecule in a given state of motion" (p.\ 568).} 

The strength $P$ of the polarization is  given by:
\begin{equation}
P =N\overline{\overline{p_z}}.
\label{statistical link}
\end{equation}
The electric susceptibility, $\chi$, is defined as the ratio of the strengths of the polarization and the external field:
\begin{equation}
\chi \equiv \frac{P}{E}= \frac{N}{E} \, \overline{\overline{p_z}}.
\label{susceptibility definition}
\end{equation}
When it comes to the derivation of expressions for $\chi$, the various theories differ only in how  $\overline{p_z}$  and $\overline{\overline{p_z}}$ are obtained.
%\citep[p.\ 4]{VanVleck_1932-the-theory}.  

We first go through the calculation in the classical theory, covered elegantly in Ch. II of Van Vleck's book, ``Classical Theory of the Langevin-Debye Formula" \citep[pp.\ 27--41]{VanVleck_1932-the-theory}. Consider a multiply-periodic system with $f$ degrees of freedom, which, in its unperturbed state, is described by the Hamiltonian $H^0$, and which is subjected to a small perturbation coming from an external electric field $\bold{E}$ in the $z$-direction.  The Hamiltonian for the perturbed system can then be written as the sum $H^0 + H^{\rm int}$, where $H^{\rm int} < \! \! < H^0$. In this case, the full Hamiltonian is given by:
\begin{equation}
\label{f}
H = H^0 - E p_z.
\end{equation}
As in his NRC {\it Bulletin},  \citet[p.\ 38]{VanVleck_1932-the-theory} used action-angle variables $(w^0_k, J^0_k)$ ($k = 1, \dots, f$) for the {\it unperturbed} Hamiltonian $H^0$, even when dealing with the full Hamiltonian (see also Van Vleck, 1927b, p.\ 50; cf.\ our discussion in sec.\ 3.2).

The $z$-component of the polarization of the system, $p_z$, can be written as a Fourier expansion. For a system with only one degree of freedom the expansion is given by:\footnote{\citet[p.\ 38]{VanVleck_1932-the-theory} writes $p^{(z)}_\tau$ for the complex amplitudes $(p_z)_\tau$ and suppresses the argument $J^0$ in his notation.}
\begin{equation}
\label{d}
p_z(w^0,J^0) = \sum_{\tau = 0, \pm 1, \pm 2, \ldots}^\infty (p_z)_\tau(J^0) \,
e^{\displaystyle 2 \pi i \tau w^0}.
\end{equation}
Essentially the same Fourier expansion is  the starting point both for the derivation of the Kramers dispersion formula discussed in sec.\ 3.2 and for Heisenberg's (1925a) {\it Umdeutung} paper \citep[p.\ 592--594]{Duncan_2007-on-the-verge}.\footnote{As Dennison put it in the introduction of the paper that \citet{VanVleck_1926_magnetic} used for his note in {\it Nature} on the susceptibility of a gas of rigid rotators (see sec.\ 4): ``According to [matrix mechanics] the coordinates of a multiply periodic system which may be expressed classically by means of multiple Fourier series in the time, are to be replaced by infinite matrices of the Hermite type of which each member is a harmonic component in time" \citep[p.\ 318]{Dennison_1926-the-rotation}.}

To ensure that $p_z$ in Eq.\ \eqref{d}  is real, the complex amplitudes $(p_z)_\tau$ must satisfy $(p_z)_\tau = (p_z)^*_{- \tau}$. Eq.\ \eqref{d} also gives the expansion for a system with $f$ degrees of freedom, if, following Van Vleck, we introduce the abbreviations $J^0 \equiv J^0_1 \ldots J^0_f$, $w^0 \equiv  w^0_1 \ldots w^0_f$, $\tau \equiv \tau_1 \ldots \tau_f$, and $\tau w^0 = \sum_{k = 1}^f \tau_k \, w_k^0$ \citep[p.\ 38]{VanVleck_1932-the-theory}. Through $p_z$, the full Hamiltonian $H$ in Eq.\ \eqref{f} depends on $w^0$, so the action-angle variables $(w^0,J^0)$ are {\it not} action-angle variables for $H$. The phase space element, however, is invariant under the transformation from action-angle variables for $H$ to action-angle variables for $H^0$, i.e., $dJ^0dw^0 = dJdw$ (ibid., p.\ 39).  

Using the standard formula for the canonical ensemble average, we find for $\overline{\overline{p_z}}$ (ibid., p.\ 38):\footnote{One can think of the integration of $p_z$ over one period of the angle variable $w_0$ for a fixed value of $J_0$ as giving $\overline{p_z}$ and of the subsequent integration over $J_0$ as turning $\overline{p_z}$ into $ \overline{\overline{p_z}}$ (ibid., note 11). In this case, averaging over a thermal ensemble of identical systems is replaced by taking a weighted average over different states of one system, where the weight factor is given by the usual Boltzmann factor $e^{-H/kT}$, in which $H \approx H_0(J_0)$.}
\begin{equation}
\label{g}
\chi = \frac{N}{E} \frac{\int \!\! \int p_z \, e^{-H/kT} \, dJ^0 dw^0}{\int \!\! \int e^{-H/kT} \, dJ^0 dw^0}.
\end{equation}
To first order in the field $E$, the Boltzmann factor is given by:
\begin{equation}
\label{h}
e^{-H/kT} \approx e^{-H^0/kT} \left( 1 + \frac{E p_z}{kT} \right).
\end{equation}
Assuming there is no residual polarization in the absence of an external field (which is true for gases if not always for solids), i.e., $\overline{\overline{p_z}} = 0$ for $E=0$, we have
\begin{equation}
\label{i}
\int \!\! \int p_z \, e^{-H^0/kT} \, dJ^0 dw^0 = 0.
\end{equation}
Using Eqs.\ (\ref{h}) and (\ref{i}), we can rewrite Eq.\ (\ref{g}) as (ibid., p.\ 39)
\begin{equation}
\label{j}
\chi = \frac{N}{kT} \frac{\int \!\! \int p_z^2 \, e^{-H^0/kT} \, dJ^0 dw^0}{\int \!\! \int e^{-H^0/kT} \, dJ^0 dw^0}.
\end{equation}
For $p_z^2$ we insert its Fourier expansion
\begin{equation}
\label{k}
p_z^2(w^0,J^0) = \sum_{\tau = 0, \pm 1, \pm 2, \ldots}^\infty (p^2_z)_\tau(J^0) \, e^{\displaystyle 2 \pi i \tau w^0}.
\end{equation}
Only the $\tau=0$ terms on the right-hand side will contribute to the integral of $p_z^2$ over $w^0$ in Eq.\ (\ref{j}). All $\tau \neq 0$ terms are periodic functions of $w^0$, which vanish when integrated over a full period of these functions. Hence,
\begin{equation}
\label{l}
\int p^2_z \, dw^0 = (p^2_z)_0.
\end{equation}
\noindent In other words, $(p^2_z)_0$ is  the time average $\overline{p^2_z}$ of $p^2_z$. It follows from Eq.\ \eqref{l} that the integrals over $w^0$ in numerator and denominator of Eq.\ (\ref{j}) cancel. Eq.\ \eqref{j} thus reduces to (ibid., pp.\ 39--40):
\begin{equation}
\label{n}
\chi = \frac{N}{kT} \, \frac{\int \overline{p_z^2} \, e^{-H^0/kT} \, dJ^0}{\int e^{-H^0/kT} \, dJ^0} = \frac{N}{kT} \, \overline{\overline{p_z^2}} = \frac{N}{3kT} \, \overline{\overline{p^2}},
\end{equation}
where in the last step we used that 
\begin{equation}
\overline{\overline{p^2_z}} = {\textstyle \frac{1}{3}} \,\overline{\overline{p^2}}. 
\label{average 1/3}
\end{equation}
This relation holds both in the classical theory and in quantum mechanics. That it does not hold in the old quantum theory is central, as we will see, to that theory's failure to reproduce the Langevin-Debye formula. Van Vleck thus took great care explaining this relation:

\begin{singlespacing}
\begin{quotation}
\noindent
$\overline{\overline{p^2_z}}$ denotes the statistical mean square of $p_z^2$ in the absence of the field $E$, i.e. the average over only the $J^0$ part of the phase space, weighted according to the Boltzmann factor, of the time average value of $p_z^2$ [in our notation: $\overline{p^2_z}$]  for a molecule having given values of the $J^0$'s [recall that $J^0$ short-hand for $ J^0_1 \ldots J^0_f$]. Now if the applied electric field $E$ is the only external field, all spacial orientations will be equally probable when $E=0$, and the mean squares of the $x$, $y$, and $z$ components of moment will be equal [i.e., $\overline{\overline{p^2_x}} = \overline{\overline{p^2_y}} = \overline{\overline{p^2_z}}$]. This will also be true even when there are other external fields (e.g. a magnetic field) besides the given electric field[,] provided, as is usually the case, these other fields do not greatly affect the spacial distribution. We may hence replace $\overline{\overline{p^2_z}}$ by one-third the statistical mean square of the vector momentum $\bold{p}$ of the molecule  \citep[pp.\ 39--40]{VanVleck_1932-the-theory}.
\end{quotation}
\end{singlespacing}

\noindent
In the old quantum theory, as pointed out by \citet{Pauling_1927-the-influence}, the susceptibility is sensitive to the presence of a magnetic field (see sec.\ 5.2.2). In classical theory and in quantum mechanics it is not. This is undoubtedly why Van Vleck emphasized this feature.

 \citet{VanVleck_1932-the-theory}  called Eq.\ \eqref{n} ``a sort of generalized Langevin-Debye formula'' (p.\ 40). No particular atomic model need be assumed for its derivation.  
To obtain the familiar Langevin-Debye formula \eqref{langevin debye formula} with terms corresponding to permanent and induced electric moments, we need to adopt a model for the molecule of the gas similar to that underlying the classical dispersion theory of von Helmholtz, Lorentz, and Drude involving harmonically-bound charges (see sec.\ 3.2). As \citet{VanVleck_1932-the-theory} wrote: ``This na\"ive depicture of an atom or molecule as a collection of harmonic oscillators is not in agreement with modern views of atomic structure as exemplified in the Rutherford atom, but yields surprisingly fruitful results" (p.\ 30).\footnote{The same can be said about classical dispersion theory \citep[pp.\ 576--577]{Duncan_2007-on-the-verge}.} Let $s$ be the number of degrees of freedom with which these bound charges can vibrate, then with a set of normal coordinates $\xi_1, \xi_2, \ldots,\xi_s$, we can write the component of the electric moment  $\bold{p}$ along the principal axis of inertia, labeled $x$, as a linear function of these normal coordinates (ibid., p.\ 33):
\begin{equation}
p_x=\mu_x+\sum_{i=1}^s c_{xi}\xi_i,\label{function of normal coordinates}
\end{equation}
where $\mu_x$ is the $x$-component of the permanent electric dipole moment of the molecule and where the coefficients $c_{xi}$ are real positive numbers. Similar expressions obtain for the $y$- and $z$-components of $\bold{p}$.

Since positive and negative displacements will cancel during the averaging process, $\overline{\overline{\xi_i\xi_j}}=0$ for $i\neq j$ (ibid., p.\ 40).  If we associate a `spring constant' $a_i$ with the linear force binding the $i^{\text{th}}$ charge, then, by the equipartition theorem, we get: $\frac{1}{2}a_i\overline{\overline{\xi_i^2}}=\frac{1}{2}kT$.  Inserting Eq.\ \eqref{function of normal coordinates} for $p_x$ and similar equations for $p_y$ and $p_z$  for the components of $\bold{p}$ in Eq.\ \eqref{n} and using the relations for  $\overline{\overline{\xi_i\xi_j}}$ and $\overline{\overline{\xi_i^2}}$, we find (ibid., p.\ 37):
\begin{equation}
\chi = \frac{N\mu^2}{3kT}+\frac{N}{3}\sum_i\frac{c_{xi}^2+c_{yi}^2+c_{zi}^2}{a_i}.
\end{equation}
As desired, the first term gives us the contribution of the permanent moment with a factor of $1/3$, and the second is of the form $N\alpha$, where $\alpha$ is independent of temperature.  

Unfortunately, the assumption that electrons can be thought of as harmonically-bound charges in the atom had to be discarded as the old quantum theory began to shed light on atomic structure. This is the same development that was responsible for the old quantum theory's Kuhn loss in dispersion theory (see sec.\ 3.2). Expanding on the comment quoted above, Van Vleck concluded Ch.\ II by emphasizing the limitations of the classical theory:

\begin{singlespacing}
\begin{quotation}
\noindent
A model such as we have used, in which the electronic motions are represented by harmonic oscillators, is not compatible with modern knowledge of atomic structure
%We know that actually the electrons are subject to inverse square rather than linear restoring forces, and move in approximately Keplerian orbits instead of executing simple harmonic vibrations about positions of static equilibrium. 
%In fact Earnshaw's theorem in electrostatics tells us that there are no such positions for all charges.
\ldots\ 
Inasmuch as we have deduced a generalized Langevin-Debye formula for any multiply periodic system, the question naturally arises whether [Eq.\ \eqref{n}] cannot be specialized in  a fashion appropriate to a real Rutherford atom instead of to a fictitious system of oscillators mounted on a rigid rotating framework. This, however, is not possible \citep[p.\ 41]{VanVleck_1932-the-theory}.
\end{quotation}
\end{singlespacing}

\noindent
The reason Van Vleck gave for this is that in the Rutherford(-Bohr) atom, the energy of the electron ranges from $0$ to $-\infty$ causing the Boltzmann factors $e^{-H/kT}$ to diverge. Hence, he concluded, ``the practical advantages of the [general formula \eqref{n}] are somewhat restricted because of the inherent limitations in classical theory" (ibid.)

%As we will show in the next section, attempts to derive a formula for susceptibility in the old quantum theory similar to the one in classical theory given above ran afoul of some of the old quantum theory's most striking yet little-known inconsistencies.

\subsubsection{Susceptibilities in the Old Quantum Theory}

Attempts to derive a formula for susceptibility in the old quantum theory similar to the one in classical theory given above ran afoul of some of the old quantum theory's most striking yet little-known inconsistencies. The old quantum theory was at its best when physicists could be agnostic about the details of the multiply-periodic motion in atoms or molecules (as in the case of the Kramers dispersion formula [see sec.\ 3.2]).  As soon as they were forced to take these details seriously, new problems emerged that could not easily be dealt with.  Van Vleck had run into such problems in his work on helium. Similar problems arose
%The situation was even worse: I think Clayton would take issue with that (MJ)
in molecular physics, where the details of rotational and vibrational motion of specific models for various molecules had to be taken into account.  The derivation of a formula for susceptibility hinges on detailed consideration of rotational motion, in particular on the question of how to quantize angular momentum. Unlike modern quantum mechanics, the old quantum theory did not provide clear instructions on how to do this.
%a question for which the old quantum theory, unlike modern quantum mechanics, did not have a general satisfactory answer. The problems this caused in calculations of susceptibilities go a long way to justifying Van Vleck's claim in his 1932 book, in the opening paragraph of Ch.\ V on the old quantum theory, that ``there is perhaps no better field than that of electric and magnetic susceptibilities  to illustrate the inadequacies of the old quantum theory and how they have been removed by the new mechanics" \citep[p.\ 105]{VanVleck_1932-the-theory}. 
As a result, as Van Vleck wrote in Part I of his 1927--28 trilogy,

%Detailed considerations of rotational motion also enter into calculations of susceptibilities and the difficulties of the old quantum theory manifest themselves glaringly in t
%The sensitivity of the results to the models used is what lies behind Van Vleck's claim in 
%It is for these reasons that 

\begin{singlespacing}
\begin{quotation}
\noindent
the old quantum theory replaced the factor $1/3$ [in the Langevin-Debye formula (\ref{langevin debye formula})] by a constant $C$ whose numerical value depended rather chaotically on the type of model employed, whether whole or half quanta were used, whether there was ``weak" or ``strong" spacial quantization, etc.$^{[}$\footnote{At this point a  footnote is inserted with references to \citet{Pauli_1921-zur-theorie} and \citet{Pauling_1926-the-quantum}.}$^{]}$ This replacement of $1/3$ by $C$ caused an unreasonable discrepancy with the classical theory at high temperatures, and in some instances the constant $C$ even had the wrong sign \citep[p.\ 728]{VanVleck_1927_on-dielectric1}.
\end{quotation}
\end{singlespacing}
 
The issue of `weak' or `strong' quantization mentioned in this passage has to do with  the question of how to quantize the {\it unperturbed} motion in the old quantum theory. Consider a rotating molecule.  If a strong enough electric field is present, it makes sense to quantize the molecule's rotation with respect to the direction of the field.  But how to quantize in the absence of an external field?  In that case, there is no reason to assume a preferred direction in space and it seems arbitrary to preclude entire classes of rotational states. 
Yet one had to proceed somehow.  Two different kinds of quantization could be assumed \citep[p.\ 106]{VanVleck_1932-the-theory}.  In the first, called `strong spatial quantization', rotation was assumed to be quantized with respect to the field {\it even when there was, as yet, no field}. In the other, called `weak spatial quantization', molecules were assumed to be in some intermediate state between `strong quantization' and a classical distribution of rotational states.\footnote{In Part II of his 1927--28 trilogy, \citet[p.\ 37]{VanVleck_1927_on-dielectric2} referred to his NRC {\it Bulletin} for a discussion of `weak' versus `strong' quantization  \citep[p.\ 165]{VanVleck_1926-quantum}. In the {\it Bulletin} the same distinction is also made in terms of `diffuse' versus `sharp' quantization (ibid., pp.\ 171--178).} Van Vleck  highlighted this conceptual conundrum:
%took this conceptual conundrum to be among the most glaring of the old quantum theory's flaws:

\begin{singlespacing}
\begin{quotation}
\noindent
Spacial quantization cannot be effective unless it has some axis of reference.  In the calculation of Pauli and Pauling \ldots\ the direction of the electric field is taken as such an axis  \ldots\ [I]n the absence of all external fields \ldots\ there is no reason for choosing one direction in space rather than another for the axis of spacial quantization \citep[p.\ 108]{VanVleck_1932-the-theory}.
\end{quotation}
\end{singlespacing}

We need to take a closer look at these calculations by \citet{Pauli_1921-zur-theorie} and \citet{Pauling_1926-the-quantum}. They considered the special case in which the rigid rotator is used to model diatomic molecules such as HCl.  Its rotational states are specified by two angular coordinates, the azimuthal coordinate $\vartheta$ and the polar coordinate $\varphi$, and their conjugate angular momenta $p_{\vartheta}$ and $p_{\varphi}$.  The angle $\vartheta$ is measured from the $z$-axis chosen in the direction of the external field $\bold{E}$. The Hamiltonian for the system in this field is:
\begin{equation}
H=\frac{1}{2I} \left( p_{\vartheta}^2+\frac{p_{\varphi}^2}{\sin^2\vartheta} \right)-\mu E\cos\vartheta, \label{old quantum hamiltonian}
\end{equation}
where $I$ is the molecule's moment of inertia \citep[p.\ 321]{Pauli_1921-zur-theorie}.\footnote{Pauli used $A$ and $F$ and Pauling used $I$ and $F$ for what in our notation are $I$ and $E$, respectively.} 

Implicitly assuming strong spatial quantization, Pauli (ibid., p.\ 324)
%\citet[p.\ 324]{Pauli_1921-zur-theorie} 
quantized the angular momentum of the rigid rotator with respect to the direction of the field even when the field has not been switched on yet. Keep in mind that Pauli wrote this paper the year before Otto Stern and Walther Gerlach published what appeared to be strong evidence for spatial quantization \citep{Gerlach_1922-das-magnetische}. When Pauling redid Pauli's calculation with half rather than whole quanta in early 1926, he likewise assumed strong spatial quantization, but, unlike Pauli, was quite explicit about it and devoted the final subsection of his paper to a discussion of the issue of strong versus weak spatial quantization \citep[p.\ 576]{Pauling_1926-the-quantum}.

\citet[p.\ 321, p.\ 324]{Pauli_1921-zur-theorie} introduced the quantities $K$ and $J$, defined as (a sum of) action variables subject  to Sommerfeld-Wilson-type quantum conditions (cf.\ Eq.\ \eqref{Bohr-Sommerfeld} in sec.\ 3.2):
\begin{equation}
K \equiv \oint p_{\vartheta}d\vartheta + \oint p_{\varphi}d\varphi=lh, \quad J \equiv \oint p_{\varphi}d\varphi=2\pi p_{\varphi}=mh.
\label{quantum conditions}
\end{equation}
\citet[p.\ 570]{Pauling_1926-the-quantum} did not use the designations $K$ and $J$ for these quantities and changed the first condition to
 \begin{equation}
\oint p_{\vartheta}d\vartheta + \left| \oint p_{\varphi}d\varphi \right| =lh. 
\label{quantum conditions 2}
\end{equation}
Both Pauli and Pauling actually used $m$ instead of $l$ and $n$ instead of $m$. We use $l$ and $m$ because it turns out that these quantum conditions boil down to setting the norm and the $z$-component of the angular momentum $\bold{L}$, both averaged over periods of $\vartheta$ and $\varphi$, equal to $l\hbar$ and $m\hbar$, respectively ($\hbar \equiv h/2 \pi$). The reason Pauling modified Pauli's first quantum condition was probably because he realized that  $L$ could never be smaller than $L_z$. For the purposes of reconstructing the calculation (cf.\ note \ref{haphazard}), the quantum conditions \eqref{quantum conditions}--\eqref{quantum conditions 2} can be replaced by:
\begin{equation}
L = l\hbar, \quad L_z = m\hbar.
\label{quantum conditions sanitized}
\end{equation}
Pauli used integer quantum numbers, which means that $l = 1, 2, 3, \ldots$; Pauling used half-integers, which means that $l = \frac{1}{2}, \frac{3}{2}, \frac{5}{2}, \ldots$. In both cases, $m$ runs from $-l$ to $l$. The state $l=m=0$ was forbidden in the old quantum theory.

The equations on the blackboard behind Van Vleck in the picture in Fig.\ \ref{fig2} may serve as a reminder that even this sanitized version \eqref{quantum conditions sanitized} of the quantum conditions \eqref{quantum conditions}--\eqref{quantum conditions 2}  is {\it not} how angular momentum is quantized in modern quantum mechanics.\footnote{For a concise modern discussion of angular momentum in quantum mechanics, see, e.g., \citet[Ch.\ 6]{Baym_1969-lectures}.} 
%As we will see below, 
This modern treatment of angular momentum underlies the calculations of susceptibilities by \citet{Mensing_1926-uber-die-dielektrizitatkonstante}, \citet{VanVleck_1926_magnetic}, and others  in the new quantum theory. It is precisely because of the dubious way in which it quantized angular momentum---the conditions \eqref{quantum conditions sanitized} in conjunction with spatial quantization---that the old quantum theory came to grief in its treatment of susceptibilities.

To find the susceptibility of a gas of rigid rotators in the old quantum theory, Pauli and Pauling first calculated the time average (indicated by the single overbar, cf.\ sec.\ 5.2.1) of the component of the electric moment in the direction of $\bold{E}$ in a particular state of the rigid rotator characterized by the quantum numbers $l$ and $m$:
%the values of $K$ and $J$ in Eqs.\ \eqref{quantum conditions}--\eqref{quantum conditions 2}:
\begin{equation}
\mu \overline{\cos \vartheta} = \frac{\mu}{T} \int_0^T \cos\vartheta \, dt,
\label{T}
\end{equation}
where $T$ is the period of rotation. Substituting the classical equation $p_{\vartheta}=I(d\vartheta/dt)$ into the Hamiltonian in Eq.\ \eqref{old quantum hamiltonian} and using Eq.\ \eqref{quantum conditions} to set $p_{\varphi}=m\hbar$, Pauli and Pauling derived an equation relating $dt$ to $d\vartheta$:
\begin{equation}
dt=\frac{2\pi Id\vartheta}{\sqrt{8\pi^2IW -
{\displaystyle \frac{m^2h^2}{\sin^2\vartheta}}
+8\pi^2I\mu E\cos \vartheta}},\label{time differential}
\end{equation}
where $W$, the value of $H$, is the total energy of the molecule (Pauli, 1921, p.\ 322; Pauling, 1926, p.\ 570). Using Eq.\ \eqref{time differential}, Pauli and Pauling could  replace integration over $t$ in Eq.\ \eqref{T} by integration over $\vartheta$ at the cost of a rather more complicated expression. 

In the evaluation of $ \overline{\cos \vartheta}$, a distinction needs to be made between two energy regimes \citep[p.\ 322]{Pauli_1921-zur-theorie}.  In the first, the molecules have energies $W$ much smaller than $\mu E$, the energy of the interaction between the electric moment and the field.  In the second, $W$ is much larger than $\mu E$. The calculations of Pauli and Pauling only apply to the latter $W > \!\! > \mu E$ regime. In that case, we can take $\mu E$ to be a small perturbation of a purely rotational Hamiltonian and expand the denominator on the right-hand side of Eq.\ \eqref{time differential} in the small dimensionless parameter $\mu E/W$, keeping only first-order terms.

\citet[p.\ 324]{Pauli_1921-zur-theorie} and \citet[p.\ 570]{Pauling_1926-the-quantum} eventually arrived at:
%expression for $\overline{\overline{\cos \vartheta}}$
%the average  of $\cos\vartheta$ over a single state characterized by the values of $l$ and $m$:
\begin{equation}
\overline{\cos \vartheta}=\frac{\mu EI}{2 \hbar^2 l^2}\left(\frac{3m^2}{l^2} - 1\right)
\label{cos average}.
\end{equation}
The ratio $(m^2/l^2)$ on the right-hand side corresponds to the {\it time} average $\overline{(L_z/L)^2} = \overline{\cos^2 \vartheta}$ for the {\it unperturbed} system. In the {\it classical} theory,\footnote{\citet[sec.\ 4, pp.\ 322--324]{Pauli_1921-zur-theorie}, in fact, first showed that, in classical theory, $\overline{\cos \vartheta} = (\mu EI/2 \overline{L^2})(3\overline{L_z^2/L^2} - 1)$ (in our notation, where the overbars on the left- and the right-hand sides refer to time averages for the {\it perturbed} and the {\it unperturbed} system, respectively). He then set  $\overline{L}=l\hbar$ and $\overline{L_z}= m\hbar$ (cf.\ Eq.\ \eqref{quantum conditions sanitized}) to turn this classical equation into Eq.\ \eqref{cos average} in the old quantum theory.\label{haphazard}} 
but {\it not} in the old quantum theory, the {\it ensemble} average,  $\overline{\overline{\cos^2 \vartheta}}$, of this time average, $\overline{\cos^2 \vartheta}$ (both for the {\it unperturbed} system) is equal to 1/3. 
This is the same point that \citet[pp.\ 39--40]{VanVleck_1932-the-theory} made in one of the passages we quoted in sec.\ 5.2.1: $\overline{\overline{p^2_z}} = {\textstyle \frac{1}{3}} \,\overline{\overline{p^2}}$ (see Eq.\ \eqref{average 1/3}). 
It thus follows from the classical counterpart of Eq.\ \eqref{cos average} (see note \ref{haphazard}) that the ensemble average, $\overline{\overline{\cos \vartheta}}$, of the time average, $\overline{\cos \vartheta}$ (now both for the {\it perturbed} system) vanishes. 

According to the classical theory, in other words, there is no contribution to the susceptibility {\it at all} from molecules in the energy regime $W > \!\! > \mu E$ for which the classical counterpart of Eq.\ \eqref{cos average} (see note \ref{haphazard})  was derived. As \citet[p.\ 324]{Pauli_1921-zur-theorie} noted, this fits with the conclusion drawn earlier by \citet{Alexandrow_1921-eine-bemerkung} that it is only the molecules in the lowest energy states that contribute to the susceptibility. Pauli also noted, however, that the lowest energy state in the old quantum theory $(l =m=0)$ is forbidden. In the old quantum theory, we thus have the paradoxical situation that there are ``{\it only such orbits present that according to the classical theory do not give a sizable contribution to the electrical polarization}" \citep[p.\ 325; emphasis in the original]{Pauli_1921-zur-theorie}.  

Pauli went on to show that, contrary to the situation in the classical theory, the ensemble average, $\overline{\overline{\cos \vartheta}}$, of the time average, $\overline{\cos \vartheta}$, given by Eq.\ \eqref{cos average} does {\it not} vanish in the old quantum theory (where both averages are for the perturbed system). Hence, he concluded, in the old quantum theory the susceptibility does not come from molecules in the low energy states but from those in the high energy states of the $W > \!\! > \mu E$ regime in which Eq.\ \eqref{cos average} holds. It therefore should not surprise us, Pauli argued, that the old quantum theory does not reproduce the factor $1/3$ of the Langevin-Debye formula \citep[p.\ 325]{Pauli_1921-zur-theorie}.

%The derivation of Eq.\ \eqref{cos average} is the place where the old quantum theory breaks down.  In modern quantum mechanics, this average can be derived generally with degenerate perturbation theory (see, e.g., Shankar, 1994, sec.\ 17.3).  The old quantum theory was missing a suitably generalized degenerate perturbation theory, which in this case manifests itself in the form of the need for an assumption about spatial quantization.

Before calculating $\overline{\overline{\cos \vartheta}}$, 
%the ensemble average of $\overline{\cos \vartheta}$, 
Pauli rewrote the factor multiplying the expression in parentheses in Eq.\ \eqref{cos average}. Elementary Newtonian mechanics and the quantum condition, $L^2=\hbar^2l^2$, tell us that the energy $W_0$ of the molecule in the absence of the field is given by:\footnote{Using that $L=I \omega$ (with $\omega$ the angular frequency), we can rewrite the rotational energy $W_0 = \frac{1}{2} I \omega^2$ as $W_0 = L^2/2I$.\label{elementary mechanics}}
\begin{equation}
W_0= \frac{\hbar^2 l^2}{2I}.
\label{W0}
\end{equation}
This energy, in turn, can be expressed in terms of a new quantity $\sigma$ \citep[p.\ 326]{Pauli_1921-zur-theorie}: 
\begin{equation}
\sigma \equiv \frac{\hbar^2}{2IkT} = \frac{\Theta}{T},
\label{sigma}
\end{equation}
where $\Theta$ is a ``temperature characteristic for the quantum drop in specific heat associated with the rotational degree of freedom" (ibid.).  Combining Eqs.\ \eqref{W0} and \eqref{sigma}, we see that 
\begin{equation}
W_0 =  \sigma kT l^2.
\label{W0 1}
\end{equation}
From Eqs.\ \eqref{W0} and \eqref{W0 1}, it follows that $I/2\hbar^2l^2 = 1/4W_0 = 1/4 \sigma kT l^2$. Using this relation, we can rewrite Eq.\ \eqref{cos average} as:
\begin{equation}
\overline{\cos \vartheta}=\frac{\mu E}{4  \sigma kT l^2}
\left(\frac{3m^2}{l^2}- 1\right).
\label{cos average prime}
\end{equation}

The ensemble average of $\overline{\cos \vartheta}$ is given by \citep[p.\ 325]{Pauli_1921-zur-theorie}:
\begin{equation}
\overline{\overline{\cos\vartheta}}=\frac{\sum_{l>0} \sum_m\overline{\cos \vartheta}\,e^{-W_0/kT}}{\sum_{l>0}\sum_m e^{-W_0/kT}},
\label{ensemble average 1}
\end{equation}
where we used that, in the $W > \!\! > \mu E$ regime, $W$ can be replaced by $W_0$ in the Boltzmann factors. Inserting Eq.\ \eqref{cos average prime} for $\overline{\cos \vartheta}$ and using Eq.\ \eqref{W0 1} for $W_0$, we arrive at:
\begin{equation}
\overline{\overline{\cos \vartheta}}=\frac{\mu E}{4\sigma kT} \frac{\sum_{l>0} \sum_m {\displaystyle \frac{1}{l^2}\left(\frac{3m^2}{l^2}-\ 1\right)e^{-\sigma l^2}}}{\sum_{l>0} \sum_m e^{-\sigma l^2}}
\label{ensemble average 2}
%=\frac{\mu E}{kT}C,
\end{equation}
(Pauli, 1921, p.\ 326; Pauling, 1926, p.\ 571). Evaluating these sums for integer and half-integer quantum numbers, respectively, and multiplying by $N\mu$, both \citet[p.\ 327]{Pauli_1921-zur-theorie} and \citet[pp.\ 571--572]{Pauling_1926-the-quantum} arrived at an expression of the general form $C(N\mu/kT)$ for the temperature-dependent term in the formula for electric susceptibilities. 
%\citep[p.\ 326]{Pauli_1921-zur-theorie}. Of the form $C(\mu E/kT)$. 
Using whole quanta, Pauli found $C=1.5367$, which is 4.6 times the classical value of $1/3$.  Half-quanta---first introduced, at Einstein's suggestion, by Reiche in 1920 \citep[p.\ 158]{Gearhart_2010-astonishing}---typically led to better agreement with the data in the old quantum theory. In this case, however, it did not help matters at all (see note \ref{puzzle} for an explanation). Pauling found even more troubling departures from $C = 1/3$ with half-quanta than Pauli had with whole quanta.  For low temperatures ($T\approx\Theta$), Pauling calculated $C$ to be 1.578.  In his theory, however, $C$ increases with  temperature and in the limit of $T >\! \! >\Theta$ (a limit in which his calculation should have been entirely valid) takes on the value 4.570, over 13 times the classical value. As we saw in sec.\ 1.2.1, reliable experimental data to rule out values other than $C=1/3$ only became available after Pauling's paper was published, but it certainly was odd
%, to say the least,
that $C$ would increase with temperature in this way.
%depend on temperature in such a sensitive way. 

As we mentioned in sec.\ 4, Pauli revisited the problem of the susceptibility in diatomic dipole gases such as HCl shortly after the advent of matrix mechanics in a paper he co-authored with Lucy Mensing. Mensing had just obtained her doctorate in Hamburg, where  Wilhelm Lenz and Pauli had been her advisors. She was now working as a postdoc with Born and Jordan in G\"ottingen \citep[p.\ 188]{Mehra_1982-the-historical}. In an earlier paper, \citet{Mensing_1926-die-rotations} had already applied the new matrix mechanics to the rigid rotator, taking the treatment of angular momentum in the {\it Dreim\"annerarbeit} \citep[Ch.\ 4, sec.\ 1]{Born_1926-zur-quantenmechanik} as her point of departure.\footnote{See \citet{Cassidy_2007-Oppenheimer} for discussion of this paper.} Instead of the {\it ad hoc} quantization conditions in Eqs.\ \eqref{quantum conditions}--\eqref{quantum conditions 2} that Pauli and Pauling
%\citet{Pauli_1921-zur-theorie} and \citet{Pauling_1926-the-quantum} 
had used earlier, \citet{Mensing_1926-uber-die-dielektrizitatkonstante} based their calculation on quantum conditions for the angular momentum of the rigid rotator systematically derived from the fundamental principles of the new theory. \citet{VanVleck_1926_magnetic} did the same in his note on susceptibilities in {\it Nature} (see sec.\ 4), citing both \citet{Mensing_1926-die-rotations} and \citet{Dennison_1926-the-rotation} for the ``matrices of the rotating dipole" (p.\ 227).
%\footnote{We will cover Van Vleck's more elegant treatment of the rigid rotator in sec.\ 5.2.3.}
%\footnote{In his note on susceptibilities in {\it Nature}, \citet[p.\ 227]{VanVleck_1926_magnetic} cited both \citet{Mensing_1926-die-rotations} and \citet{Dennison_1926-the-rotation} for the ``matrices of the rotating dipole" (cf.\ sec. 4).}
%who had already applied matrix mechanics to diatomic molecules \citep[p.\ 188]{Mehra_1982-the-historical}. Drawing on this earlier paper by \citet{Mensing_1926-die-rotations}, \citet{Mensing_1926-uber-die-dielektrizitatkonstante} redid Pauli's calculation of 1921 for the rigid rotator in the new theory. 

The new theory replaced Eqs.\ \eqref{quantum conditions}--\eqref{quantum conditions 2} for the quantization of the rigid rotator's angular momentum in the old quantum theory by relations familiar to the modern reader:\footnote{We will continue to use the letter $l$ even though \citet{Mensing_1926-die-rotations}, \citet{Mensing_1926-uber-die-dielektrizitatkonstante}, and \citet[sec.\ 37, pp.\ 147--152; see sec.\ 5.2.3 below]{VanVleck_1932-the-theory} all used $j$ instead. To a modern reader, the letter $j$ may suggest a combination of orbital angular momentum and spin, whereas in the case of the rigid rotator we only have the former, $\bold{L} = \bold{x} \times \bold{p}$.}
\begin{equation}
L^2 =  l(l+1)\hbar^2, \quad L_z =  m\hbar,
\label{quantum conditions new}
\end{equation}
where $l = 0, 1, \ldots$ and $-l \leq m \leq l$ (see, e.g., Mensing, 1926, p.\ 814).
%\citep[p.\ 814]{Mensing_1926-die-rotations}. 
Eq. \eqref{W0} for the molecule's rotational energy $W_0$ in the absence of a field  accordingly changes to \citep[p.\ 510]{Mensing_1926-uber-die-dielektrizitatkonstante}:\footnote{\citet[p.\ 166]{Gearhart_2010-astonishing} discusses this same formula in the context of work on the specific heat of hydrogen and work on molecular spectra in the early 1920s, which likewise involved rotating dumbbells and half-quanta (cf.\ note \ref{gearhart}).\label{clayton} }
\begin{equation}
W_0 = \frac{\hbar^2}{2 I} \, l(l+1) =  \frac{\hbar^2}{2 I} [(l+ {\textstyle \frac{1}{2}})^2 - {\textstyle \frac{1}{4}} ].
\label{W0 new}
\end{equation}
Hence, up to an additive constant, the energy is given by  squares of half-integers rather than integers, as Pauli had assumed in 1921. In this respect, matrix mechanics thus vindicated Pauling's use of half-quanta (ibid., p.\ 511). 
%\citep[p.\ 511]{Mensing_1926-uber-die-dielektrizitatkonstante}. 

Mensing and Pauli now considered the average value $\overline{\mu_z} = \mu \, \overline{\cos \vartheta}$
%\begin{equation}
%\overline{\mu_z} = \mu \, \overline{\cos \vartheta}
%\end{equation}
of the component of the dipole moment of the molecule in the direction of the field (cf.\ Eq.\ \eqref{T}). They wrote this in the form
\begin{equation}
\overline{\mu_z} = \alpha(l, m) \, E.
\label{alpha l m}
\end{equation}
In the old quantum theory, 
%as \citet[p.\ 511]{Mensing_1926-uber-die-dielektrizitatkonstante} pointed out, 
$\alpha(l,m)$ would be given by $(\mu/E)$ times the expression on the right-hand side of Eq.\ \eqref{cos average} for $ \overline{\cos \vartheta}$. In the new quantum theory, $\alpha(l,m)$ is given by
\begin{equation}
\frac{2\mu^2 I}{3\hbar^2}, \quad \quad  \frac{2\mu^2 I}{\hbar^2} \frac{1}{(2l -1)(2l+3)} \left\{ \frac{3m^2}{l(l+1)} - 1 \right\},
\label{mu in z dir}
\end{equation}
for $l =0$ and $l \neq 0$, respectively (ibid., p.\ 512).\footnote{Note that, for $l > \!\! > 1$, the second expression in Eq.\ \eqref{mu in z dir} reduces to $(\mu/E)$ times on the right-hand side of Eq.\ \eqref{cos average}, the corresponding expression in the old quantum theory. In sec.\ 5.2.3, we will cover the corresponding step in Van Vleck's (1932b, pp.\ 151--152) calculation for the rigid rotator in more detail.\label{promises}}
%\citep[p.\ 512]{Mensing_1926-uber-die-dielektrizitatkonstante}. 

These results can be used to calculate the ensemble average $\overline{\overline{\mu_z}}$ (cf.\ Eqs.\ \eqref{ensemble average 1}--\eqref{ensemble average 2} for $\overline{\overline{\cos \vartheta}}$ in the old quantum theory). Setting $W = W_0$ in the Boltzmann factors as in Pauli's earlier calculation (see Eq.\ \eqref{ensemble average 1}),\footnote{As we will see in sec.\ 5.2.3, \citet[p.\ 182]{VanVleck_1932-the-theory} was more careful with these Boltzmann factors.} we find 
\begin{equation}
\overline{\overline{\mu_z}} = \frac{\sum_l \sum_m \overline{\mu_z} \, e^{-W_0/kT}}{\sum_l \sum_m \, e^{-W_0/kT}}
= E \frac{\sum_l \sum_m \alpha(l,m) \, e^{-\sigma l(l+1)}}{\sum_l (2l+1) \, e^{-\sigma l(l+1)}},
\label{emsemble average new}
\end{equation}
where in the second step we used the relation $W_0 = \sigma kT l(l+1)$, the analogue in the new theory of the relation $W_0 = \sigma kT l^2$ in the old one (see Eq.\ \eqref{W0 1}), and evaluated the sum over $m$ in the denominator (ibid., p. 510).

When Eq.\ \eqref{mu in z dir} is for $\alpha(l,m)$ is substituted into Eq.\ \eqref{emsemble average new}. we find that only the $(l = 0)$-term  in the sum over $l$ in the numerator contributes to $\overline{\overline{\mu_z}}$ 
%\citep[p.\ 512]{Mensing_1926-uber-die-dielektrizitatkonstante}
(ibid., p.\ 512).\footnote{Whereas the sum in Eq.\ \eqref{emsemble average new 1} for the new quantum theory vanishes, the corresponding sum in Eq.\ \eqref{ensemble average 2} for the old quantum theory (in which $l = 0$ is forbidden) does not. It is because of this key difference between the calculation based on the modern quantum conditions \eqref{quantum conditions new} and the calculation based on the old quantum conditions \eqref{quantum conditions sanitized} that the switch from whole to half quanta did nothing to bring the value for the electric susceptibility closer to what we now know to be the empirically correct one.\label{puzzle}}  The contributions coming from $l \neq 0$ can be written as:
\begin{equation}
\overline{\overline{\mu_z}} =  \frac{2E\mu^2 I/\hbar^2}{ \sum_l (2l+1) \, e^{-\sigma l(l+1)}}  
\sum_{l \neq 0} \left( \frac{ e^{-\sigma l(l+1)} }{ (2l -1)(2l+3) } \sum_m  \left\{ \frac{3m^2}{l(l+1)} - 1 \right\} \right).
\label{emsemble average new 1}
\end{equation}
The well-known sum-of-squares formula tells us that
\begin{equation}
3 \sum_{m=-l}^l m^2 = 6 \sum_{m=1}^l m^2 = l(l+1)(2l+1).
\label{sum of squares}
\end{equation}
Using this formula to evaluate the sum over $m$ in Eq.\ \eqref{emsemble average new 1} for any fixed non-zero value of $l$, we find:
\begin{equation}
\sum_m  \left\{ \frac{3m^2}{l(l+1)} - 1\right\} = \frac{3 \sum_m m^2}{l(l+1)} - (2l+1) = 0.
\label{sum m gives 0}
\end{equation}
This shows that none of the $(l \neq 0)$-terms in the sum over $l$ in the numerator of Eq.\ \eqref{emsemble average new} contribute to $\overline{\overline{\mu_z}}$. 
As \citet{Mensing_1926-uber-die-dielektrizitatkonstante} commented with obvious relief: ``{\it Only the molecules in the lowest state {\rm [$l=0$]} will therefore give a contribution to the temperature-dependent part of the dielectric constant}" (p.\ 512; emphasis in the original). The new quantum theory thus reverted to the classical theory in this respect.\footnote{In his note on susceptibilities in {\it Nature}, \citet[p.\ 227]{VanVleck_1926_magnetic} made the same point: ``The remarkable result is obtained that only molecules in the state {\rm [$l=0$]} of lowest rotational energy make a contribution to the polarisation. This corresponds very beautifully to the fact that in the classical theory only molecules with energy less than {\rm [$\mu E$]} contribute to the polarisation." Like \citet[p.\ 324]{Pauli_1921-zur-theorie}, \citet{VanVleck_1926_magnetic} cited \citet{Alexandrow_1921-eine-bemerkung} for this result in the classical theory. So did \citet[p.\ 491]{Kronig_1926-the-dielectric}, who also drew attention to this analogy between classical theory and quantum mechanics.\label{alexandrow}}

Substituting Eq.\ \eqref{mu in z dir} for $\alpha(0,m)$ into Eq.\ \eqref{emsemble average new}, we find that 
\begin{equation}
\overline{\overline{\mu_z}} = \frac{2\mu^2 I E}{3\hbar^2} \frac{1}{\sum_l (2l+1) \, e^{-\sigma l(l+1)}}.
\label{emsemble average new 2}
\end{equation}
Using the relation $\chi = (N/E) \overline{\overline{\mu_z}}$ (see Eq.\ \eqref{susceptibility definition}) in combination with the expression $NC\mu^2/kT$ for the temperature-dependent term in $\chi$ (see Eq.\ \eqref{C}), we can write $C$ as:
\begin{equation}
C = \frac{kT}{\mu^2 E} \overline{\overline{\mu_z}}.
\label{C2}
\end{equation}
Inserting Eq.\ \eqref{emsemble average new 2} for $\overline{\overline{\mu_z}}$ and using that $\sigma \equiv \hbar^2/2I kT$ (see Eq.\ \eqref{sigma}), we find:
\begin{equation}
C
%=\frac{kT}{\mu^2 E} \overline{\overline{\mu_z}} 
%= \frac{8\pi^2 I kT}{h^2} \frac{1}{3} \frac{1}{\sum_l (2l+1) \, e^{-\sigma l(l+1)}} 
= \frac{1}{3\sigma \sum_l (2l+1) \, e^{-\sigma l(l+1)}}. 
\end{equation}
%\citet[p.\ 512]{Mensing_1926-uber-die-dielektrizitatkonstante}  showed that for small values of $\sigma$, i.e., 
For sufficiently high temperatures, $l \approx l+1$ in most terms of the sum over the $l$ in the denominator and the sum can be replaced by an integral: 
\begin{equation}
\sum_l (2l+1) \, e^{-\sigma l(l+1)} \approx \int_0^\infty  2l \, e^{- \sigma l^2} \, dl = \frac{1}{\sigma},
\label{int}
\end{equation}
in which case $C= 1/3$. Mensing and Pauli concluded:

\begin{singlespacing}
\begin{quotation}
\noindent
This result is completely opposite to the results that were obtained on the basis of the earlier quantum theory of periodic systems according to which the coefficient $C$ \ldots\ should have a numerical value substantially different from $1/3$ even in the limiting case of high temperatures.$^[$\footnote{Here the authors cite \citet{Pauli_1921-zur-theorie} and \citet{Pauling_1926-the-quantum}.}$^]$ This shows that here, as in many other cases, the new quantum mechanics follows classical mechanics more closely than the earlier quantum theory when it comes to statistical averages \citep[p.\ 512]{Mensing_1926-uber-die-dielektrizitatkonstante}.
\end{quotation}
\end{singlespacing}

\noindent
And thus Mensing and Pauli recovered the Kuhn loss in susceptibility theory, at least for the special case of a gas consisting of rotating dumbbells. The authors, however, did not explain how the new calculation gets around the choice of a preferred axis of quantization. Mensing and Pauli, in other words, avoided the thorny problem of spatial quantization. As we will see in sec.\ 5.2.3, the solution to that problem boils down to the proof that the sum $\sum_m m^2$, and thereby the vanishing of Eq.\ \eqref{emsemble average new 1}, does not depend on the choice of the $z$-axis for the quantization of $L_z$. Van Vleck already indicated this in his brief note in {\it Nature} in 1926. Translated into our notation, he wrote: 

\begin{singlespacing}
\begin{quotation}
\noindent
The average value of $L_z^2$ is then 
$$
\frac{1}{2l+1} \sum_{m=-l}^l m^2 \hbar^2 = {\textstyle \frac{1}{3}} l(l+1) \hbar^2 = {\textstyle \frac{1}{3}} L^2,
$$
which is obviously the same result as with random orientations \citep[p.\ 227]{VanVleck_1926_magnetic}.\footnote{\citet[Vol.\ 2, 34--11]{Feynman_1964-the-feynman} used this same relation as an argument for why one should set $L^2 = l(l+1)\hbar^2$, if one sets $L_z = m \hbar$ with $m = 0, \pm 1, \pm 2, \ldots, \pm l$. It is only natural  to demand that the average value of $L^2$ be three times the average value of $L_z^2$. The average value of $L^2$ is then given by $3\hbar^2(\sum_m m^2)/(2l+1)$, which the sum-of-squares formula tells us is equal to $l(l+1)\hbar^2$.\label{serge}} 
\end{quotation}
\end{singlespacing}

\noindent
Note that this relation does not hold if the quantum-mechanical relation $L^2 =  l(l+1) \hbar^2$ is replaced by the relation $L^2 = l^2 \hbar^2$ of the old quantum theory (see Eq.\ \eqref{quantum conditions sanitized}). This is one way to understand the difficulties the old quantum theory ran into with susceptibilities. In quantum mechanics, $\overline{L_z^2/L^2} = 3m^2/l(l+1)$. In that case, the sum-of-squares formula tells us that  the ensemble average $\overline{\overline{L_z^2/L^2}} = 1/3$ (see Eqs.\ \eqref{mu in z dir}--\eqref{sum m gives 0}). In the old quantum theory, $\overline{L_z^2/L^2} = 3m^2/l^2$ and $\overline{\overline{L_z^2/L^2}} \neq 1/3$ (see Eqs.\ \eqref{cos average prime}--\eqref{ensemble average 2}).

In subsequent publications, \citet{VanVleck_1927_on-dielectric1, VanVleck_1932-the-theory} explained in more detail and with greater generality how the new quantum theory dispensed with the need for spatial quantization. This is precisely what is provided by the elusive notion of ``spectroscopic stability" (cf.\ the quotations in the introduction to sec.\ 5.2). Because of this general property of quantum mechanics, Van Vleck showed,  it is true for a broad class of models and regardless of the axis along which one chooses to quantize that the only contribution to the susceptibility comes from the lowest energy states (the term $l=0$ in Eq.\ \eqref{emsemble average new} for the special case of the rigid rotator). This was true in classical theory as well, but not in the old quantum theory. As he explained in \emph{The Theory of Electric and Magnetic Susceptibilities}:

\begin{singlespacing}
\begin{quotation}
\noindent
[C]lassically the susceptibility arises entirely from molecules which possess so little energy that they would oscillate rather than rotate through complete circles \ldots\ As the temperature is increased, the fraction of molecules which are located in the `lazy' states that contribute to the susceptibility will steadily diminish, and hence we can see qualitatively why the susceptibility due to permanent dipoles decreases with increasing temperature \ldots\ In the old quantum theory the susceptibility did not arise uniquely from the lowest rotational state \ldots\ and this is perhaps one reason why the old theory gave such nonsensical results on the dielectric constants \citep[p. 184]{VanVleck_1932-the-theory}.\footnote{That the only contribution to the susceptibility comes from the lowest state is a special feature of the rigid rotator. It is true much more generally, however, that the bulk of the susceptibility comes from the lower energy states (ibid.).}
\end{quotation}
\end{singlespacing}

\noindent In blaming the ``nonsensical results" of the old quantum theory on this unusual feature, Van Vleck ignored that, without it, the temperature-dependent term of the susceptibility could not be derived \emph{at all}. In the case of the rigid rotator, the state $l=0$ was forbidden in the old quantum theory. The susceptibility thus had to come from the $l \neq 0$ states. The preferred direction introduced by spatial quantization ensured that the sum over $l \neq 0$ in Eqs.\ \eqref{ensemble average 1}--\eqref{ensemble average 2} for $\overline{\overline{\cos \vartheta}}$ does {\it not} vanish, thus producing a non-zero contribution to the susceptibility. Without spatial quantization, all orientations would be equiprobable and the average moment in the direction of the field would be zero. We would then be stuck with the absurd conclusion that a permanent electric moment contributes nothing to the susceptibility! This is why, at the end of his paper, \citet[p.\ 577]{Pauling_1926-the-quantum} suggested that `strong spatial quantization'  itself was the mechanism responsible for polarization. 

While spatial quantization thus offered make-shift solutions to some problems in the old quantum theory, it also introduced new ones.  If one took it seriously, one was faced with a question about the quantization process itself.  If it was somehow caused by the presence of a field, did it happen all at once or gradually as the field was applied? Either way, there would be physical consequences.  Indeed, the experimentalist August Glaser claimed to have observed such an effect, a transition from `weak' to `strong' spatial quantization as the strength of the field was increased.  Van Vleck was not fond of the ``unphysical  \ldots\ bugbear of weak and strong spacial quantization'' \citep[p.\ 110]{VanVleck_1932-the-theory}. He had already expressed his displeasure about this ``bugbear"  in Part I of his 1927--28 trilogy \citep[p.\ 37]{VanVleck_1927_on-dielectric1}. In the 1932 book, he ends his discussion of it on a reassuring note:

\begin{singlespacing}
\begin{quotation}
\noindent
If the reader has felt that our presentation of weak and strong quantization in the old quantum theory was somewhat mystifying (as indeed it had to be, as physicists themselves were hazy on the details of the passage from one type of quantization to another), he need now no longer feel alarmed, as the new mechanics gives no susceptibility effects without some analogue in classical theory \citep[p.\ 111]{VanVleck_1932-the-theory}.
\end{quotation}
\end{singlespacing}

Spatial quantization also led to problems in the old quantum theory's treatment of the effect of magnetic fields on the dielectric constant. It was Pauling 
%\citet{Pauling_1927-the-influence} 
who drew attention to that problem. As he explained in the abstract of a paper submitted in September 1926:

\begin{singlespacing}
\begin{quotation}
\noindent
The investigation of the motion of a diatomic dipole molecule in crossed magnetic and electric fields shows that according to the old quantum theory there will be spatial quantization \ldots\ with respect to the magnetic field \ldots\ As a result of this the old quantum theory definitely requires that the application of a strong magnetic field to a gas such as hydrogen chloride produce a very large change in the dielectric constant of the gas. \ldots [T]he new quantum theory, on the other hand, requires the dielectric constant not to depend upon the direction characterizing the spatial quantization, so that no effect of a magnetic field would be predicted. The effect is found experimentally not to exist; so that it provides an instance of an apparently unescapable and yet definitely incorrect prediction of the old quantum theory \citep{Pauling_1927-the-influence}.
%[p.\ 145]
\end{quotation}
\end{singlespacing}

\noindent
By late 1926, as this passage shows, Pauling had come to recognize the ``wonderful nonsense" of the old quantum theory for what it was. Pauli had recognized this even earlier. This makes it understandable how both of them could be
%Seeing how both Pauli and Pauling realized shortly after the advent of matrix mechanics that the results they had found for susceptibilities in the old quantum theory were indeed nonsensical, one can understand better how they could be 
so magnanimous in their reviews of Van Vleck's 1932 book (see the quotations at the end of sec.\ 5.1), even though the author pounced on their earlier work.

Van Vleck devoted a section of Ch.\ V of his book to the issue raised by Pauling \citep[sec.\ 31, ``Effect of a Magnetic Field on the Dielectric Constant"]{VanVleck_1932-the-theory}. As with the anomalous values for $C$ found by \citet{Pauli_1921-zur-theorie} and \citet{Pauling_1926-the-quantum} (cf.\ our discussion in sec.\ 1.2.1), Van Vleck
%---deliberately or inadvertently---
left the reader with the impression that physicists had been well aware of the discrepancy between the old quantum theory's prediction of the effect and reliable experimental data. If we look more carefully, we see that \citet[p.\ 114]{VanVleck_1932-the-theory} credited \citet{Pauling_1927-the-influence} with having been the first to derive the prediction and that, like Pauling, he only cited papers published in 1926 or later for its experimental refutation. The way in which Van Vleck used this spurious effect to lambast the theory that predicted them makes it easy to forget that the prediction was not made, let alone tested, until after  the theory's demise: 

\begin{singlespacing}
\begin{quotation}
\noindent
The influence of a magnetic field on the dielectric constant \ldots\ was ludicrously large in the old quantum theory because of spacial quantization \ldots\ a crossed magnetic field would make the constant $C$ in [Eq.\ \eqref{C}] negative, an absurdity. Only a comparatively feeble magnetic field would be required \ldots\ An innocent little magnetic field of only a few gauss should thus in the old quantum theory change the sign of the temperature coefficient of the dielectric constant
and make the electric susceptibility negative in so far as the orientation rather than induced polarization is concerned. 
This is what one might term extreme spectroscopic instability. Needless to say, such a cataclysmic influence of a magnetic field on the dielectric constant is not found experimentally \ldots\ In the new quantum mechanics the choice of the axis of spacial quantization is no longer of importance, and so a magnetic field should be almost without effect on the dielectric constant, in agreement with the experiments \citep[p.\ 113--115]{VanVleck_1932-the-theory}.
\end{quotation}
\end{singlespacing}

In light of all this, it is no mystery that Van Vleck was so impressed by the way in which quantum mechanics dispensed with spatial quantization and, in the process, restored the factor of $1/3$ in the Langevin-Debye formula in full generality. ``The new mechanics," he wrote, ``always yield [sic] $C = 1/3$ without the necessity of specifying the details of the model, and the generality of this value of $C$ is one of the most satisfying features of the new theory" \citep[pp.\ 107--108]{VanVleck_1932-the-theory}. This then is one of the ``less heralded successes" and ``great achievements" of the new quantum theory that Van Vleck was referring to in the preface of his book (see the quotation in sec.\ 1.1). The following subsection explores this achievement in greater detail.

\subsubsection{Susceptibilities in the New Quantum Mechanics}

In this subsection, we present Van Vleck's derivation in his 1932 book of the electric susceptibility of a diatomic gas such as HCl with the rigid rotator as the model for its molecules. The most important difference between this derivation and the one by \citet{Mensing_1926-uber-die-dielektrizitatkonstante} discussed in sec.\ 5.2.2 is that Van Vleck's starts from a much more general approach to the calculation of the susceptibility in gases, one that he used 
%as his starting point 
in calculations for a variety of models for the gas molecules. The first, more general steps of this derivation run in parallel to the classical calculation we outlined in sec.\ 5.2.1.  While we will not go into the details of the general derivation, at the end of this subsection we will comment on one of its crucial components---Van Vleck's proof of spectroscopic stability and the elimination of spatial quantization. 

%What we do want to note right away is that this general derivation establishes that 
The Langevin-Debye formula holds under very general conditions in quantum mechanics. One assumption identified by Van Vleck is that ``the medium is sufficiently rarefied so that one may use the Boltzmann instead of the Fermi statistics" (p.\ 181).\footnote{Unless noted otherwise, all page references in sec.\ 5.2.3 are to the book by \citet{VanVleck_1932-the-theory}.} This assumption becomes critical only when Van Vleck tried to extend his approach from gases to solids. For gases (in weak fields), we only need two assumptions (p.\ 187): first, that the constituent molecules have a permanent dipole moment; second, that all possible transitions are such that the energy jumps $h\nu_{i \rightarrow f}$ are either much greater or much smaller than $kT$. Quantum mechanics thus solves the problem one runs into in the classical theory that the Langevin-Debye formula only obtains for unrealistic models of matter (see the quotation at the end of sec.\ 5.2.1). In other words, Van Vleck's  quantum-mechanical theory of susceptibilities can be seen as another instance of what \citet[p.\ 105]{Kuhn_1996_the-structure} described as a ``reversion (which is not the same as a retrogression)" to an older theory.

Van Vleck gave the general quantum-mechanical derivation of the Langevin-Debye formula for the electric susceptibility in gases in Ch.\ VII of his book (secs.\ 44--47, pp.\ 181--202). In this chapter, he used several results of Ch.\ VI, ``Quantum-Mechanical Foundations" (secs.\ 32--43, pp.\ 122--180), especially from the sections on perturbation theory (secs.\ 34--36, pp.\ 131--147).\footnote{A footnote appended to the title of sec.\ 34, ``Perturbation Theory," acknowledges that perturbation theory in quantum mechanics was first developed in the {\it Dreim\"annerarbeit} \citep{Born_1926-zur-quantenmechanik} and in the third communication on wave mechanics by Erwin \citet{Schroedinger_1926-Quantisierung3}.}   Moreover, in sec.\ 37, he had already derived the susceptibility for the special case of the rigid rotator (pp.\ 147--152). He briefly revisited this special case in Ch.\ VII (sec.\ 45, pp.\ 183--185).
%\footnote{This is the model with which the factor $1/3$ in the Langevin-Debye formula was first recovered in quantum mechanics, by  \citet{Mensing_1926-uber-die-dielektrizitatkonstante}, \citet{Kronig_1926-the-dielectric}, \citet{Manneback_1926-die-elektrizitaetskonstante}, and  \citet{VanVleck_1926_magnetic} himself, papers Van Vleck all cited in a footnote at the beginning of this section (p.\ 147). Cf.\ sec.\ 4).} 
Our discussion combines elements from these sections of Chs. VI and VII.
 
Following Van Vleck, we first derive an expression for the susceptibility of a gas without assuming a special model for its molecules. Let 
\begin{equation}
H= H^0 -Ep_{E},
\label{H with p_E}
\end{equation}
be the Hamiltonian for a gas molecule, represented by some multiply-periodic system, in an external electric field $\bold{E}$. $H^0$ is the Hamiltonian of the unperturbed system, $p_{E}$ the electric moment of the system in the direction of the field. The quantities $H$, $H^0$, and $p_E$ are now operators, $E$ is still just a real number. The electric moment $p_E$ can be extracted from the Hamiltonian by taking the derivative with respect to the field strength:
\begin{equation}
p_{E}=-\frac{\partial H}{\partial{E}}\label{electric moment}.
\end{equation}
This relation is crucial for the calculation of the matrix elements of $p_{E}$ (p.\ 143, p.\ 181).

In general, Van Vleck wrote the Hamiltonian of a system subject to a small perturbation as $H= H^0 + \lambda H^{(1)} + \lambda^2 H^{(2)} + \ldots$ with the parameter $\lambda < \!\! < 1$ (p.\ 132). For the Hamiltonian in Eq.\ \eqref{H with p_E}, $\lambda = E$ and $\lambda H^{(1)}$ is the only term in the expansion.  Perturbation theory allowed Van Vleck to compute the energy of the perturbed system as a series of corrections to the energy of the unperturbed system, each term corresponding to a different power of the expansion parameter:
\begin{equation}
W_n=W_n^0+EW_n^{(1)}+E^2W_n^{(2)}+\mathcal{O}(E^3).\label{perturbation series}
\end{equation}
To second order, we have (p.\ 133):
\begin{equation}
W_n^{(1)} = \braket{ n^0|H^{(1)}|n^0 }, \quad \quad
W_n^{(2)} = \sum_{n' \neq n}\frac{|\braket{ n'^0|H^{(1)}|n^0 }|^2}{W_n^0-W_{n'}^0}, 
 \label{W1 and W2}
\end{equation}
where the $\ket{n^0}$'s are the eigenvectors of the unperturbed Hamiltonian.\footnote{We use modern Dirac notation both because it is more familiar to the modern reader and because it is the notation Van Vleck  adopted when he began revising his 1932 book for a second edition (cf.\ sec.\ 1.2.3). He wrote what in our notation would be $\langle n |H|n' \rangle$ as $H(n; n')$. He also typically used two or three quantum numbers to label the (degenerate) energy eigenstates, writing, for instance, $H(nm; n'm')$ or $H(njm; n'j'm')$. We will follow his example 
%when we turn to the special 
in the case of the rigid rotator (see Eq.\ \eqref{W1 and W2 l m}).}
  Combining Eqs.\ \eqref{electric moment} and \eqref{perturbation series}, we obtain an expression for the matrix elements of the electric moment in eigenstates of the full Hamiltonian with eigenvectors $| n \rangle$ (ibid., p.\ 144):
\begin{equation}
\braket{n|p_{E}|n}=\braket{n|\left( -\frac{\partial H}{\partial{E}} \right) |n}= -\frac{\partial}{\partial{E}} \braket{n| H |n} = -W_n^{(1)}-2EW_n^{(2)}+\mathcal{O}(E^2).
\end{equation}
Inserting the expressions in Eq.\ \eqref{W1 and W2} for $W_n^{(1)}$ and $W_n^{(2)}$, using that $E H^{(1)}=- Ep_{E}$, we find, to first order in $E$ (p.\ 144):
\begin{equation}
\braket{n|p_{E}|n}=\braket{n^0|p_{E}|n^0} - 2E\sum_{n' \neq n}\frac{|\braket{ n'^0|p_{E}|n^0 }|^2}{W_n^0-W_{n'}^0}.\label{electric moment matrix}
\end{equation}
Van Vleck used the Bohr frequency condition (p.\ 133) to write $W_n^0-W_{n'}^0 = -h \nu_{n' \rightarrow n}$.\footnote{When he first published this formula,  Van Vleck commented: ``This is, of course, the same result as given by extrapolation of the Kramers dispersion formula to infinitely long impressed wavelengths" \citep[p.\ 734]{VanVleck_1927_on-dielectric1}. \citet[p.\ 511]{Mensing_1926-uber-die-dielektrizitatkonstante} and \citet[p.\ 490]{Kronig_1926-the-dielectric} had made that same connection.  In Ch.\ XIII of his book, Van Vleck gave a formula for the index of refraction $n$ of some transparent material as an ensemble average of the polarization of its constituents, given by the Kramers dispersion formula  (p.\ 361):
$$
n^2 - 1 = \frac{ 8\pi N }{ \sum_l{e^{-W^0_l/kT}} } \sum_{l,l'}{ \frac{ \nu_{l' \rightarrow l} |\braket{ l' | p_{E}| l} |^2 }{ h ( \nu^2_{l' \rightarrow l} - \nu^2 )} e^{-W^0_l/kT} },
$$
where $\nu$ is the frequency of the incident light wave and $\nu_{l' \rightarrow l} = W_{l'}^0-W_l^0$ (cf.\ Duncan and Janssen, 2007, p.\ 658). For $\nu = 0$, the sums over $l'$ for fixed $l$ have the same form (modulo the Boltzmann factor) as the second term on the right-hand side of Eq.\ \eqref{electric moment matrix}. This underscores the relation between dispersion and susceptibility that we drew attention to in sec.\ 1.2.3 and at the beginning of sec.\ 5.2.\label{1927 Kramers dispersion formula}}

To find an expression for the susceptibility $\chi$ we need to take two averages (cf.\ the discussion leading up to Eq.\ \eqref{statistical link} in sec.\ 5.2.1): (1) the expectation value $\overline{p_E} = \braket{n|p_{E}|n}$ of the electric moment of an individual molecule in a given state; (2) the average $\overline{\overline{p_E}}$ of this expectation value over a thermal ensemble of $N$ such molecules.  Both steps are captured in the following formula (p. 181):\footnote{Even though in the modern view, the expectation value cannot be viewed as a time average, in 1932 Van Vleck considered it to be something very similar: ``A diagonal Heisenberg matrix element $\bra{n}f\ket{n}$ has the physical significance of being the average value of $f$ over all the phases of motion in a given stationary state'' \citep[p. 129]{VanVleck_1932-the-theory}.}
\begin{equation}
\chi=\frac{N}{E} \, \overline{\overline{p_E}} = \frac{N}{E} \, \frac{\sum_n\braket{n|p_{E}|n}e^{-W_n/kT}}{\sum_n e^{-W_n/kT}}.
\label{suscept def 2}
\end{equation}
The Langevin-Debye formula is applicable only in regimes for which we can neglect saturation effects, which means that the susceptibility must be independent of the field strength $E$.  Accordingly, we will assume the numerator in Eq.\ \eqref{suscept def 2} to be linear in $E$, and the denominator to be independent of $E$.  

To first order, the Boltzmann factors in Eq.\ \eqref{suscept def 2} are given by (p.\ 182; cf.\ Eq.\ \eqref{h} in the classical calculation in sec.\ 5.2.1):
\begin{equation}
e^{-W_n/kT} = e^{-W_n^0/kT} e^{-EW_n^{(1)}/kT} = e^{-W_n^0/kT} \left(1 + \frac{E}{kT} \braket{n^0|p_{E}|n^0} \right),
\label{exponential approx}
\end{equation}
where in the last step we used Eq.\ \eqref{W1 and W2} for $W^{(1)}_n$ (with $H^{(1)} = - p_E$). We now substitute Eqs.\  \eqref{electric moment matrix} and \eqref{exponential approx} into Eq.\ \eqref{suscept def 2}, keeping only terms to first order in the numerator and  terms of zeroth order in the denominator. This gives us:
\begin{equation}
\chi=\frac{B}{E}\sum_n{\left(\braket{n^0|p_{E}|n^0}+ \frac{E}{kT} \braket{n^0|p_{E}|n^0}^2 - 2E\sum_{n'\neq n}{\frac{| \braket{ n'^0|p_{E}|n^0 }|^2}{W_n^0-W_{n'}^0}} \right)e^{-W_n^0/kT}},
\end{equation}
where $B \equiv N/\sum_n{e^{-W_n^0/kT}}$ (p.\ 190).  The first term, $B \sum_n{\braket{n^0|p_{E}|n^0}e^{-W_n^0/kT}}$, represents the average electric moment in the absence of an external field.  This kind of `hard' polarization is nonexistent in gases, so the term must be zero (p.\ 182).  We are then left with (p.\ 189):
\begin{equation}
\chi=B\sum_n{\left(\frac{\braket{n^0|p_{E}|n^0}^2}{kT} - 2\sum_{n'\neq n}{\frac{|\braket{ n'^0|p_{E}|n^0 }|^2}{W_n^0-W_{n'}^0}} \right)e^{-W_n^0/kT}},
\label{general suscept 0}
\end{equation}
or equivalently, in terms of the energy corrections (p. 182):
\begin{equation}
\chi=B\sum_n{\left(\frac{{W_n^{(1)}}^2}{kT} - 2 W_n^{(2)} \right)e^{-W_n^0/kT}}.
\label{general suscept}
\end{equation}
Eqs.\ \eqref{general suscept 0}--\eqref{general suscept} hold for any model of the constituent molecules of a gas.  Van Vleck used it as a starting point for all of his electric susceptibility calculations, including the most general derivation of the Langevin-Debye formula. However, from this point onward, we will focus on the special case of the rigid rotator (sec.\ 37, pp.\ 147--152). In sec.\ 5.2.2, we covered the calculations for this special case by \citet{Pauli_1921-zur-theorie} and \citet{Pauling_1926-the-quantum} in the old quantum theory and by \citet{Mensing_1926-uber-die-dielektrizitatkonstante} in the new quantum theory.\footnote{\citet{VanVleck_1932-the-theory} cited \citet{Mensing_1926-uber-die-dielektrizitatkonstante}, \citet{Kronig_1926-the-dielectric}, \citet{Manneback_1926-die-elektrizitaetskonstante}, and  \citet{VanVleck_1926_magnetic} at the beginning of sec.\ 37 (p.\ 147) and mentioned them again at the beginning of sec.\ 45 (p.\ 183). Cf.\ sec.\ 4 and note \ref{nature note}.}

The Hamiltonian for a rigid rotator in an external electric field $\bold{E}$ is given by (cf.\ Eq.\ \eqref{old quantum hamiltonian} in sec.\ 5.2.2):
\begin{equation}
H = \frac{L^2}{2I} - \mu E \cos \vartheta,
\label{new quantum hamiltonian}
\end{equation}
where $\bold{L}$  is the angular momentum and $I$ is the moment of inertia (cf.\ note \ref{elementary mechanics}). Consider the vectors $\ket{l, m}$, which are simultaneous eigenvectors of $L^2$ and $L_z$:
\begin{equation}
L^2 \ket{l, m} = \hbar^2 l(l+1) \ket{l, m}, \quad \quad L_z \ket{l, m} = \hbar m \ket{l, m},
\label{eigenvectors L and L_z}
\end{equation}
with  $l = 0, 1, \ldots$ and $-l \leq m \leq l$. Since $H^0 = L^2/2I$, these are also eigenvectors of the unperturbed Hamiltonian
\begin{equation}
H^0 \ket{l, m}=W_l^0\ket{l,m},
\label{rigid rotator hamiltonian}
\end{equation}
with ($(2l+1)$-fold degenerate) eigenvalues:
\begin{equation}
W_l^0  =\frac{\hbar^2}{2I} \, l(l+1)
\label{W0 lm}
\end{equation}
(cf.\ Eq.\ \eqref{W0 new} in sec.\ 5.2.2). The vector $\ket{l,m}$ corresponds to the wave functions $\psi_{lm}^0(\vartheta,\varphi)  \equiv \braket{\vartheta, \varphi | l, m}$ \citep[p.\ 160]{Baym_1969-lectures} given by (sec.\ 37, p.\ 149):
\begin{equation}
\psi_{lm}^0(\vartheta,\varphi)=\sqrt{ \frac{(2l+1)(l-m)!}{4\pi(l+m)!}}P_l^m(cos\,\vartheta)e^{im\varphi},\label{angular momentum wave equation}
\end{equation}
where the $P_l^m(x)$ are associated Legendre functions.  

The susceptibility for a gas of rigid rotators is given by (p.\ 182):
\begin{equation}
\chi= \frac{N}{\sum_l{\sum_m{e^{-W_l^0/kT}}}} \sum_l{\sum_m{\left( \frac{{W_{lm}^{(1)}}^2}{kT} - 2 W_{lm}^{(2)} \right) e^{-W_l^0/kT}}},
\label{general suscept lm}
\end{equation}
which is just the general Eq.\ \eqref{general suscept} for $\chi$ derived above with $l$ and $m$ rather than $n$ labeling the (degenerate) energy eigenstates. To find $\chi$, we need to find the first- and second-order energy corrections $W_{lm}^{(1)}$ and $W_{lm}^{(2)}$ to $W^0_l$. Replacing subscripts $n$ by $lm$ and vectors $\ket{n^0}$ by $\ket{l, m}$  in Eq.\  \eqref{W1 and W2} and substituting $H^{(1)} = - \mu \, \cos \vartheta$, we find (p.\ 152):
\begin{equation}
W_{lm}^{(1)}  = - \mu \braket{l, m | \cos \vartheta | l, m}, \quad \quad W_{lm}^{(2)} = \mu^2 \sum_{l'm' \neq lm}\frac{|\braket{ l', m' | \cos \vartheta | l, m }|^2}{W_l^0-W_{l'}^0}.
\label{W1 and W2 l m}
\end{equation}
These expressions can be evaluated with the help of the following characteristic recursion formula for associated Legendre functions (p.\ 151):
\begin{equation}
(2l+1)\cos \vartheta P_l^m(\cos \vartheta)=(l+m)P_{l-1}^m(\cos \vartheta)+(l-m+1)P_{l+1}^m(\cos \vartheta).
\label{recursion}
\end{equation}
Combining this recursion formula with Eq.\ \eqref{angular momentum wave equation}, we find (ibid.)
\begin{equation}
\cos \vartheta \, \psi^0_{lm}(\vartheta, \varphi) = A_{l-1, m} \; \psi^0_{l-1,m}(\vartheta, \varphi) + B_{l+1, m} \; \psi^0_{l+1,m}(\vartheta, \varphi),
\label{linear combination wave functions}
\end{equation}
where we introduced the abbreviations:
%the coefficients $A_{l-1, m}$ and $B_{l+1, m}$ are given by:
\begin{equation}
A_{l-1, m} \equiv \sqrt{\frac{l^2-m^2}{(2l-1)(2l+1)}}, \quad \quad
B_{l+1, m} \equiv \sqrt{\frac{(l+1)^2-m^2}{(2l+3)(2l+1)}}.
\label{A and B}
\end{equation}
For $l=0$, only the $B_{l+1, m}$ term is present. In terms of the corresponding state vectors, Eq.\ \eqref{linear combination wave functions} expresses that the vector obtained by letting the operator $\cos \vartheta$ act on $\ket{l,m}$ can be written as a linear combination of $\ket{l-1, m}$ and $\ket{l+1, m}$:\footnote{Taking the inner product with an arbitary vector $\ket{l', m'}$ on both sides of Eq.\ \eqref{linear combination state vectors}, we find: 
$$
\bra{l',m'} \cos \vartheta\ket{l, m}=  A_{l-1, m} \braket{l',m' | l-1,m}+ B_{l+1, m} \braket{l',m' | l+1,m}.
$$ 
In coordinate space, these inner products turn into integrals:
$$
\int d\omega \, \psi^{0\,*}_{l'm'} \cos \vartheta \, \psi^0_{lm} = A_{l-1, m} \int d\omega \, \psi^{0\,*}_{l'm'} \psi^0_{l -1 \, m}+ B_{l+1, m}  \int d\omega \, \psi^{0\,*}_{l'm'} \psi^0_{l + 1 \, m}
$$
where $d\omega \equiv d\vartheta d\varphi$ and where we suppressed the argument $(\vartheta, \varphi)$ of the various $\psi$ functions. Since $\psi^{0}_{l'm'}$ is arbitrary, this last relation implies Eq.\ \eqref{linear combination wave functions}, the form in which Van Vleck gave Eq.\ \eqref{linear combination state vectors}.}
\begin{equation}
\cos \vartheta \, \ket{l, m}=  A_{l-1, m} \, \ket{l-1,m}+ B_{l+1, m} \, \ket{l+1,m}.
\label{linear combination state vectors}
\end{equation}
Since $\ket{l', m'}$ is orthogonal to $\ket{l, m}$ as soon as $l' \neq l$ or $m' \neq m$, it follows immediately from Eq.\ \eqref{linear combination state vectors} that $W_{lm}^{(1)}$ in Eq.\ \eqref{W1 and W2 l m} vanishes, and that the only contributions to $W_{lm}^{(2)}$ come from terms with $(l' = l - 1, m' =m)$ and $(l' = l +1, m' = m)$, for which we have:
\begin{equation}
\braket{l-1,m | \cos \vartheta | l,m} = A_{l-1, m}, \quad \quad  \braket{l+1,m | \cos \vartheta | l,m} = B_{l+1, m}.
\label{cos expectation}
\end{equation}
For $l > 0$, the expression for $W_{lm}^{(2)}$ in Eq.\ \eqref{W1 and W2 l m} thus reduces to:
\begin{equation}
W_{lm}^{(2)} = \mu^2 \left( \frac{A^2_{l-1, m}}{W_l^0-W_{l-1}^0} + \frac{B^2_{l+1, m}}{W_l^0-W_{l+1}^0} \right).
\label{W2 lm 2}
\end{equation}
For $l=m=0$, only the second term is present. Eq.\ \eqref{A and B} tells us that $B^2_{0+1, 0} = 1/3$ and Eq. \eqref{W0 lm} that $W_0^0-W_1^0 = - \hbar^2/I$, which means that, for $l=m=0$, Eq.\ \eqref{W2 lm 2} gives (p.\ 152, p.\ 183):
\begin{equation}
W_{00}^{(2)} = \mu^2 \frac{B^2_{0+1, 0}}{W_0^0-W_1^0} = - \frac{I\mu^2}{3\hbar^2}.
\label{W2 lm zero}
\end{equation}
Using Eq.\ \eqref{A and B} for $A_{l-1, m}$ and $B_{l+1, m}$ and Eq.\ \eqref{W0 lm}  for $W_l^0$, Van Vleck showed that, for arbitrary non-zero values of $l$ and $m$, Eq.\ \eqref{W2 lm 2} becomes (ibid.):
\begin{equation}
W_{lm}^{(2)} = \frac{I\mu^2}{\hbar^2}\frac{l(l+1)-3m^2}{l(l+1)(2l-1)(2l+3)}.
\label{W2 lm non-zero}
\end{equation}

We now substitute these results for the energy corrections into Eq.\ \eqref{general suscept lm} for $\chi$. Since $W_{lm}^{(1)} =0$, the equation reduces to:
\begin{equation}
\chi=  \frac{-2N \sum_l{\sum_m{ W_{lm}^{(2)} e^{-W_l^0/kT}}}}{\sum_l{\sum_m{e^{-W_l^0/kT}}}}.
\label{general suscept lm simplified}
\end{equation}
Carrying out the sum over $m$ in th denominator, we can rewrite this as (p.\ 183):
\begin{equation}
\chi=\frac{-2N \sum_l{e^{-W_l^0/kT}} \sum_m{W_{lm}^{(2)} } }{\sum_l{(2l+1) e^{-W_l^0/kT} }}.
\label{simplified suscept}
\end{equation}
As we already saw in sec.\ 5.2.2, where we covered Mensing and Pauli's (1926) calculation for the rigid rotator (see Eqs.\ \eqref{emsemble average new 1}--\eqref{emsemble average new 2}), only the $l=0$ term in the summation over $l$ in the numerator gives a contribution to $\chi$. The terms for all other values of $l$ vanish. To verify this, we insert Eq.\ \eqref{W2 lm non-zero} for $W_{lm}^{(2)}$ ($l \neq 0$) in the sum over $m$ in Eq.\ \eqref{simplified suscept}:
\begin{equation}
\sum_m W^{(2)}_{lm} = \frac{I \mu^2}{\hbar^2} \; \frac{(2l+1)l(l+1) -3\sum_m m^2}{l(l+1)(2l-1)(2l+3)}.
\end{equation}
%\begin{eqnarray}
%\sum_m W^{(2)}_{lm}&=& \frac{I \mu^2}{\hbar^2} \sum_m \frac{l(l+1)-3m^2}{l(l+1)(2l-1)(2l+3)} \nonumber \\
%&=& \frac{I \mu^2}{\hbar^2} \; \frac{(2l+1)l(l+1) -3\sum_m m^2}{l(l+1)(2l-1)(2l+3)}.
%\end{eqnarray}
As Van Vleck noted (p.\ 183), the numerator in this last expression vanishes on account of the formula $3  \sum_m m^2 =l(l+1)(2l+1)$ (p.\ 152; cf.\ the sum-of-squares formula \eqref{sum of squares}). The entire susceptibility thus comes from the $l=0$ term. This fits with the classical theory for which  \citet{Alexandrow_1921-eine-bemerkung} had already shown that the susceptibility is due entirely to molecules with energies less than $\mu E$.   \citet{Mensing_1926-uber-die-dielektrizitatkonstante}, \citet{Kronig_1926-the-dielectric}, and \citet{VanVleck_1926_magnetic} had all noted with satisfaction earlier that the new quantum theory reverted to the classical theory in this respect (cf.\ sec.\ 5.2.2, especially note \ref{alexandrow}).  

Eq.\ \eqref{simplified suscept} thus reduces to the $(l=0)$-term (p.\ 184):
\begin{equation}
\chi=\frac{-2N \, e^{-W_0^0/kT} \, W_{00}^{(2)} }{\sum_l{(2l+1)e^{-W_l^0/kT}}}
= \frac{2NI \mu^2}{3\hbar^2}\frac{e^{-W_0^0/kT}}{\sum_l(2l+1)e^{-W_l^0/kT}},
\label{simplified suscept 00}
\end{equation}
where in the last step we used Eq.\ \eqref{W2 lm zero} for $W_{00}^{(2)}$. Since $kT>\!\!> W_0^0$ at the temperatures of interest, the Boltzmann factor in the numerator in this expression can be replaced by 1.  In the denominator, we use Eq.\ \eqref{W0 lm} for $W_l^0$. At sufficiently high temperatures $l\approx l+1$ for most terms in the sum, which can then be replaced by an integral (p.\ 185; cf. Eq.\ \eqref{int}):
\begin{equation}
\sum_{l}(2l+1)e^{-l(l+1)\hbar^2/2IkT}\approx \int_0^{\infty} 2 l \, e^{-l^2\hbar^2/2IkT} \, dl =\frac{2IkT}{\hbar^2}.
\end{equation}
With these approximations, Eq.\ \eqref{simplified suscept 00} becomes (p.\ 185):
\begin{equation}
\chi=\frac{N\mu^2}{3kT},
\label{end result}
\end{equation}
which is just the temperature-dependent term in the Langevin-Debye formula of the classical theory (see Eqs.\ \eqref{langevin debye formula}--\eqref{C}). Though the derivation above is for the special case of a gas of rigid rotators, \citet{VanVleck_1927_on-dielectric1, VanVleck_1932-the-theory} showed that the result holds under very general conditions in the new quantum theory {\it and does not involve spatial quantization}. And thus was the Kuhn loss in susceptibility theory recovered.\footnote{As \citet[pp.\ 222--223]{Born_1930-elementare} noted in their textbook, this ``was one of the first ``practical" successes of the new quantum mechanics. The methods of the old quantum theory [here a footnote is inserted citing \citet{Pauli_1921-zur-theorie}], in which a ``directional quantization" of the axes of the molecules had to be imposed, lead to a wrong numerical factor at high temperatures."}

Van Vleck did not bother to show explicitly that, despite appearances to the contrary, this derivation of the susceptibility of a gas of rigid rotators does not involve the choice of a preferred $z$-axis for the quantization of $L_z$. For Van Vleck this was just an instance of his  general theorem of spectroscopic stability (pp.\ 137--143). To bring out the role of this theorem in this specific case, we prove that the susceptibility of a gas of rigid rotators is indeed independent of our choice of a $z$-axis. In the calculation above, we used the orthonormal basis $\{ | l,m \rangle \}_{m=-l}^l$ to span the $(2l+1)$-dimensional subspace corresponding to the $(2l+1)$-fold degenerate energy eigenvalue  $W^0_l$ (see Eq.\ \eqref{W0 lm}). The number $m$ labels the different values of $L_z$ with respect to a $z$-axis chosen in the direction of the applied field $\bold{E}$. We can span that same subspace with a different orthonormal basis $\{ | l,r \rangle \}_{r=-l}^l$, where $r$ labels the different values of $L_z$ with respect to a  $z$-axis in some arbitrary direction. The vectors in the old basis can be written in terms of the new one:
\begin{equation}
| l, m \rangle = \sum_{r=-l}^l | l, r \rangle \langle l, r | l, m \rangle.
\label{onb}
\end{equation}
What we need to show is that the derivation of Eq.\ \eqref{end result} for the susceptibility of a gas of rigid rotators does not depend on whether we use $m$ or $r$ to label the degeneracy. More specifically, we need to check whether $\sum_m W^{(2)}_{lm}$ in Eq.\ \eqref{simplified suscept} is invariant under rotation of the $z$-axis, i.e., under switching from the orthonormal basis $\{ | l,m \rangle \}_{m=-l}^l$ to the orthonormal basis $\{ | l, r \rangle \}_{r=-l}^l$. Using Eq.\ \eqref{W1 and W2 l m}, we can write:
\begin{equation}
\sum_m W^{(2)}_{lm} =  \sum_{l' \neq l} \frac{\mu^2}{W_l^0-W_{l'}^0} \left( \sum_{m,m'} |\braket{ l', m' | \cos \vartheta | l, m }|^2 \right).
\label{sum W2 m}
\end{equation}
It is easy to show that $m$ and $m'$ in the expression in parentheses can be replaced by $r$ and $r'$:\footnote{The sum over $m$ and $m'$ in Eq.\ \eqref{sum W2 m} for fixed values of $l$ and $l'$ can be written as:
$$
\sum_{m,m'} |\braket{ l', m' | \cos \vartheta | l, m }|^2 = \sum_{m,m'} \braket{ l', m' | \cos \vartheta | l, m }\braket{ l, m | \cos \vartheta | l', m'}.
$$
With the help of Eq.\ \eqref{onb} we can write the vectors $| l, m \rangle$ in terms of the vectors $| l, r \rangle$:
$$
\sum_{m,m',r,r',\hat{r}, \hat{r}'} 
\braket{l',m' | l',r' } \braket{ l', r' | \cos \vartheta | l, r } \braket{l, r | l,m } \braket{ l,m | l,\hat{r}} \braket{ l, \hat{r} | \cos \vartheta | l', \hat{r}'} \braket{l', \hat{r}' | l', m'},
$$
where  $m, r, \hat{r}$ run from $-l$ to $l$ and $m', r', \hat{r}'$ run from $-l'$ to $l'$. Reordering the various factors, we find
$$
\sum_{m,m',r,r',\hat{r}, \hat{r}'}  \braket{l, r | l,m } \braket{ l,m | l,\hat{r}} \braket{l', \hat{r}' | l', m'} \braket{l',m' | l',r' } \braket{ l', r' | \cos \vartheta | l, r }  \braket{ l, \hat{r} | \cos \vartheta | l', \hat{r}'}. 
$$
Since $\sum_m \braket{l, r | l,m } \braket{ l,m | l,\hat{r}}  =  \braket{l, r | l,\hat{r}} = \delta_{r\hat{r}}$ and $\sum_{m'} \braket{l', \hat{r}' | l', m'} \braket{l',m' | l',r' } = \delta_{\hat{r}'r'}$, this reduces to
$$
\sum_{r,r'} \braket{ l', r' | \cos \vartheta | l, r }  \braket{ l, r | \cos \vartheta | l', r'} = \sum_{r,r'} |\braket{ l', r' | \cos \vartheta | l, r }|^2,
$$
which is what we wanted to prove.}
\begin{equation}
\sum_{m,m'} |\braket{ l', m' | \cos \vartheta | l, m }|^2 = \sum_{r,r'} |\braket{ l', r' | \cos \vartheta | l, r }|^2.
\label{sum W2 r}
\end{equation}
The derivation of the susceptibility thus does not depend on how the degeneracy in the energy levels $W^0_l$ is resolved.

To conclude this section, we consider some features of Van Vleck's more general derivation of the Langevin-Debye formula and how they relate to the hated ``bugbear'' of spatial quantization.  First, recall Eq.\,\eqref{n}, what Van Vleck called a ``sort of generalized Langevin-Debye formula" (p.\ 40).  The last step in obtaining this formula is the assumption that $\overline{\overline{p_z^2}}=\frac{1}{3} \overline{\overline{p^2}}$ (Eq.\ \eqref{average 1/3}), i.e., the mean square average of the unperturbed electric moment in the $z$-direction (the direction of the field \emph{even when the field is turned off}) is $1/3$ the mean square average of the total moment.  In the classical theory, this is exactly what one would expect. When the field is turned off, there should be equal contributions to the mean square of the moment for each spatial dimension. 
%(see Eq.\ \eqref{average 1/3} in sec.\ 5.2.1). 
This is exactly the feature, however, that was \emph{eliminated} by spatial quantization in the old quantum theory. This made it possible for molecules in high-energy states to contribute to the temperature-dependent term in the Langevin-Debye formula (see our discussion in sec.\ 5.2.2).

In the general quantum-mechanical derivation of the Langevin-Debye formula, Van Vleck ultimately produced a quantum-theoretical analogue of Eq.\,\eqref{n} (pp.\ 186--194).  This generalized formula hinges on an assumption analogous to Eq.\ \eqref{average 1/3} in the classical theory. In quantum mechanics, it takes the form (p.\ 140):\footnote{Instead of the angular momentum $\bold{L}$, Van Vleck considered a general vector quantity $\bold{A}$.}
\begin{equation}
\sum_{m,m'} | \langle l, m | L_z | l',m' \rangle |^2 = \frac{1}{3} \sum_{m,m'} | \langle l, m| L | l', m' \rangle |^2.
\end{equation}
As Van Vleck emphasized and as we showed explicitly in the case of Eq.\ \eqref{sum W2 r} above, relations such as these are clearly, as Van Vleck put it somewhat awkwardly, ``invariant of the choice of axis of quantization" (p.\ 140). This relation is just one example of the more general theorem of spectroscopic stability that Van Vleck was able to prove in quantum mechanics (pp.\ 137--143).  The upshot of this proof was that, in quantum mechanics, quantities like $\overline{\overline{p^2}}$ no longer depend on an axis of quantization as they had in the old quantum theory.

The strange story of the constant $C$ in the Langevin-Debye formula can ultimately be seen as the story of spatial quantization's brief rise and rapid fall.  The factors of $1/3$ in both the classical and quantum-mechanical formulas express that mean squares of vector components do not depend on the axes with respect to which those averages are taken.  In both theories, $\overline{\overline{p_z^2}}=\frac{1}{3} \overline{\overline{p^2}}$, where the $z$-direction can be arbitrarily chosen.  The strange values of $C$ in the old quantum theory resulted from the elimination of this very feature, which was essential if one wanted to derive the temperature-dependent term of the Langevin-Debye formula at all.  Without spatial quantization there simply was no temperature-dependent term in the old quantum theory. Unfortunately, spatial quantization came with a whole raft of problems. In light of this, we can clearly see why Van Vleck used the story of $C$ to illustrate the defects of the old quantum theory and the success of matrix mechanics in restoring the predictions of the classical theory.

\section{Kuhn Losses Revisited}

Both Van Vleck's  1926 {\it Bulletin} and his 1932 book do what Kuhn said good textbooks should do: they clearly lay out the principles and the formalism of the theories they cover
%espouse 
and show how these theories can be used to solve a number of canonical problems, thus training their readers to become researchers in the relevant fields. 
%covered. 
Yet they do so without paying the price  \citet[p.\ 137]{Kuhn_1996_the-structure} suggested was unavoidable: though written in the midst or in the aftermath of a period of major conceptual upheaval, they do not ``disguise \ldots\ the role [and] the very existence" of this upheaval nor do they ``truncat[e] the scientist's sense of his discipline's history."

This is especially striking in the case of the 1932 book. Van Vleck spent roughly a third of his book (121 out of a total of 373 pages) on the classical theory (Chs.\ I--IV) and the old quantum theory (Ch.\ V). One might argue that Ch.\ V served a purely rhetorical purpose. The old quantum theory's problems with susceptibilities are a great foil for the new quantum mechanics' successes in that same area. Such use of history in a textbook can readily be reconciled with Kuhn's views. There are two further considerations regarding this chapter that would seem to be in Kuhn's favor. First, the history recounted in Ch.\ V  is somewhat misleading in that Van Vleck, inadvertently or deliberately, made it sound as if there had been reliable experimental evidence disproving the ``wonderful nonsense" produced by the old quantum theory all along.  In fact, such evidence had only become available around the time of the theory's demise. Second, we know that Van Vleck wanted to cut Ch.\ V to make room for new material when he began revising his book for a second edition decades later. He had no such plans, however, for Chs.\ I--IV on the classical theory.

The pedagogical goal of those early chapters was not merely to provide propaganda for the superior quantum-mechanical treatment of susceptibilities. Rather, their main function was to prepare the reader for the quantum-mechanical calculation of susceptibilities by showing how such calculations are done in the classical theory. In his biographical memoir about his teacher, \citet[p.\ 509]{Anderson_1987-john} noted that this approach might not be suited for ``a modern text for physicists poorly trained in classical mechanics" (see sec. 1.2.3). In the early 1930s, however, Van Vleck could certainly assume his intended readers to be well versed in classical mechanics. 

Using older theories for pedagogical purposes in this way is not compatible with Kuhn's picture. A new paradigm is supposed to come with its own new suite of tools for the pursuit of normal science. It is supposed to provide its own new set of {\it exemplars} to ``show [students] by example how their job is to be done" \citep[p.\ 187; discussed in sec.\ 1.2.2]{Kuhn_1996_the-structure}.

Van Vleck's book provides a clear example of such an exemplar. It gives a general recipe with many concrete illustrations of how one can calculate susceptibilities, say the electric susceptibility of a gas. First, one has to decide on a mechanical system to model the constituent molecules of the gas. This can be a specific system (e.g., a rigid rotator) or a generic one (a classical multiply-periodic system solvable in action-angle variables, a quantum system with an energy spectrum satisfying some not overly restrictive conditions). One then has to do a perturbative calculation to compute the time-average of the component of the electric dipole moment in the direction of the external field of one copy of this system in a given state. Finally, one has to take the average of this time-average for an individual system over a thermal ensemble of many such systems in all possible states. 

This general procedure works in classical theory, in the old quantum theory, and in modern quantum mechanics. The exemplar thus cuts across two paradigm shifts! Suman \citet[pp.\ 265--267]{Seth_2010-crafting} makes a similar point, contrasting a continuity of {\it problems} with a discontinuity in {\it principles} (see our discussion in sec.\ 1.2.3).

That the techniques from statistical mechanics for the calculation of ensemble averages work in all three theories does not seem to call for further comment. That this is also true for the perturbative techniques used to calculate the relevant time-averages is less obvious. Ultimately, it is because of the continuity of the underlying formalism. The perturbative techniques were originally developed in the context of celestial mechanics. They were adapted to deal with atomic mechanics, to use Born's  phrase (see note \ref{atomic mechanics}), in the old quantum theory. A large part of Van Vleck's NRC {\it Bulletin} on the old quantum theory was devoted to these techniques, which were used to derive classical expressions that could then be translated into quantum expressions under the guidance of the correspondence principle, according to which the quantum expression would have to merge with the classical one in the limit of high quantum numbers. The derivation of the Kramers dispersion formula is a prime example of this strategy (see sec.\ 2.3). In the {\it Dreim\"annerarbeit}, Born, Heisenberg, and Jordan (1926) adapted these perturbative techniques to the new matrix mechanics.

What lay behind and made possible this continuity of technique was a remarkable continuity of formalism in the transition from classical to quantum physics. Neither the old nor the new quantum theory did away with classical mechanics. The old quantum theory just added the Sommerfeld-Wilson quantum conditions to select a subset of the classically allowed motions. For some specific simple systems, notably the one-electron atom and the harmonic oscillator, this led to satisfactory results (although even the zero-point energy of a simple harmonic oscillator had to be added by sleight of hand). In other cases, multi-electron atoms or the rigid rotator, it did not. The old quantum theory actually was at its best, if generic multiply-periodic systems could be used, such as in the derivation of the Kramers dispersion formula. In those cases one could sometimes find the quantum counterpart of a classical formula through educated guesswork guided by the correspondence principle. As we saw in sec.\ 1.2, Van Vleck (1926a, p.\ 227; 1932b, p.\ 107) also appealed to the correspondence principle to {\it reject} formulas for susceptibilities produced in the old quantum theory on the grounds that they did not reduce to the Langevin-Debye formula at high temperatures, where that classical formula ought to hold. In that case, however, the correspondence principle did not suggest a better candidate for a quantum formula for susceptibilities. 

The ``wonderful nonsense" produced on this score by \citet{Pauli_1921-zur-theorie} and \citet{Pauling_1926-the-quantum} mercilessly reveals the limitations of the old quantum theory's basic approach---imposing quantum conditions on classical mechanics. Their calculations gave nonsensical results, not because the general procedure for calculating susceptibilities described above does not work in the old quantum theory, but because of the way they quantized the angular momentum of the rigid rotator, their model for polar molecules such as HCl. The problem was twofold. First, instead of the relation $L^2 = l(l+1)\hbar^2$ ($l = 0, 1, 2, \ldots$), sanctioned by modern quantum mechanics, they used $L^2 = l^2 \hbar^2$ (Pauli with integer values, Pauling with half-integer values for $l$, where $l=0$ is forbidden in both cases). As a result, the ensemble average $\overline{\overline{L^2}}$ is not equal to three times the ensemble average $\overline{\overline{L_z^2}}$ in the old quantum theory, whereas this relation does hold both  in the classical theory and in quantum mechanics. Second, they saw themselves forced to adopt what Van Vleck (1927a, p.\ 37; 1932b, p.\ 110)
%\citet[p.\ 110]{VanVleck_1932-the-theory} 
later derided as the ``bugbear" of spatial quantization.

Matrix mechanics, the incarnation of quantum mechanics that Van Vleck was most familiar and most comfortable with, retained the formalism of classical mechanics {\it without} inflicting this kind of disfigurement. This is remarkable because, unlike the old quantum theory, it radically changed the interpretation of the formalism. The basic idea of Heisenberg's {\it Umdeutung} was to conceive of the quantities related by the laws of classical mechanics as arrays of numbers.  In more mature versions of the theory, these became matrices and then operators acting in Hilbert space. Unlike the old quantum theory, the new quantum mechanics came with a systematic prescription for imposing quantum conditions. It replaced the Sommerfeld-Wilson quantum conditions by the basic commutation relations of position and momentum.  As Paul \citet{Dirac_1925-the-fundamental} first pointed out, these were the quantum analogues of Poisson brackets in classical mechanics. The recovery of the Langevin-Debye formula for the electric susceptibility in gases, a Kuhn loss of the old quantum theory, beautifully illustrates the advantages of the new quantum theory over the old. Looking at the situation from this perspective, one readily understands Van Vleck's assessment at the beginning of the chapter on the old quantum theory in his 1932 book: ``there is perhaps no better field than that of electric and magnetic susceptibilities  to illustrate the inadequacies of the old quantum theory and how they have been removed by the new mechanics" \citep[p.\ 105]{VanVleck_1932-the-theory}.

Van Vleck saw these issues clearly only in retrospect.  When he took the time to list and discuss the various flaws of the old quantum theory in his 1926 \emph{Bulletin}, he did not include its failure to give a sensible result for electric susceptibilities.  As we have seen, this was not because of ignorance (he had read the key paper by Pauli [1921] as a graduate student), but rather because of the intense focus of physicists at the time on spectroscopic phenomena. We began our paper with a quotation from an article in a chemistry journal, in which \citet[p.\ 493]{VanVleck_1928-the-new2} characterized physicists as being ``entranced by spectral lines,'' willing to ignore the peripheral phenomena of electric and magnetic susceptibilities. In 1925 Van Vleck had been such a physicist. All of this changed as he began to focus his research on electric and magnetic susceptibilities and came to understand that some of the old quantum theory's most serious flaws and some of the new quantum theory's most remarkable successes were in areas that had hardly attracted any attention before.  When Van Vleck told the chemists that physicists tend to close their eyes to phenomena other than spectra, he was also admonishing himself.

\begin{singlespacing}

\section*{Acknowledgments}

We want to thank Massimiliano Badino, Rich Bellon, Tony Duncan, Fred Fellows, Clayton Gearhart, Don Howard, David Huber, Jeremiah James, Luc Janssen, Christian Joas, Marta Jordi, David Kaiser, Sally Gregory Kohlstedt, Christoph Lehner, Chun Lin, Joe Martin, Jaume Navarro, J\"urgen Renn,  Serge Rudaz, Rob ``Ryno" Rynasiewicz, and Roger Stuewer for helpful  comments, discussion, and references. We are grateful to John Comstock for permission to use the pictures in Figs.\ 2 and 3. Work on this paper was supported by J\"urgen Renn's department of the {\it Max-Planck-Institut f\"ur Wissenschaftsgeschichte}.

\end{singlespacing}

%\nocite{*}

%Note on the Postulates of the Matrix Quantum Dynamics. PNAS 1926 12 (6) 385--388

%The Correspondence Principle in the Statistical Interpretation of Quantum Mechanics. PNAS 1928 14 (2) 178--188

\bibliographystyle{apalike}
\begin{singlespacing}
\bibliography{QuantumTextbooksVanVleck}
\end{singlespacing}

\end{document}